\documentclass[preprint]{aastex}
\shorttitle{GMBCG cluster catalog for SDSS DR7}

\shortauthors{Hao et al.}

\usepackage[usenames]{color}
\usepackage{url}
\usepackage{color}

\begin{document}

\title{A GMBCG Galaxy Cluster Catalog of 55,424 Rich Clusters from SDSS DR7}

\author{Jiangang Hao\altaffilmark{1}, Timothy A. McKay\altaffilmark{2,3}, Benjamin P. Koester\altaffilmark{4}, Eli S. Rykoff\altaffilmark{5,6}, Eduardo Rozo\altaffilmark{7}, James Annis\altaffilmark{1}, Risa H.
Wechsler\altaffilmark{8}, August Evrard\altaffilmark{2,3}, Seth R. Siegel\altaffilmark{2}, Matthew Becker\altaffilmark{10}, Michael
Busha\altaffilmark{8}, David Gerdes\altaffilmark{2}, David E.
Johnston\altaffilmark{1} and Erin Sheldon\altaffilmark{11}}

\altaffiltext{1}{Center for Particle Astrophysics, Fermi National
  Accelerator Laboratory, Batavia, IL 60510}
\altaffiltext{2}{Department of Physics, University of Michigan, Ann
  Arbor, MI 48109} \altaffiltext{3}{Department of Astronomy,
  University of Michigan, Ann Arbor, MI 48109}
\altaffiltext{4}{Department of Astronomy and Astrophysics, The
  University of Chicago, Chicago, IL 60637} 
\altaffiltext{5}{TABASGO  Fellow, Physics Department, University of California at Santa
  Barbara, Santa Barbara, CA 93106}
\altaffiltext{6}{Physics Division, Lawrence Berkeley National Laboratory, Berkeley, CA 94720}
\altaffiltext{7}{Einstein and KICP Fellow, Kavli Institute for
  Cosmological Physics, The University of Chicago, Chicago, IL 60637}
\altaffiltext{8}{Kavli Institute for Particle Astrophysics \&
  Cosmology, Physics Department, and SLAC National Accelerator
  Laboratory, Stanford University, Stanford, CA 94305}
\altaffiltext{9}{Department of Physics, California Institute of
  Technology, Pasadena CA 91125} 
\altaffiltext{10}{Department of
  Physics, The University of Chicago, Chicago, IL 60637}
\altaffiltext{11}{Brookhaven National Laboratory, Upton, New York
  11973}

\begin{abstract}
  We present a large catalog of optically selected galaxy clusters from the application of a new Gaussian Mixture Brightest Cluster Galaxy (GMBCG) algorithm to SDSS Data Release 7 data. The algorithm detects clusters by identifying the red sequence plus Brightest Cluster Galaxy (BCG) feature, which is unique for galaxy clusters and does not exist among field galaxies. Red sequence clustering in color space is detected using an Error Corrected Gaussian Mixture Model. We run GMBCG on 8240 square degrees of photometric data from SDSS DR7 to assemble the largest ever optical galaxy cluster catalog, consisting of over 55,000 rich clusters across the redshift range from $0.1 < z <  0.55$. We present Monte Carlo tests of completeness and purity and perform cross-matching with X-ray clusters and with the maxBCG sample at low redshift. These tests indicate high completeness and purity across the full redshift range for clusters with 15 or more members. The catalog can be accessed from the following website: \color{blue}\url{http://home.fnal.gov/~jghao/gmbcg_sdss_catalog.html}. 
\end{abstract}

\keywords{Galaxies: clusters, Catalog- Cosmology: observations - Methods:
Data analysis, Gaussian Mixture}

\section{Introduction}

One of the most exciting discoveries in physics and astronomy over the
past decade is the accelerating expansion of the
Universe~\citep{perlmutter98,riess98}, which has been more recently
confirmed by a series of independent
experiments~\citep{spergel03,spergel07,tegmark04,eisenstein05}. This
cosmic acceleration cannot be explained without exotic physics, for
example, modifications to General Relativity (GR), a cosmological
constant, or an additional energy component with negative pressure
adequate to drive acceleration. Perhaps the simplest possibility, a
cosmological constant, is consistent with all available data, although
the theoretical challenges with this explanation have not been
resolved. If the framework of GR is retained without a cosmological
constant, something like dark energy must exist. In an effort to
distinguish between these possibilities, studies of expansion history
and the growth of structure have become central research topics in
physics and astronomy.

One way to test theories of expansion and the growth of structure is
to measure the abundance and properties of galaxy clusters. Clusters
are the largest peaks in the density field. Their abundance and
spatial distribution encode rich information about the
Universe~\citep{evrard89,oukbir92}, making them sensitive probes for
cosmology~\citep{mohr04self,hu04self,lima04,lima05}. Cosmological
constraints from optically selected galaxy clusters have been carried
out recently by ~\citet{gladders07} based on the RCS cluster
catalog~\citep{gladders05rcs}, by~\citet{rozoa07,rozob07, rozo10},
based on the maxBCG catalog~\citep{koester07cat,koester07alg} and by~\citet{wen10} based on a cluster catalog assembled based on photometric redshift~\citep{wen09}.

Galaxy clusters are observationally rich as well.  They can be
detected based on their properties determined using a number of different
observables, including X-ray emission from and the Sunyaev-Zeldovich
decrement caused by hot intracluster gas, optical and NIR emission
from stars in cluster galaxies, and the gravitational lensing
distortions imposed on background galaxy images by the total cluster
gravitational potential.  Each probe relies on different aspects of
cluster physics and provides different, though often correlated,
information about cluster mass and structure. For cluster detection,
the different probes have complementary virtues. Cluster X-ray
emission and the SZ decrement both require the presence of very hot
intracluster gas. This can only be present in very deep potential
wells, so these methods only detect the highest mass systems, but are
consequently relatively free from projection contamination.
Unfortunately, neither very naturally provides information about
cluster redshift, so optical follow-up is required.  Cluster searches
using optical data are more able to identify clusters in three
dimensions, obtaining distances as part of cluster detection.  Optical
selection can identify systems corresponding to much lower mass dark
matter halos than methods based on the intracluster gas, but this also
results in more serious projection effects.  Cluster detection in the
optical also benefits from the high signal to noise for individual
galaxy detection and large data volumes available in optical surveys.

The existence of a uniformly old stellar population in many cluster galaxies gives them remarkably similar spectral energy distributions which include a strong 4000~\AA break. As a result, galaxies within clusters are tightly clustered in color as well as space. When the cluster redshift increases, this break shifts across the optical filters, creating a strong correlation between cluster galaxy color and redshift. It has been shown that red-sequence galaxies exist in clusters of varied richness
and extend to redshift $z \sim 1.6$~\citep{bower92,smail98,barrientos99,mullis05,eisenhardt05,papovich10}. Red
sequence galaxies are a very prominent feature of galaxy clusters and
thus provide a very powerful means for removing projected field
galaxies during cluster detection. As these red sequence galaxies have
mostly E and S0 morphologies, dominate the bright end of the cluster
luminosity function ~\citep{sandage85,barger98}, and exhibit narrow
color scatter ( $\sim 0.05$ in $g-r$ and $r-i$ colors in the redshift range we probe), they are also referred to as the E/S0 ridgeline~\citep{visvanathan77,annis99}. For reviews of red sequence
galaxies in clusters, refer to ~\citet{gladders00}, ~\citet{haoecgmm}
and references therein.

In this paper, we extend the use of red sequence galaxies and
brightest cluster galaxies (BCG) for cluster detection, and develop an
efficient cluster finding algorithm which we name the Gaussian Mixture
Brightest Cluster Galaxy (GMBCG) method. The algorithm uses the Error
Corrected Gaussian Mixture Model (ECGMM) algorithm~\citep{haoecgmm} to
identify the BCG plus red sequence feature and convolves the
identified red sequence galaxies with a spatial smoothing kernel to
measure the clustering strength of galaxies around BCGs.  We apply
this technique to the Data Release 7 of Sloan Digital Sky Survey and
assemble a catalog of over 55,000 rich galaxy clusters in a redshift
range extending from $0.1 < z < 0.55$. The catalog is approximately
volume limited up to redshift $z \sim 0.4$  and shows high purity and
completeness when tested against a mock catalog. The algorithm is very
efficient, producing a cluster catalog for the full SDSS DR7 data
($\sim$ 8,000 ${\rm deg}^2$) within 23 hours on a single modern
desktop computer.

Cluster finding algorithms are closely related to the properties of
the data they are applied to.  Therefore, we begin with a general
description of the GMBCG algorithm, then add additional features that
are particular to its application to the SDSS data. The paper is
organized as follows: in \S~\ref{deproject}, we review
de-projection, the major challenge of optical cluster detection,
summarizing the de-projection methods used in previous cluster finding
algorithms and demonstrating why red sequence color outperforms the
others. In \S~\ref{gmbcg}, we introduce the major steps of the
GMBCG algorithm and compare it with the maxBCG algorithm. In
\S~\ref{catdr7}, we introduce the cluster catalog we constructed
from the SDSS DR7 using the GMBCG algorithm. In \S~\ref{cattest},
we evaluate this new DR7 catalog by matching it to catalogs of known
X-ray clusters and previously published maxBCG clusters. The
completeness and purity of the GMBCG catalog are then also tested
against a mock catalog.  We conclude with a summary of the properties
of the GMBCG catalog, along with a discussion of the prospects for
using this method on future optical surveys.

By convention, we use a $\Lambda$CDM cosmology with $h=1$,
$\Omega_m=0.3$ and $\Omega_{\Lambda}=0.7$ throughout this paper. Also,
we will omit the $h^{-1}$ when describing distances, i.e., we will use
Mpc directly instead of $h^{-1}$Mpc.

\section{Optical Galaxy Cluster Detection and De-projection}\label{deproject}

Our goal is to detect galaxies clustered in three spatial dimensions,
but we have precise information in only two: RA and DEC. Large
uncertainties in galaxy position along the line of sight leads to
projections which contaminate richness estimates for all clusters and
confuse cluster detection at low richness. Therefore, every optical
cluster finding algorithm needs to effectively de-project field
galaxies before calculating overdensities in the RA/DEC plane.

The ability to locate the positions of galaxies along the line of
sight is limited by the technology available. Over the past 60 years,
various algorithms for optical galaxy cluster detection based on
photometric data have been employed
~\citep{abell57,huchra82,davis85,shectman85,efstathiou88,couch91,lidman96,
  postman96,kepner99,annis99,gladders00,gladders05,gal00,gal03,kim02,
  goto02,ramella02,lopes04,botzler04,koester07alg,li08,wen09}.\footnote{When
  spectroscopic redshifts are available, other algorithms have been
  developed, for example,~\citet{berlind06,yang07, miller05}. In this
  paper, we will mainly consider the algorithms based on photometric
  data.}  For a recent review of the cluster finding algorithms,
see~\citet{gal06}. Though these methods differ in many detailed
respects, we can roughly classify them according to the de-projection
methods they use. In Table.~\ref{table:algorithms}, we list the
cluster finding algorithms for photometric data of the past two decades and the
de-projection methods used.

\begin{table}
\caption{Summary of optical cluster finding algorithms for photometric data}\label{table:algorithms}
\begin{minipage}[b]{1\linewidth}
\footnotesize
\begin{tabular}{l l l}
\hline\hline Algorithm & Type of data applied & De-projection method\\
\hline
   Percolation\footnote{~\citet{huchra82,davis85,efstathiou88,ramella02}}& Single band/Simulation &   Magnitude/photo-$z$ \\
   Smoothing Kernels\footnote{~\citet{shectman85}} & Single band&  Magnitude\\
   Adaptive Kernel\footnote{~\citet{gal00,gal03}} &   Single band &   Magnitude\\
   Matched Filter \footnote{~\citet{postman96}} & Single band & Magnitude\\
   Hybrid and Adaptive Matched Filter\footnote{~\citet{kepner99,kim02,dong08}} & Single band & Magnitude/photo-$z$\\
   Voronoi Tessellation\footnote{~\citet{kim02,lopes04}} & Single band & Magnitude\\
   Cut-and-Enhance \footnote{~\citet{goto02}} & Single band & Magnitude\\
   Modified Friends of Friends\footnote{~\citet{li08}} & Multi-band & Photo-$z$\\
   C4\footnote{~\citet{miller05}} & Multi-band & All Colors\\
   Percolation with Spectroscopic
   redshift\footnote{~\citet{berlind06}} & Multi-Band &
   Spectroscopic Redshift \\
   Cluster Red Sequence\footnote{~\citet{gladders00,gladders05}} & Multi-band & Red sequence\\
   MaxBCG \footnote{~\citet{annis99,koester07cat,koester07alg}}& Multi-band & Red sequence\\
   WHL\footnote{~\citet{wen09}}& Multi-band & Photo-$z$\\
   GMBCG\footnote{~\citet{haogmbcg1,haogmbcg2,haothesis}} & Multi-band & Red sequence\\
\hline
\end{tabular}
 \vspace{-7pt}\renewcommand{\footnoterule}{}
\end{minipage}
\end{table}

The de-projection method used by each algorithm is often determined by
the properties of the data for which the algorithm was developed. When
only single band data were available the major de-projection methods
were all magnitude based. However, the broad luminosity function of
galaxies makes magnitude a poor indicator of galaxy position along the
line of sight. Even so, these methods are quite effective for
detecting massive clusters. Unfortunately, they cannot maintain good
purity and completeness for clusters with low or intermediate
richness. Moreover, the contamination of cluster richness induced by
projection also creates large scatter in the richness-mass relations
derived from these methods.

Multi-band digital imaging technology greatly alleviates the
projection effects that plagued optical galaxy cluster detection for
decades. In a precise multi-band sky survey, we have magnitude
information from more than one band, allowing better reconstruction of
the galaxy spectra. Even the crude Spectral Energy Distribution (SED)
information provided by colors provides very effective information for
locating galaxies along the line of sight.  

The red sequence, or E/S0 ridgeline, which defines cluster galaxies,
has a very narrow color scatter ( $\sim 0.05$ in $g-r$ and $r-i$ colors) and a slightly tilted color magnitude relation, the study of which has a long history,
e.g.~\citep{visvanathan77, bower92, gladders98, lcyee04, blakeslee03,
  blakeslee06, delucia07, stott09, mei09, haoecgmm}.  This color
information is the primary tool to determine the position of galaxies
along the line of sight.

There are basically two ways to de-project galaxies using multi-color
data: use the colors to obtain photometric redshifts and then
de-project using these redshifts, or use the red sequence to detect
clustering directly in color space. The first approach is
straightforward in principle, but more complex in practice. There are
many machine learning algorithms~\citep{oyaizu07,gerdes09} that can be
used to assign photo-$z$s based on the multi-band
colors/magnitudes. However, these methods are limited by the available
training set of spectroscopic redshifts.  For galaxies that are
similar to the training set, reconstructed photo-$z$s can reach a
precision of $\sim 0.03$~\citep{oyaizu07}. However, for galaxies that
are not represented in the training set, photo-$z$s can be very
imprecise and biased.

\begin{figure*}
\begin{center}
\includegraphics[width=3 in, height=2.5in]{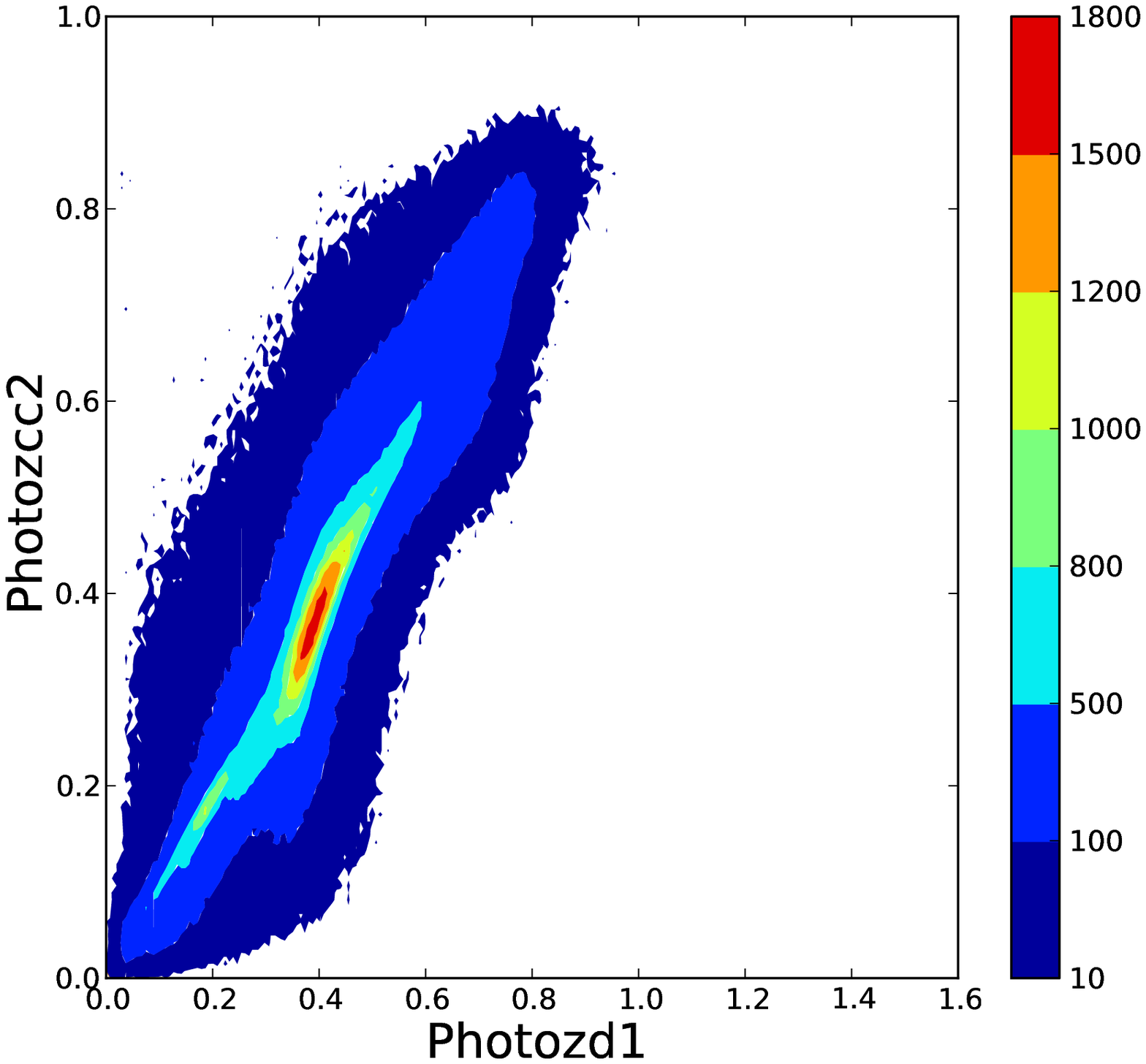}
\includegraphics[width=2.5 in, height=2.5in]{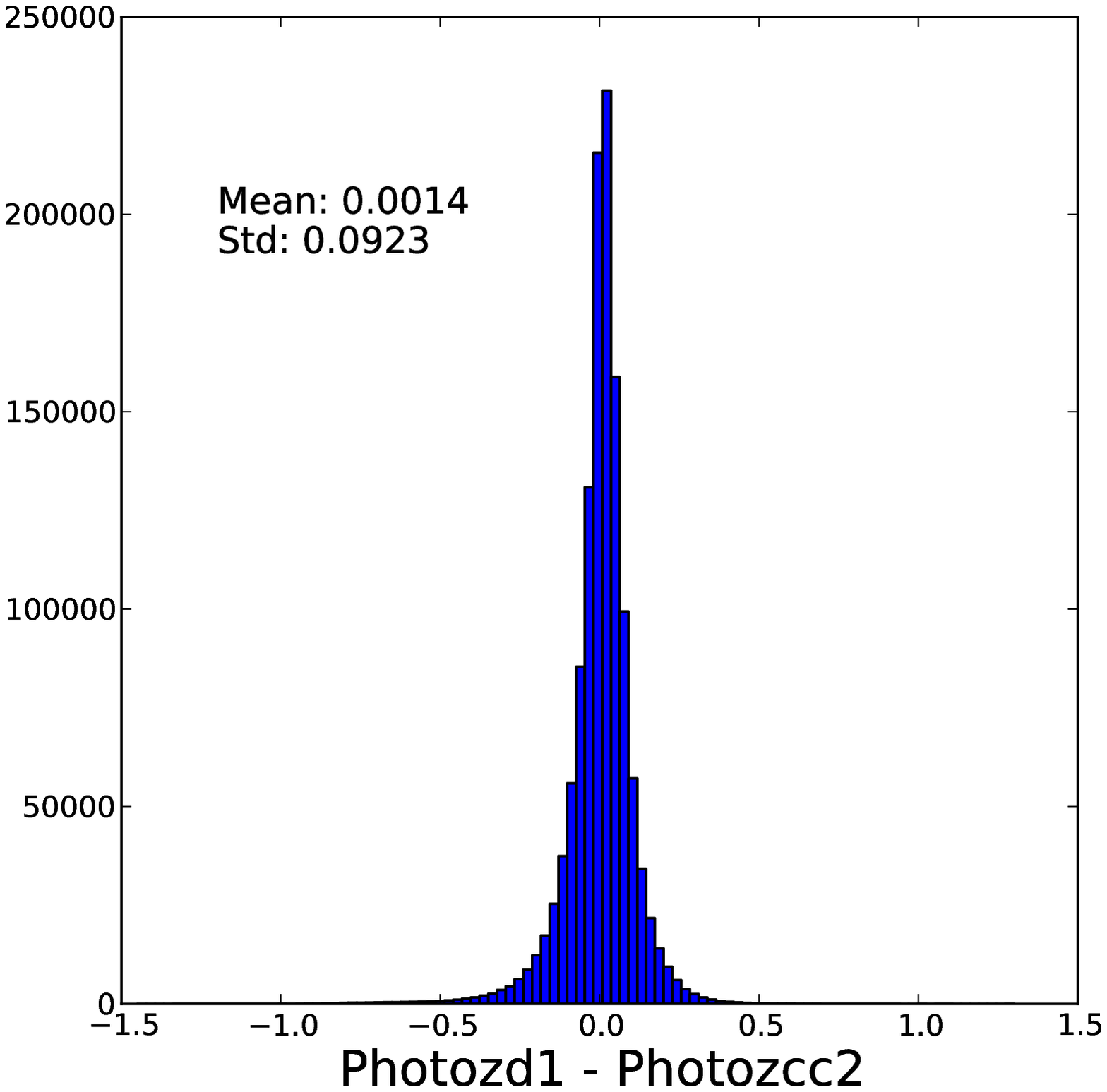}
\caption{Scatter between two well-trained photo-$z$ estimators.  Two
  neural network algorithms from ~\cite{oyaizu07} (photo-$z$d1, which
  uses magnitudes only, and photo-$z$cc2, which uses galaxy colors)
  applied to SDSS DR6 data are compared.  Left panel plots the two
  estimators against each other for the full sample, right panel shows
  the scatter between the two estimators.  Although the algorithms are
  well tuned using existing spectroscopic data, the two photo-$z$s
  have an rms difference of $\sim 0.1$.}
\label{fig:photozbias}
\end{center}
\end{figure*}

To get a sense of how photo-$z$s perform for all galaxies (up to 21 magnitude in I-band), we can simply compare the results of two different estimators. Take the neural network photo-$z$s for SDSS data~\citep{oyaizu07} as an
example. There are two well-tested estimators provided in the SDSS
catalogs, labeled {\tt photo-$z$d1} and {\tt photo-$z$cc2}.  The {\tt
  photo-$z$d1} is obtained by training only on magnitudes, while {\tt
  photo-$z$cc2} is obtained by training only on colors. In
Figure~\ref{fig:photozbias}, we compare photo-$z$s based on these two
estimators. The difference of the two photo-$z$s has a standard
deviation of $\sim 0.1$.  For a typical cluster, with a velocity
dispersion of 900 km $s^{-1}$, the dispersion between galaxy redshifts
is $\pm 0.003$, much smaller than the precision possible from
photo-$z$s alone. Therefore, though it is a lot better than the
magnitude based de-projection, photo-$z$ de-projection will still be
insufficient to remove projection effects, especially when we probe slightly fainter cluster populations.

As an alternative, we may stay closer to the data and look for
clustering directly in color space. Red sequence galaxies in low
redshift clusters display a scatter in $g-r$ color of $\sim
0.05$. Most importantly, a tight cluster red sequence accompanied by a
BCG presents a pattern exhibited only by clusters and not found in
field galaxies. Therefore, directly looking for the red sequence plus
BCG feature provides a powerful way to improve cluster detections. It
is this approach which we follow in the GMBCG method. \footnote{One may wonder why do red sequence colors do better than photo-$z$s that are essentially derived from colors. In particular, photo-$z$s are obtained by using multi-color/magnitudes while the ridgeline color is only one color. This would suggest that photo-$z$s should do better than red sequence colors. However, looking at the problem closely, one can immediately realize that there are two additional information associated with red sequence color de-projection. The first is the spatial proximity/clustering and the second is the discrimination of red and blue galaxies. For this combination of reasons, de-projection using red sequence colors out-performs de-projection using photo-$z$s. As a result, we can push the cluster detection to lower richness limits than we can do using photo-$z$s. For very big clusters, one can find them with any means. But for lower richness systems, appropriate de-projection is crucial for detection and richness measurement. For cosmology, clusters with a wide mass and redshift ranges will provide substantially more leverage on the constraints on cosmological parameters.}

\section{Details of the GMBCG Algorithm for Optical Cluster Detection}\label{gmbcg}
\subsection{Overview}

\begin{figure*}
\begin{center}
\includegraphics[width=5in, height=2.5in]{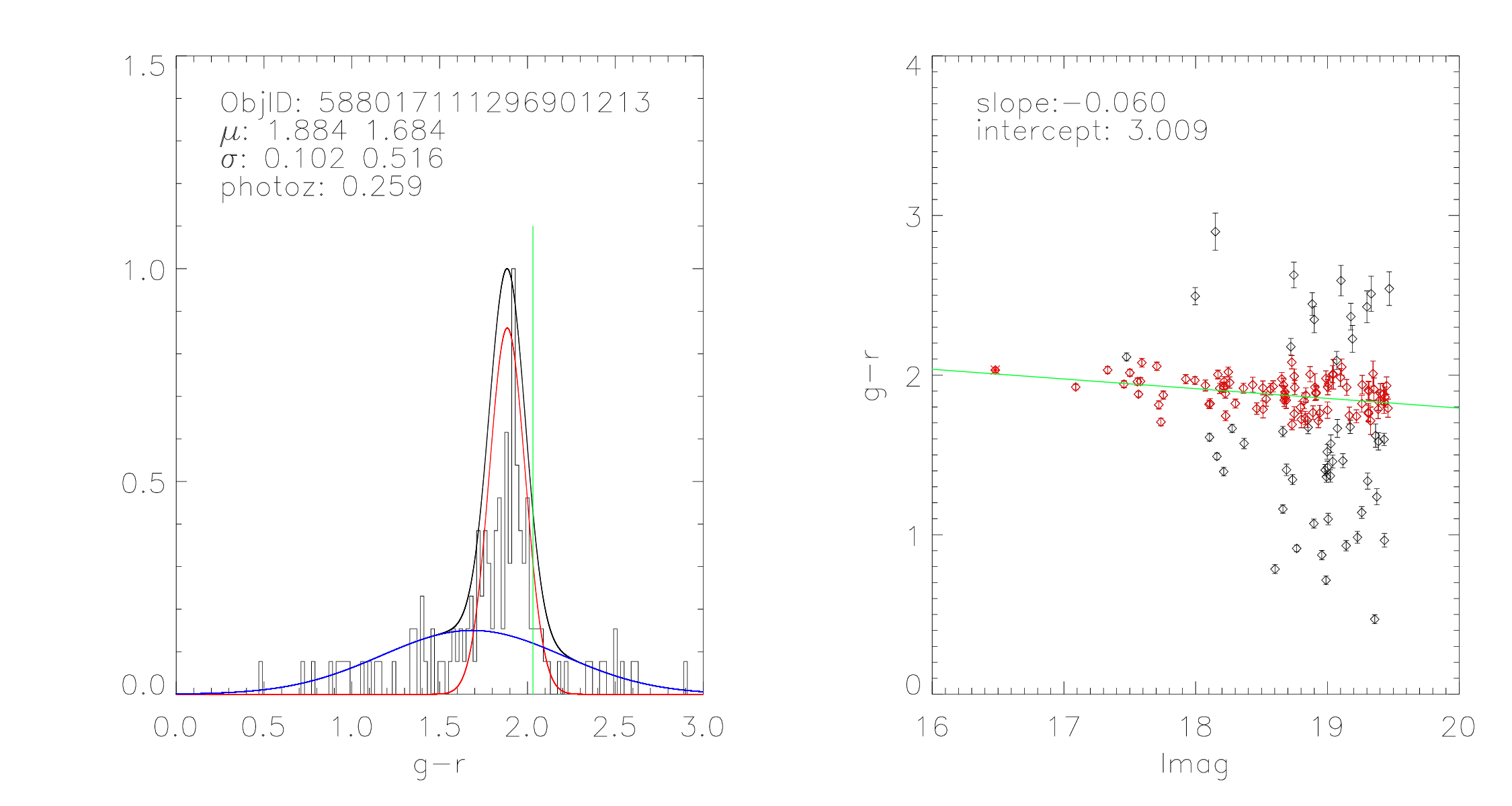}
\includegraphics[width=5in, height=2.5in]{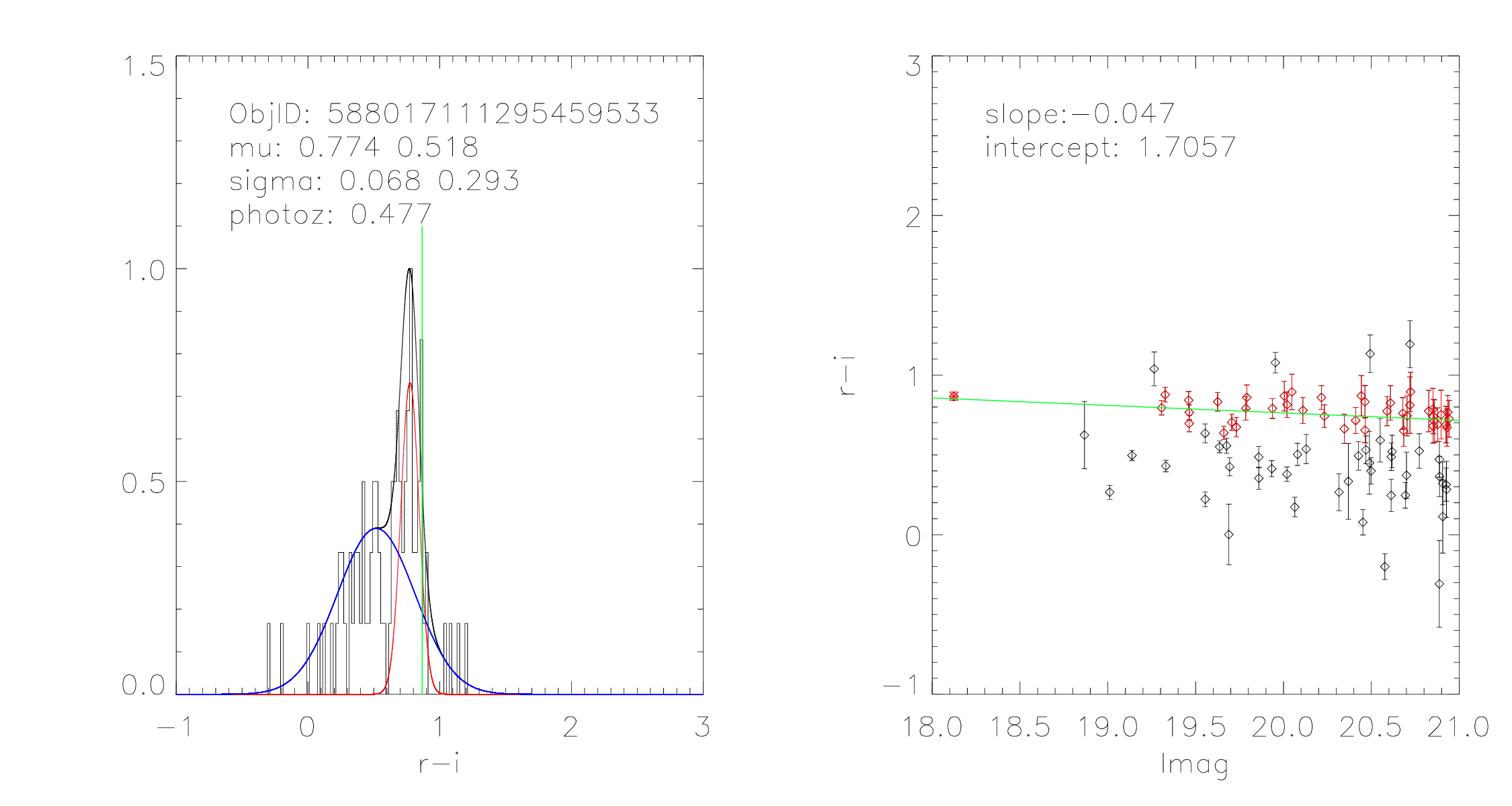}
\caption[color distribution around a cluster]{Color distributions and
  color-magnitude relations around two representative clusters. {\em
    Top Left} Galaxy $g-r$ color distribution around a cluster
  overlaid with a model constructed of a mixture of two Gaussian
  distributions. The red curve corresponds to the red sequence
  component while the blue one corresponds to the sum of background
  galaxies and blue cluster members. The green vertical line indicates
  the color of the BCG. $\mu$ and $\sigma$ are the means and standard
  deviations of the two Gaussian components. {\em Top right}
  Color-magnitude relation for the same galaxies.  Galaxies within the
  2$\sigma$ clip of the red sequence component are shown with red
  points; the green line indicates the best fit slope and intercept of
  this red sequence.  The left most red point is the BCG.  The bottom
  two panels shown the same plots for a second, higher redshift
  cluster, where the color used is $r-i$ instead of $g-r$.} \label{fig:gmr_color_space}
\end{center}
\end{figure*}

As pointed out in the previous section, the BCG plus red sequence
pattern is a unique feature of galaxy clusters. We therefore make
identifying this feature a key step in our cluster finding
algorithm. The distribution of galaxy colors in a cluster can be well
approximated by a mixture of two Gaussian
distributions\citep{haoecgmm}. The redder and narrower Gaussian
distribution corresponds to the cluster's red sequence, while the
bluer and wider one includes both foreground and background galaxies
along with the ``blue cloud'' cluster members. In
Figure~\ref{fig:gmr_color_space}, we show the galaxy color
distribution around two real clusters and the corresponding color
magnitude relation. If there is no cluster, then the color
distribution in a given patch of sky will be well represented by a
single Gaussian with a wide width. Fitting the color distribution with
mixture of Gaussian distributions is well suited for our purpose. A
complication in our case is that the measurement errors of the colors
are not negligible and proper modelling of them is essential for the
detection of red sequence. The traditional Gaussian Mixture Model
(GMM) does not consider the measurement errors and we therefore use an
error corrected GMM to developed in our earlier work~\citep{haoecgmm}.

As long as we effectively isolate red sequence galaxies, we reduce the
problem of cluster finding to a clustering analysis on the ra/dec
plane. One can then use either parametric~(such as convolving with a
model kernel) or non-parametric~(such as Voronoi Tessellation) methods
to analyze the strength of the clustering signal. When we apply such a
scheme to data spanning a wide redshift range there are three other
complications to consider.

First, as redshift increases the red sequence shows up in different
colors. This is mainly a result of the 4000 \AA~ break shifting across
the filters. Because of this effect, the most informative color will
vary as redshift increases. For the set of SDSS filters, the relation
between red sequence color and redshift is given in Table
~\ref{table:colorz}.

\begin{table}
\begin{center}
\caption{Red sequence color in different redshift ranges for SDSS
filters}\label{table:colorz}
\begin{minipage}[b]{1\linewidth}\centering
\small
\begin{tabular}{c c}
\hline \hline
Ridgeline color & Redshift range\\
\hline
$g-r$ & 0.0 $\sim$ 0.43\footnote{Although the 4000~\AA break starts shifting into SDSS r band at $z\sim 0.36$, we observed that $g-r$ color is still better than $r-i$ color for detecting red sequence up to redshift 0.43.} \\
$r-i$ & 0.43 $\sim$ 0.70  \\
$i-z$ & 0.70 $\sim$ 1.0 \\
\hline
\end{tabular}
\vspace{-7pt}\renewcommand{\footnoterule}{}
\end{minipage}
\end{center}
\end{table}

Beyond $z\sim 1.0$, one needs near infra-red color information, from
bands like Y, J, H, or K. Therefore, when detecting clusters in data
spanning a wide redshift range, it is necessary to determine which
ridgeline color we should examine. Since we will be searching for the
red sequence around candidate BCGs, we adopt the BCG's photo-$z$ as a
good estimator of cluster's redshift. BCGs are bright, making their
photo-$z$s generally much better determined than more typical
galaxies. As we discuss in \S 3.3, the precision of BCG photo-$z$s is
sufficient to determine which red sequence color should be examined
around a given BCG.  This does require a
determination of the photo-$z$ for every candidate BCG before
proceeding.

\begin{figure*}
\begin{center}
\includegraphics[width=5.5in, height=5in]{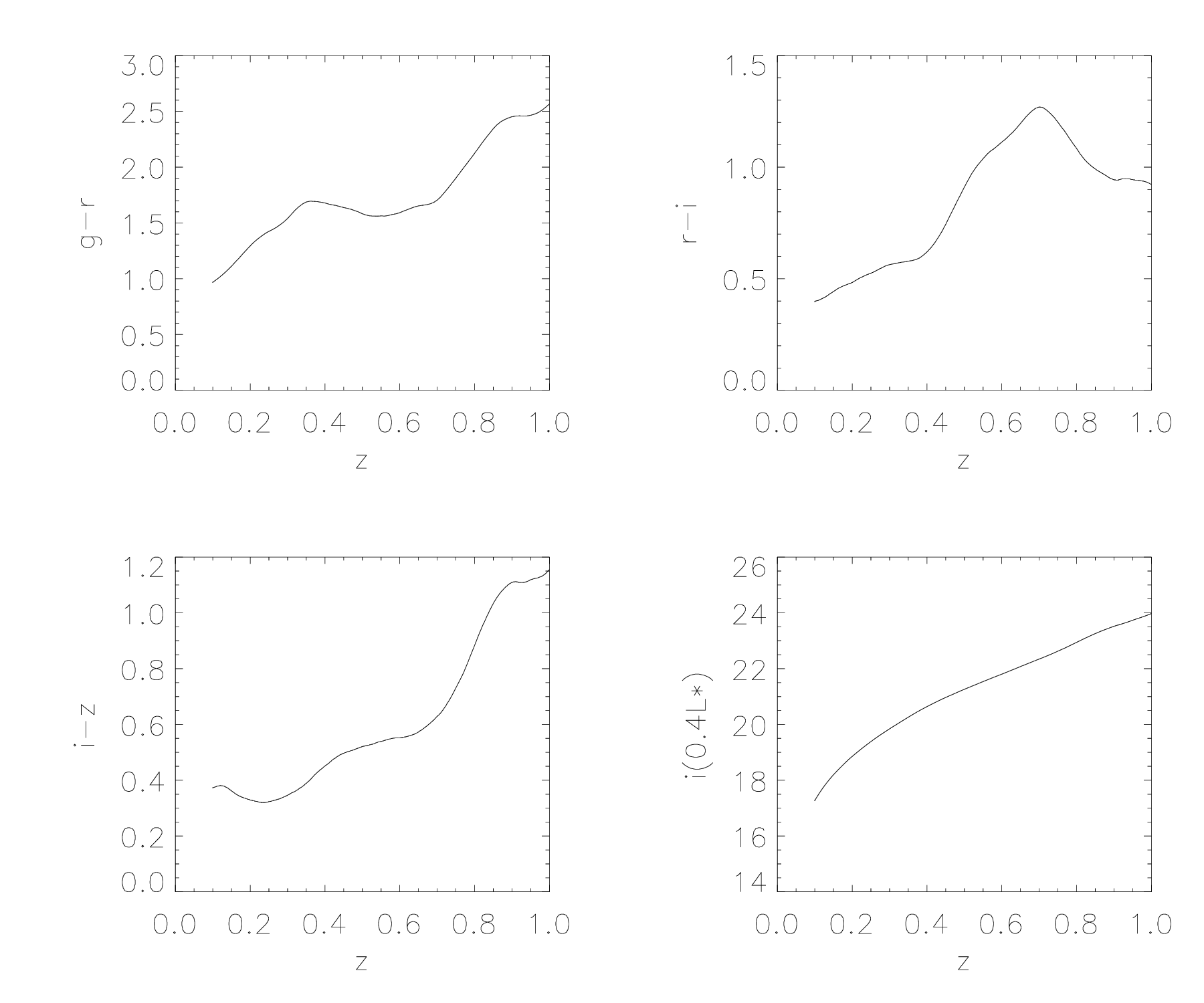}
\caption[color models]{The top two and bottom left panels are the
color evolution based on a color model of the red sequence
galaxies~\citep{koester07cat}. The bottom right panel is the I band
apparent magnitude corresponding to $0.4 L^*$ at different
redshifts.} \label{fig:color_model}
\end{center}
\end{figure*}

A second complication for cluster finding across a broad redshift
range is the increased chance of overlapping clusters, one at low
redshift and another at relatively high redshift. Such an overlap will
complicate the distribution in color space, turning it from bimodal to
tri-modal or even more. To reduce the possibility of this occurring,
we apply a broad photo-$z$ window (such as $\pm $ 0.25 in photo-$z$)
to select potential member galaxies before searching the color
distribution. The available photo-$z$ precision is adequate for this
purpose. In addition to photo-$z$ clips, we also apply luminosity cuts
and require the potential member galaxies to be brighter than 0.4 $L^*$,
where $L^*$ is the characteristic luminosity in the Schechter
luminosity function. For our application, the $i$-band apparent
magnitude corresponding to $0.4L^*$ as a function of redshift is shown
in the lower right panel of Figure~\ref{fig:color_model}. We adopted
this from~\cite{annis99} and \cite{koester07cat}. Selecting potential member
galaxies by cutting on photo-$z$ and luminosity is very effective at
simplifying the color space structure around the target galaxies. In
addition to this, the $0.4L^*$ cut allows us to measure a consistent
richness at different redshift~\footnote{For the SDSS DR7 data, the $0.4L^*$ can keep a consistent richness up to redshift 0.4.}.

The third complication concerns defining a consistent measurement of
richness across more than one color.  Red sequence galaxies selected
from different color bands have different degrees of contamination
from the background. This is a fundamental limit of all color-based
red sequence selection methods, though it has a relatively minor
effect on our cluster detection.  Once a cluster catalog is produced,
we will need to further calibrate the richness measured from different
color bands using other means, such as gravitational lensing
analysis~\citep{sheldona07, johnston07}.  In the present work, we just
adjust the richness definitions to result in a smooth transition between filters.

\subsection{Brightest Cluster Galaxies as Cluster Centers}

Brightest cluster galaxies (BCGs) reside near the cluster center of
mass, and provide important clues to other observational features of
clusters. Choosing the BCG as the center in a cluster finding
algorithm has good physical, algorithmic, and computational
motivations. The major physical motivation for focusing on the BCG is
that the central galaxy in a cluster (the one which resides near the
bottom of the cluster potential well) is very often the brightest
galaxy in the cluster. This BCG is then coincident with the region
with the deepest potential traditionally identified in theory as the center of a cluster.  To the
extent that this is true, using the BCG as the cluster center
simplifies precise comparisons between observations and theory,
although the extent to which the brightest galaxy is always at the
center, and the extent to which the most central galaxy is at the
center of the dark matter potential well, are still areas of
investigation. 

In an algorithmic sense, the BCG helps to distinguish among the bright
galaxies typically found near the cluster center. Such galaxies are
all similarly clustered, and the choice of a cluster center is thus
somewhat dominated by noise. The uniqueness of BCGs, including their
often cD-like morphologies, acts as a ``noise damper'' for positioning
cluster center.  Computationally, BCGs are bright and have
well-determined photo-$z$s, and the combination of these phenomena
boosts the efficiency of cluster detection by omitting searches around
intrinsically faint galaxies that dominate the luminosity function.
These motivating factors underscore the fact that while BCGs do not
drive the identification of clusters in the current algorithm, they
play an important fine-tuning role that minimizes the need for
downstream modelling in cosmological analyses.

\subsection{Red Sequence Color Selection}

A filter combination tuned to the selection of red-sequence galaxies
at a given redshift is of utmost importance. In our algorithm, we use
the photo-$z$ of the BCG to determine which color to choose. For SDSS
filters, we list the corresponding red shift ranges for different
colors in Table \ref{table:colorz}. In principle, the wrong color can
be chosen for a cluster due to an inaccurate BCG photo-$z$. In practice
this is not a serious problem; the photo-$z$s for BCGs are usually
well determined ($\sim$ 0.015 for SDSS DR7, see
\S~\ref{section:fact}).  Redshifts that place the 4000 \AA~ break
near the border of the filters are also a cause for concern, as they
can confuse the filter choice.  However, near the filter transitions,
the BCG plus red sequence pattern is apparent in both adjacent colors.
For example, for a rich cluster located at $z=0.42$, which falls in
the transition region from the SDSS $g$ band to the $r$ band, the
combined red sequence and BCG features can be still be easily captured
in either the $g-r$ or $r-i$. This ambiguity can impact the richness
estimates for clusters near the transition between filters (see
\S~\ref{section:rescaling_richness}), but does not result in
issues for cluster detection for the richness range considered in the
current work.

\subsection{Red Sequence Detection}

\subsubsection{Cluster Member Galaxy Selection}

The sizes of clusters are varied, increasing substantially with mass. Therefore, using a scaled aperture
is preferred for keeping a consistent richness estimation.  Ideally
for a candidate cluster, a series of different aperture radii should
be examined and chosen by maximizing S/N. However, this can be
computationally expensive. As a substitute, we take a two-step
approach similar to ~\citet{koester07alg}, which attempts to deal with
this fact: first, we measure the richness of the cluster using a fixed
metric aperture; then we scale the radius based on our measured fixed
aperture richness and remeasure everything using the scaled aperture
size.  The following describes the exact implementation for the
current work.

\subsubsection{Fixed Aperture Membership and Richness}

For a candidate BCG, we identify cluster members using a multi-step
process. We draw a 0.5 Mpc circle around the candidate BCG at its
photo-$z$\footnote{The BCG's photo-$z$ is a good estimator of the cluster redshift (see Figure \ref{fig:photoz_ngals_distribution}).
One might use the weighted average of the member galaxy photo-$z$s in the
expectation that the $\sqrt{N}$ averaging would provide a more accurate estimated
redshift. This is true, however, only when there is no systematic bias in the member 
photo-$z$s. In current practice there are often systematic errors in these members related to
their being fainter and yet just as red as the BCG.
There will always be the issue that they have lower signal/noise than the BCG.} and select all
galaxies fainter than the candidate BCG, but brighter than the
0.4$L^*$ cut at the relevant photo-$z$. Using the filter combination
relevant for the BCG, we use the Gaussian Mixture Model to fit the
distribution of the colors of all the galaxies selected above. To
remove possible overlap of two or more clusters along the line of
sight, we consider only galaxies within a photo-$z$ window of $\pm
0.25$ around the BCG. To determine the appropriate number of Gaussian
components for the fitting the color distribution, we calculate the
Akaike Information Criterion \citep[AIC][]{aic}. Around a cluster, AIC normally chooses two
Gaussian as best fit, one narrow and one broad, and the former is
chosen as the red sequence as it sits red-ward. Using the fixed 0.5
Mpc aperture it is, however, possible that the field of view is
dominated by a large cluster and therefore the best fit to the color
distribution is a single Gaussian representing the red-sequence. This
highlights the need for a scaled aperture (in this case, enlarged)
which would include more background galaxies and push the fitting
towards two color components. 

Next, for the two mixture case, we need to determine to which Gaussian
component the candidate BCG belongs. We compare its corresponding
likelihoods of the candidate BCG's color belonging to each of the two
Gaussian components and assign the most likely Gaussian component to
the candidate BCG. If this Gaussian component is wider than than the other Gaussian component, we flag this candidate
BCG as a field galaxy and remove it from the searching list for next
steps. For the case where there is only one Gaussian component, we
impose a threshold on its width, beyond which we do not deem it
suggestive of a red sequence and remove the corresponding candidate
BCG from consideration. Extensive testing on rich clusters in the SDSS
sets a color width of 0.16 (about twice the intrinsic width) as an
appropriate threshold in both the $g-r$ and $r-i$ colors.

Following this process, we consider only the candidate BCGs with an
appropriate red sequence measured. All the galaxies whose colors are
within $\pm$ 2 standard deviations of the mean of the corresponding
Gaussian component are flagged as members. The number of member
galaxies selected this way is denoted as $N_{gals}^{0.5Mpc}$. The
$\pm\texttt{2}\sigma$ cut corresponds roughly the level where the
background likelihood dominates over cluster likelihood. It is shown
elsewhere that indeed the two component Gaussian Mixture Model can
reliably pick up the correct peak in color space as verified by
simulations ~\citep{haoecgmm}.

\subsubsection{Scaled Aperture Size and Richness}\label{scaledaperture}
Scaled apertures are required to measure clusters of different
sizes. To select the appropriate aperture, we assume there is a
scaling relation between the aperture and the richness we measured
with 0.5 Mpc aperture, as motivated by~\citet{hansen07}.
\begin{equation}
R_{scale}=N (N_{gals}^{0.5Mpc})^{P}
\end{equation}
\noindent where $N$ and $P$ are the normalization and power
respectively, which need to be set so that the resulting $R_{scale}$
corresponds roughly to the relevant value of $R_{200}$. To determine
the scaling relation, we measure the $N_{gals}^{0.5Mpc}$ for maxBCG
clusters~\citep{koester07cat}. For every maxBCG cluster, there is a
$R_{200}^{lens}$ measured, interior to which the mean mass density of
the cluster is 200 times of the critical energy density. This
$R_{200}^{lens}$ is measured based on an exhaustive weak lensing
analysis ~\citep{johnston07,hansen07}. We find that $N_{gals}^{0.5Mpc}$
and the corresponding $R_{200}^{lens}$ follow a simple relation,
\begin{equation}
R_{scale}=0.133 \times (N_{gals}^{0.5Mpc})^{0.539},
\end{equation}
\noindent where $R_{scale}$, measured in Mpc, plays the role of the
$R_{200}^{lens}$ in \citet{johnston07} and \cite{hansen07}. Once we
have the scaled aperture, we repeat the procedure for the fixed
aperture richness measurement, substituting the corresponding scaled
aperture for 0.5 Mpc. The corresponding richness is denoted as
$N_{gal}^{scaled}$, and is used as the primary estimate of richness
for the cluster catalog.

\subsubsection{GMM vs ECGMM and Weighted Richness}\label{wrichness}

In this prescription for cluster member selection, we rely on the
detection of the red sequence as well as the measurement of its
width. The Gaussian Mixture Model (GMM) and its generalization with
error correction (ECGMM) are well-suited to detecting the red sequence
in a cluster. An unbiased measurement of the evolution of the red
sequence and its width requires the ECGMM ~\citep{haoecgmm}. However,
as the measurement errors increase, we cannot simply select member
galaxies using ECGMM with a 2$\sigma$ ($\sigma$ is the standard
deviation of the Gaussian component corresponding to the red sequence)
cut in a consistent way. GMM does give consistent membership
selection. This is mainly due to the fact that the ECGMM measures
``true" ridgeline width while our cuts are made in terms of the
observed colors. However, as the measurement error increases (e.g. at
higher redshift), GMM struggles to discern the correct number of
Gaussian components, as the measurement errors ``blur" the color
distribution. In this case, GMM will more likely favor a single
Gaussian component over two based on AIC, but ECGMM more accurately
recovers the correct number of mixtures because it properly models the
measurement errors.

On the other hand, we can also measure a weighted richness. When we
apply GMM (ECGMM) to fit the color distribution, each Gaussian
component has a weight from the fitting. This weight quantifies how
much of the total population is from the corresponding Gaussian
component. By multiplying the relative weight of the cluster component
to the total number of galaxies in the field, we measure the weighted
richness. It turns out that this weighted richness correlates better
with the true richness of the cluster when they are well
measured\footnote{Note if there is only one Gaussian component, this
  weighted richness does not make sense.}. To demonstrate this, we
performed some Monte Carlo tests. First, we generate the mock colors
from two Gaussian distribution, one corresponds to the background and
another corresponds to the cluster. We fix the number of galaxies in
the background component as 40 while vary the number of galaxies in
cluster component from 10 to 70 with increment of 5. Then, we generate
the measurement errors from a uniform distribution scaled by a noise
level (0.1 and 0.2 respectively in our case). The mock color will be
updated by adding realizations from a Gaussian distribution with the
width specified by the measurement errors.

\begin{figure*}
\begin{center}
\includegraphics[width=3in, height=1.5in]{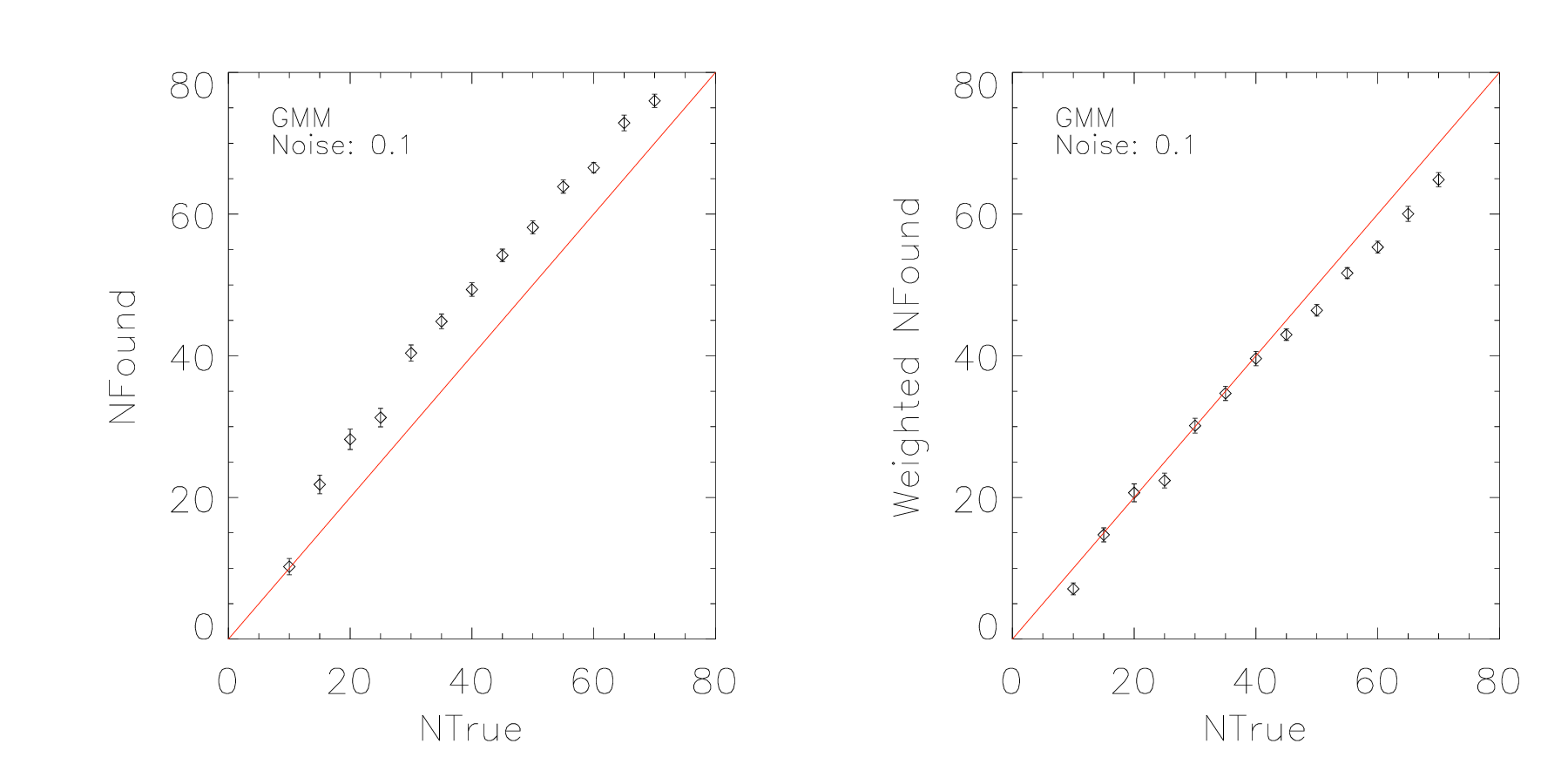}
\includegraphics[width=3in, height=1.5in]{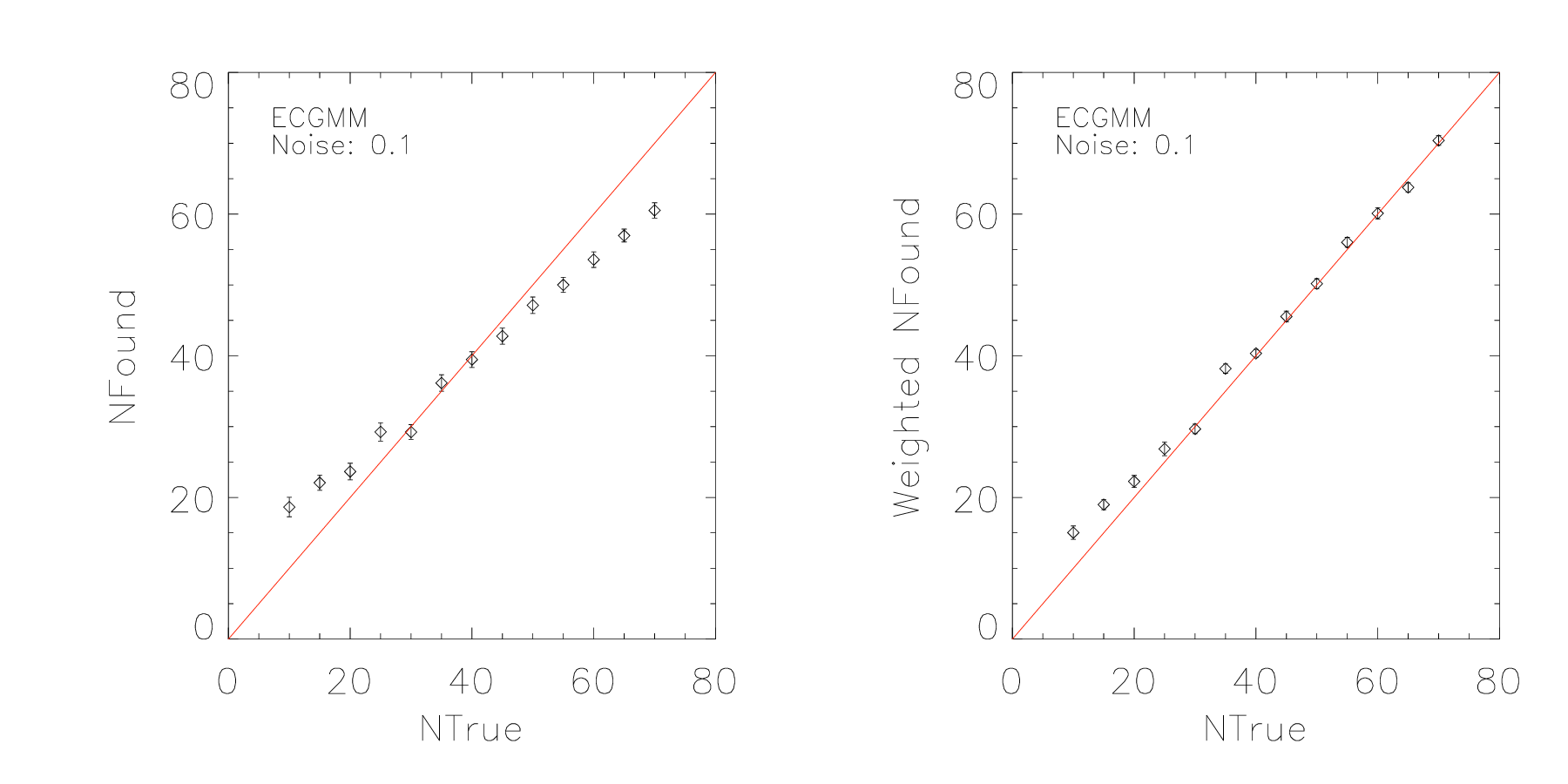}
\includegraphics[width=3in, height=1.5in]{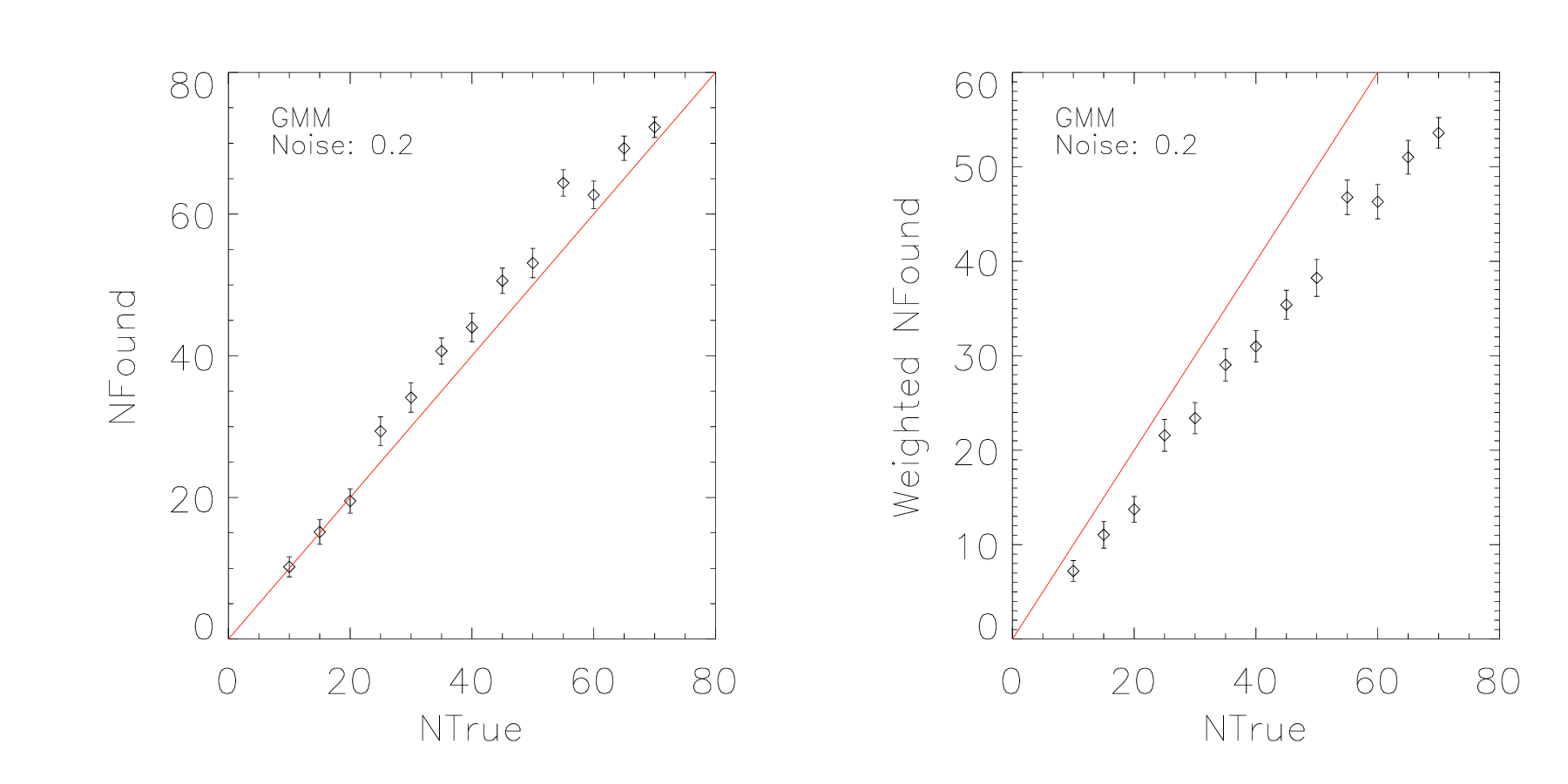}
\includegraphics[width=3in, height=1.5in]{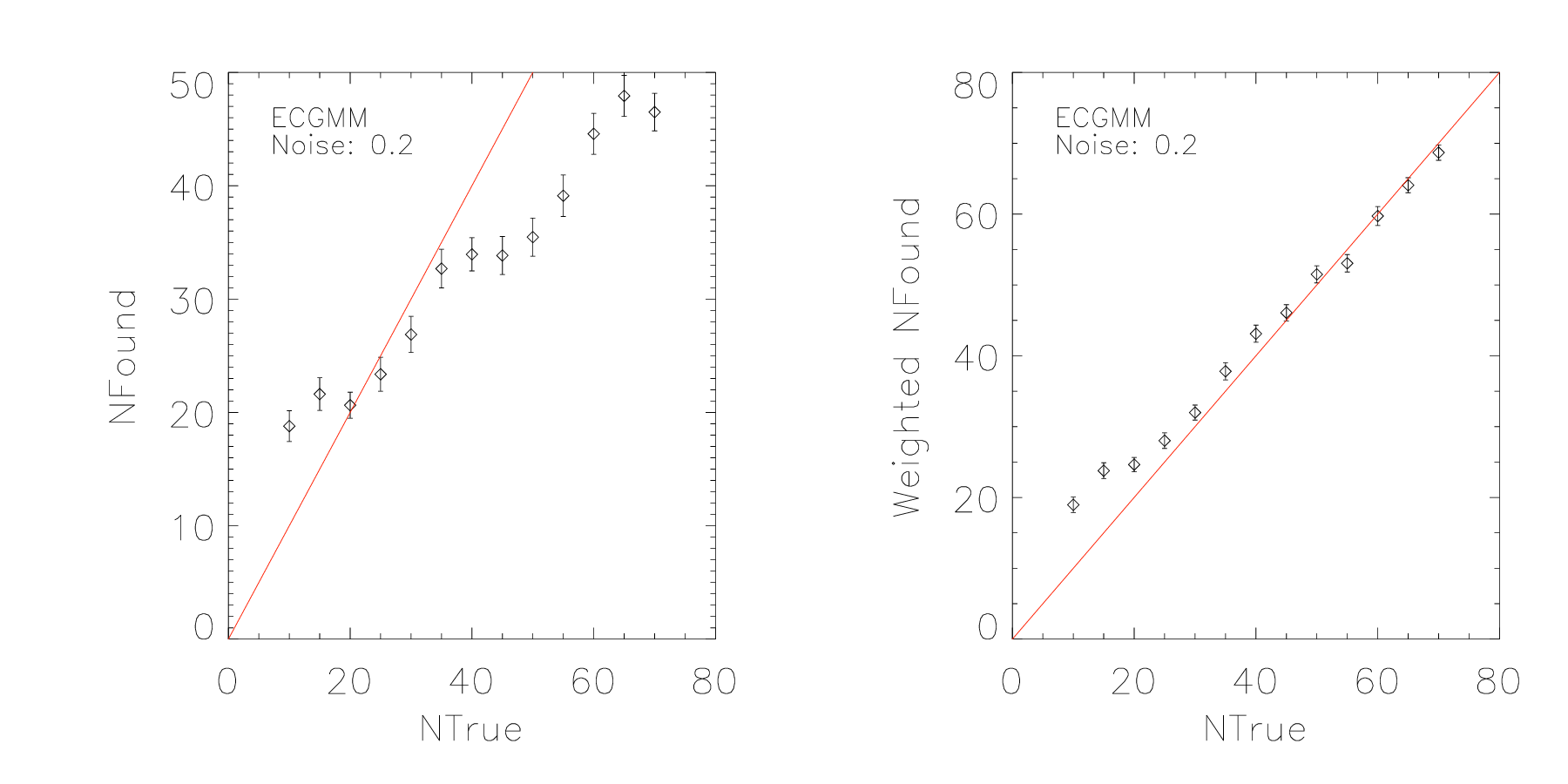}
\caption{Reconstruction of richness using GMM and ECGMM at noise level
  0.1 and 0.2. The noise on the plot indicate the scale we used to
  generate the mock measurement errors. GMM results in better number
  counts reconstruction, while ECGMM gives better weighted richness as
  measurement noise varies.}
\label{fig:rich_reconstruction}
\end{center}
\end{figure*}

For each given mock cluster richness, we repeat the above procedure
100 times and obtain a richness and weighted richness measurements
using GMM/ECGMM each time. In Figure~\ref{fig:rich_reconstruction}, we
plot our measured mean richness (NFound) and mean weighted richness
(Weighted NFound) vs the true richness (NTrue) at different
measurement noise level.

Based on these analyses, we conclude that GMM can give better richness
counts while ECGMM can give better weighted richness. Therefore, in
practice, we will use a hybrid of both GMM and ECGMM. We firstly
detect the red sequence using ECGMM and measure the weighted richness,
and then we use GMM with fixed number of mixtures (according to the
results of ECGMM) to do a follow-up measurement and select the red
sequence members.


\subsection{Clustering Strength}

We now have sufficient machinery to detect red sequence around a given
BCG candidate. If there is red sequence detected, it is still possible
that the candidate BCG is, e.g., a bright foreground galaxy, and does
not belong to the red sequence. Criteria must be chosen to determine
the association of a candidate BCG with the identified red
sequence. We thus consider it to be ``associated'' with the red
sequence if its color lies within 3 standard deviations of the peak of
the identified red sequence Gaussian.

Next, we quantify the strength of spatial clustering in the ra/dec plane
by convolving the selected members with a projected NFW~\citep{bartelmann96,nfw96,koester07alg} radial
kernel. It is worth noting that the type of kernel used is not as
important as its scale, which has been revealed by statistical
kernel density analyses~\citep{Silverman86,Scott92}. Therefore, the
specific kernel does not significantly bias the detection of
clusters that deviate from the kernel shape. 
We introduce the \textit{clustering
strength} as

\begin{equation}\label{lradial}
S_{cluster}=\sum_{k=1}^{N_g}\Sigma(x_k)
\end{equation}

\noindent where $N_g$ is the total number of member galaxies and
\begin{equation}
\Sigma(x)=\frac{2\rho_s r_s}{x^2-1}f(x),
\end{equation}
\noindent $r_s=r_{200}/c$ is the the scale radius, $\rho_s$ is the
projected critical density, $x=r/r_s$ and
\begin{equation}
  f(x) = \cases{
    1-{2 \over \sqrt{x^2-1}}\ \mbox{tan}^{-1} \sqrt{{x-1\over x+1}} & $x>1$ \cr
    1-{2 \over \sqrt{1-x^2}}\ \mbox{tanh}^{-1} \sqrt{{1-x\over x+1}} & $x<1$ \cr
    0 & $x=1$ \cr
    0 & $x>20$.
  }
\end{equation}
Similar to \citet{koester07alg}, we choose $r_s=150$ kpc, regardless of
richness.  The clustering strength parameter $S_{cluster}$ is
essentially the height of the peak of the smoothed red sequence
density field at the position of the BCG.

\subsection{Luminosity Weighted Clustering Strength}

In addition to the clustering strength parameter introduced in the preceding section,
we also measure another luminosity weighted clustering strength $S_{cluster}^{lum}$. 
The measurement is similar to $S_{cluster}^{strength}$ except that a
luminosity weight ($W_{lum}$) is attached to each galaxy. The luminosity weight
is simply defined as the ratio of each galaxy's $i$-band magnitude to the
$i$-band magnitude corresponding to 0.4$L^*$ at the candidate cluster
BCG's redshift.

\begin{equation}\label{lwradial}
S_{cluster}^{lum}=\sum_{k=1}^{N_g}\Sigma(x_k)\times W_{lum}(k)
\end{equation}

The advantage of introducing such a measure is that its ratio to the
non-luminosity weighted $S_{cluster}$ is a good indicator
of whether the candidate BCG is a contaminating bright star.
This forms an important double
check of the star/galaxy separation of the input catalog, which is a minor,
but non-negligible source of contamination.

\subsection{Implementation of the Algorithm}

With all the quantities calculated from the above definitions, the
implementation of the cluster selection is straightforward. There
are basically three steps:

\begin{enumerate}

\item For every galaxy in the catalog, evaluate the clustering
  strength $S_{cluster}$ inside a 0.5 Mpc searching aperture.  This
  $S_{cluster}$ is calculated using galaxies fainter than the
  candidate BCG and belonging to the identified red sequence.

\item Percolation procedure: rank the candidate BCGs by their
  clustering strength and remove candidates from the BCG list if they
  are identified as ``members'' of another candidate BCG with higher
  clustering strength.  Figure~\ref{fig:peak_percolation} illustrates
  the distribution of clustering strength around a candidate BCG.

\item Repeat the above process and finally obtain a
list of BCGs and their cluster members. Based on the richness
measured in 0.5 Mpc, one calculates a scaling $R_{scaled}$ for every
BCG. Then processes 1) -- 2) are repeated by changing the searching
aperture to $R_{scaled}$ from 0.5 Mpc. This concludes the search and completes the final
list of BCG members and BCGs with scaled richness
$N_{gals}^{scale}$.
\end{enumerate}

\begin{figure*}
\begin{center}
\includegraphics[width=4in,height=5in]{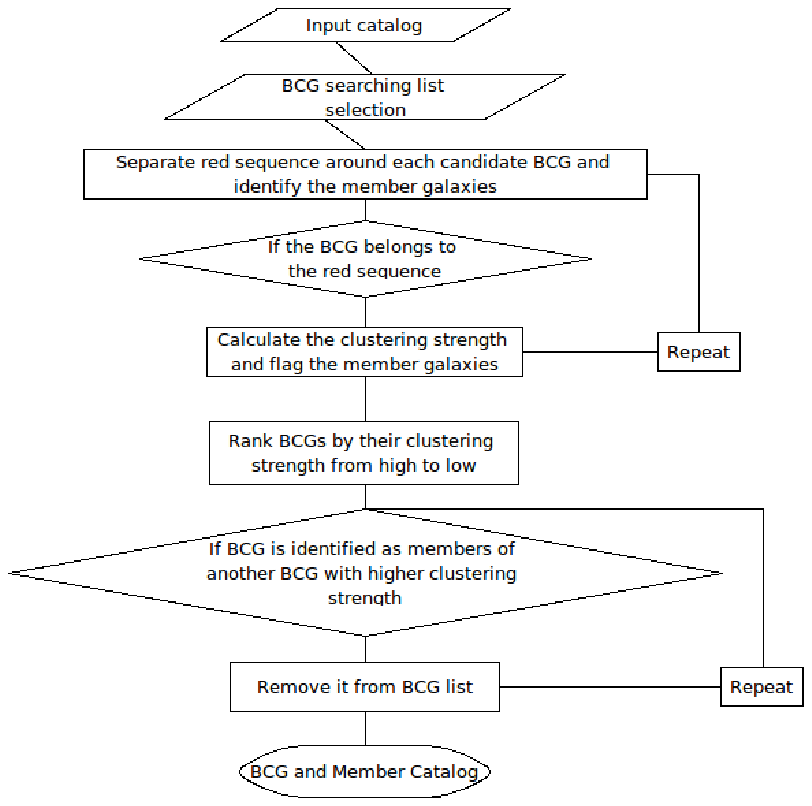}
\caption{Flowchart for the implementation of the GMBCG
algorithm} \label{fig:flowchart}
\end{center}
\end{figure*}

The procedures are summarized as a flowchart in
Figure~\ref{fig:flowchart}.

\subsection{Post Percolation Procedure}

\begin{figure*}
\begin{center}
\includegraphics[width=3in,height=3in]{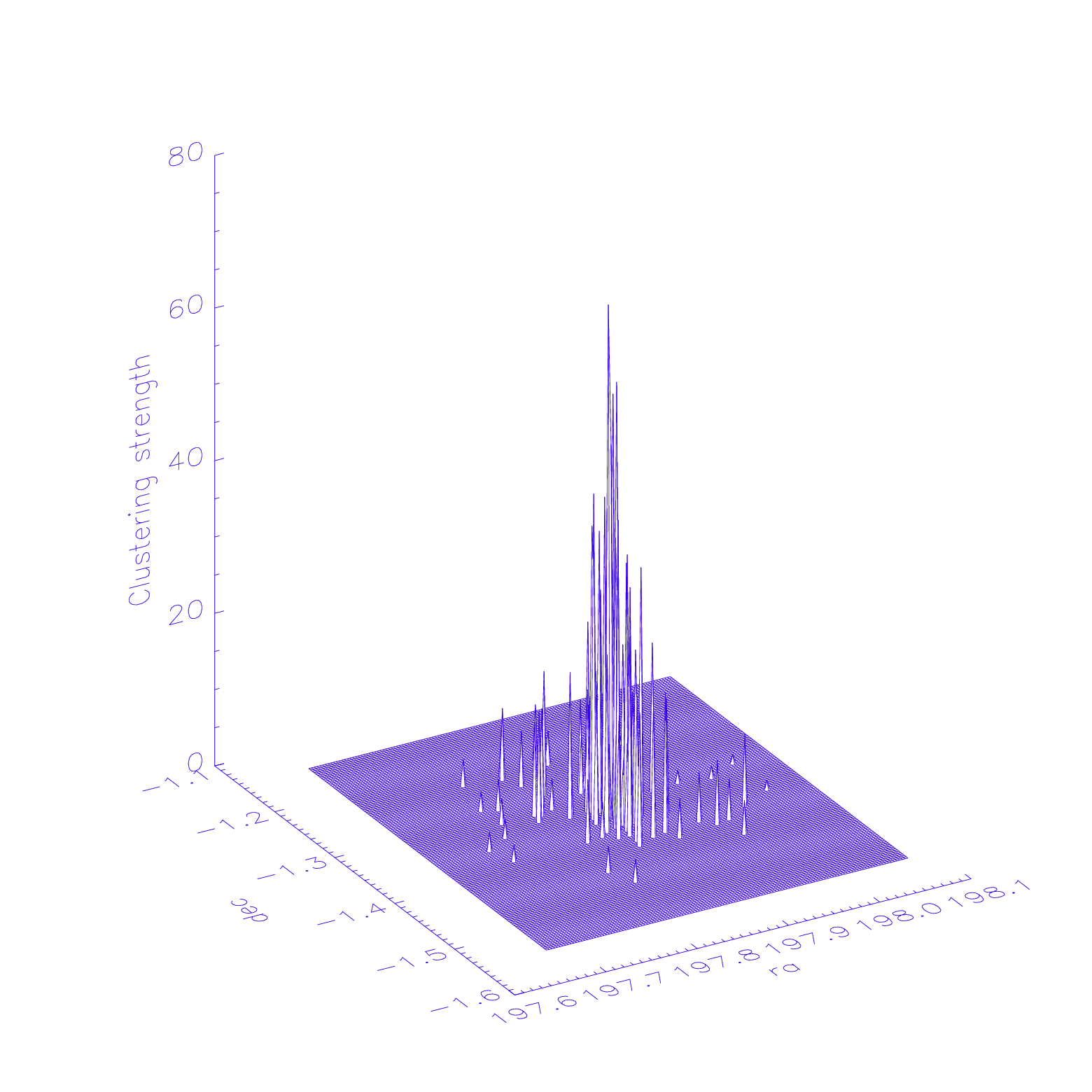}
\includegraphics[width=2.in,height=2.in]{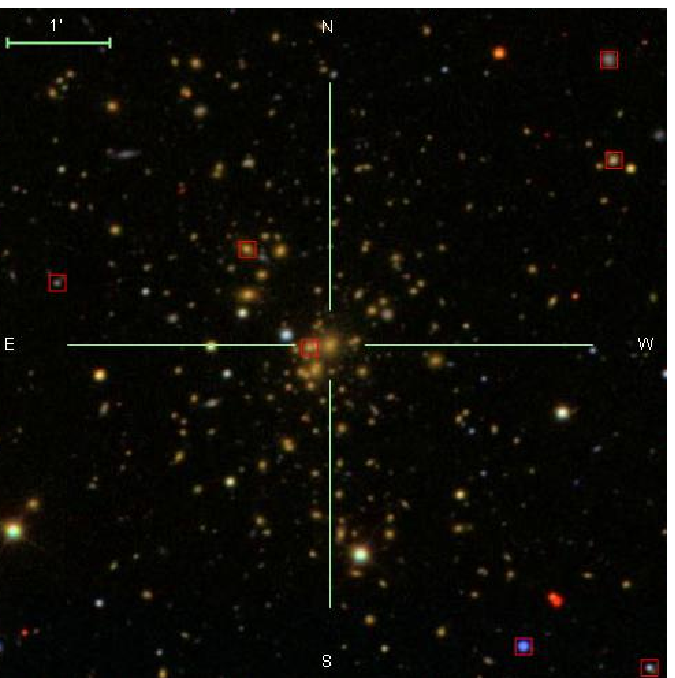}
 \caption{Left panel shows the clustering strength distribution around a galaxy cluster (Abell 1689). In this case, the
BCG is the highest peak. Right panel: SDSS image of A1689. } \label{fig:peak_percolation}
\end{center}
\end{figure*}

The above process is essentially a process of detecting the peaks of
the smoothed density field, where the height of the peaks is measured by
$S_{cluster}$. In Figure~\ref{fig:peak_percolation}, we show the
$S_{cluster}$ measured around Abell 1689.

In this cluster finding process, the center of the cluster is assumed
to be the brightest cluster galaxy. Therefore, it is possible that
several higher peaks (quantified by $S_{cluster}$) are identified in
the field of a brightest cluster galaxy and survive the previous
percolation procedure. Multiple peaks must be identified and merged
into one cluster using some criteria. This process is deemed ``post
percolation'', in contrast to the previous percolation procedure. The
major motivation for not directly blending the peaks during the
cluster finding process is the need for additional flexibility in both
merging the peaks and avoiding ``over-percolation'' the true BCGs by
some bright stars. Perhaps most importantly, the sub-peaks are
indicators of potential cluster sub-structure, and probe the internal
structures of clusters.

We settle on the following post-percolation prescription: for a given
candidate BCG (denoted as A), we identify a cylindrical region in the
ra/dec plane and redshift space around the BCG (A). The radius of the
cylinder is $R_{scale}$ of BCG (A), and the height is specified by the
BCG (A)'s photo-$z$ $\pm 0.05$. Then, if another candidate BCG
(denoted as B) falls inside this cylinder and BCG (B) is fainter than
BCG (A) but BCG (B)'s clustering strength is not more than 4 times of
that of BCG (A), we will merge BCG (B) into BCG (A). Setting the
clustering strength threshold of BCG (B) at a level of 4 times more
than that of BCG (A) avoids merging a true BCG into a very bright
galaxy. The value 4 is obtained explicitly by testing in known
situations in the SDSS, where bright foreground objects (e.g. stars)
confuse identification.

\subsection{Comparison with MaxBCG Algorithm}

It is interesting to explore the major differences between the GMBCG
and maxBCG algorithms~\citep{koester07alg}. maxBCG is a matched
filter based algorithm with an additional filter from the red
sequence colors. Using this algorithm, a large optical cluster
catalog has been created~\citep{koester07cat}, which has high purity
and completeness based on tests on both a Monte Carlo catalog and a
N-body mock catalog.

The difference between GMBCG and maxBCG can be summarized
in three major respects:

\begin{enumerate}
\item maxBCG is a generalized matched filter algorithm with the
  inclusion of a color filter in addition to radial and luminosity
  filters. It varies the filter at a grid of testing redshifts to
  maximize the match to a model filter. The redshift at which the
  model filter maximizes the match with data is selected as the
  redshift of the cluster. GMBCG does not maximize the match for a
  redshift dependent filter.  It uses a statistically well-motivated
  mixture model to identify the red sequence plus BCG feature. The

  radial NFW kernel serves as a smoothing kernel rather than a model
  filter. Therefore, GMBCG will be less biased against clusters that
  do not follow the assumed model filter in maxBCG.
  
\item maxBCG assumes an average ridgeline redshift model for all
  clusters while GMBCG does not assume any model as a priori. It uses
  the Gaussian Mixture Model to detect the red sequence and background
  in a cluster by cluster way. The advantage is that it automatically
  adjusts the cluster and background parameters across a wide redshift
  range.

\item In the maxBCG algorithm, the photo-$z$s of the clusters are
  estimated as a part of the execution of the algorithm. In GMBCG,
  photo-$z$s are obtained from other methods such as neural networks,
  nearest neighbour polynomial, etc. A photo-$z$ can also be estimated
  based on the measured red sequence colors as a by product.
\end{enumerate}

For these reasons, GMBCG is more easily extendible to a wide redshift
range and less biased against atypical clusters.

\section{GMBCG catalog For SDSS DR7}\label{catdr7}
In this section, we apply the GMBCG algorithm to the Data Release 7 of
the Sloan Digital Sky Survey (SDSS DR7), and construct an optical
cluster catalog of more than 55,000 rich clusters across
$0.1<z<0.55$. To check the quality of the cluster catalog, we
cross-match the GMBCG clusters to X-ray clusters and maxBCG
clusters. We also create a mock catalog based on DR7 data to test the
completeness and purity of the catalog. The details of the catalog
construction are covered in the following section.

\subsection{Input catalog}

The Sloan Digital Sky Survey (SDSS)~\citep{york00} is a multi-color
digitized CCD imaging and spectroscopic sky survey, utilizing a
dedicated 2.5-meter telescope at Apache Point Observatory, New
Mexico. It has recently completed mapping over one quarter of the
whole sky in $ugriz$ filters. DR7 is
a mark of the completion of the original goals of the SDSS and the
end of the phase known as SDSS-II~\citep{abazajian08}. It includes a
total imaging area of 11663 square degrees with 357 million unique
objects identified.

In this paper, we will mainly detect clusters on the so called Legacy
Survey area, which ``provided a uniform, well-calibrated map in
$ugriz$ of more than 7,500 square degrees of the North Galactic Cap,
and three stripes in the South Galactic Cap totaling 740 square
degrees''~\citep{abazajian08}. We construct the input galaxy catalog
from the CasJobs (http://casjobs.sdss.org/CasJobs/) PhotoPrimary view
of the SDSS Catalog Archive Server with type set to 3 (galaxy) and
$i$-band magnitude less than 21.0. In addition, we also apply the
following flags to keep the catalog clean: SATURATED, SATUR\_CENTER,
BRIGHT, AMOMENT\_MAXITER, AMOMENT\_SHIFT and AMOMENT\_FAINT. We
download the photo-$z$ table and cross match the objects to the galaxy
catalog to attach photo-$z$s to each galaxy we selected. In DR7, the
photo-$z$s in the photo-$z$ table are calculated based on a nearest
neighbor polynomial algorithm~\citep{abazajian08}.

In addition to the above selection requirements, we also throw away
those galaxies with bad measurements (photometric errors in $g$ and
$r$ band greater than 10 percent). In principle, we should search all
galaxies as candidate BCGs.  However, as BCG are well-known and form a
subset of the total galaxy population, the list (and computational
time) can be reduced. Based on Figure~\ref{fig:color_model}, we make
cuts in color space as shown in the red regions of
Figure~\ref{fig:red_pre_select}. Additionally, each galaxy has a
well-measured ellipticity through the SDSS data processing pipeline
based on adaptive moments~\citep{Bernstein02}. We require the
ellipticity in the $r$-band to be less than 0.7 for candidate
BCGs. This ellipticity cut helps to remove edge on spiral galaxies which, when reddened by dust, often take on the colors of much higher redshift red sequence galaxies, and hence can appear as false projected BCGs. All these cuts keep $\sim$ 70\% of the total galaxies in our candidate BCG search list, effectively eliminating only those with
quite atypical colors and morphologies.

\begin{figure*}
\begin{center}\label{fig:red_pre_select}
\includegraphics[width=2.8in,height=2.8in]{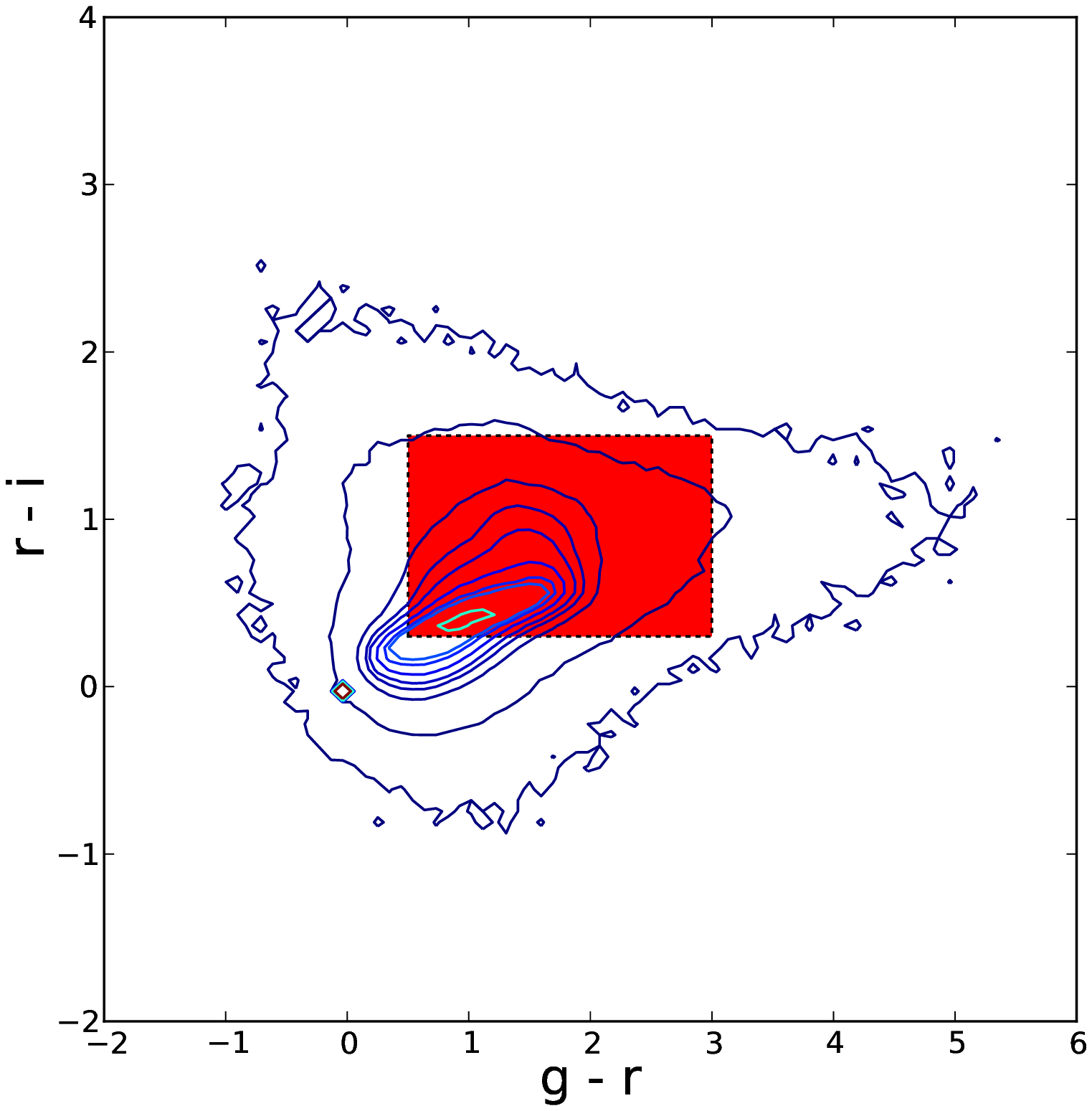}
\includegraphics[width=2.8in,height=2.8in]{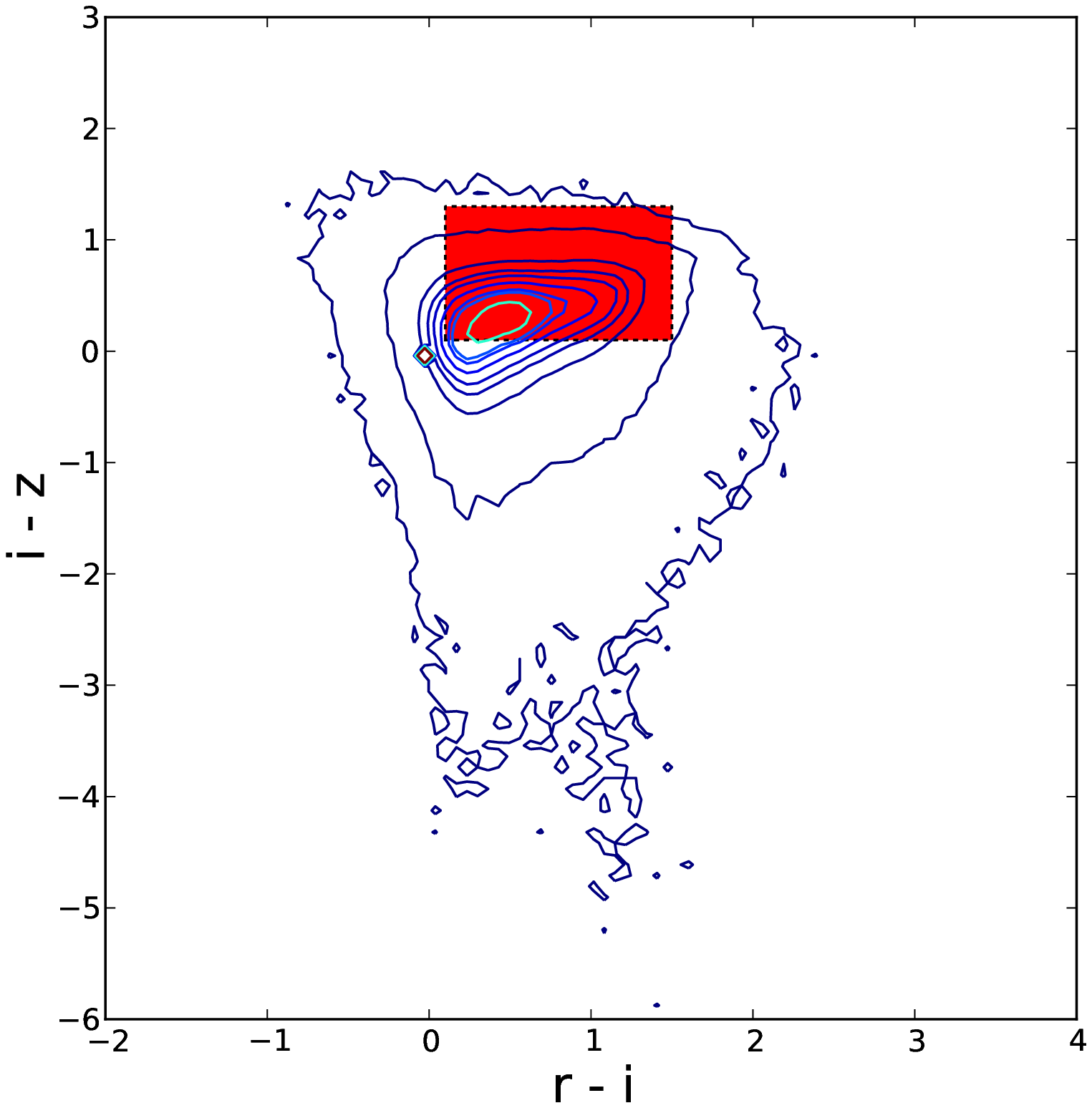}
\caption{BCG preselection in color - color space for the SDSS DR7
  data. Red regions indicate the area of $g-r$ vs. $r-i$ (left panel)
  and $r-i$ vs. $i-z$ (right panel) color-color space in which we
  preselect BCGs.  This preselection keeps $\sim$ 70\% of the total
  galaxies.} 
\label{fig:red_pre_select}
\end{center}
\end{figure*}

After the above procedures, we prepare an input catalog for our
cluster finder. It is worth noting that we did not apply any
star/galaxy separation procedures other than the ones generated by
the standard DR7 pipeline. This is a relatively tolerant selection
that may be contaminated by occasional bright stars that are not
well separated from galaxies. As described earlier, we handle these stragglers by comparing the measured luminosity 
weighted clustering strength ($S_{cluster}^{lum}$) with the non-luminosity weighted clustering strength ($S_{cluster}$) to reject those bright stars.

\subsection{Richness Re-scaling}\label{section:rescaling_richness}

\begin{figure*}
\begin{center}
\includegraphics[width=2.5in,height=2.5in]{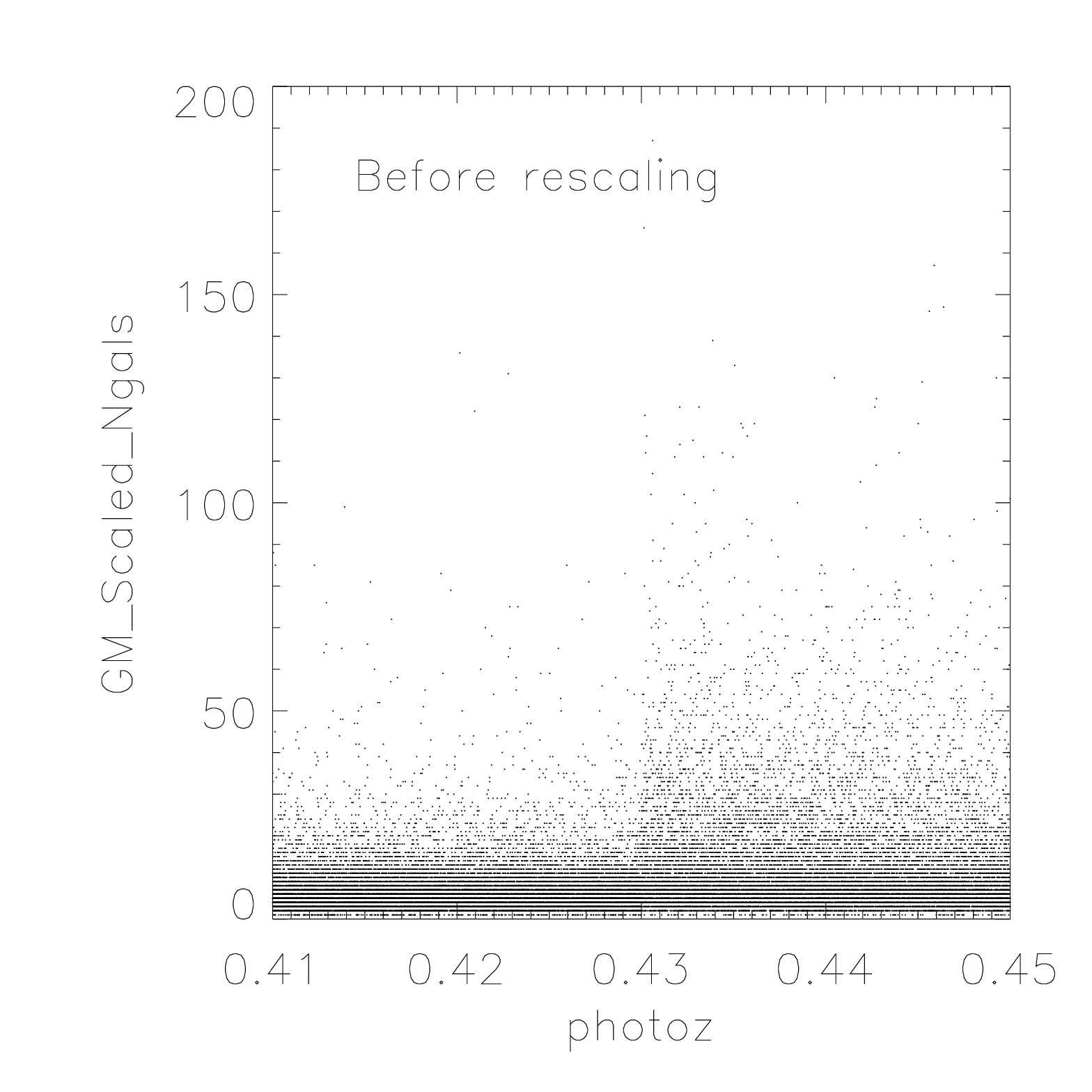}
\includegraphics[width=2.5in,height=2.5in]{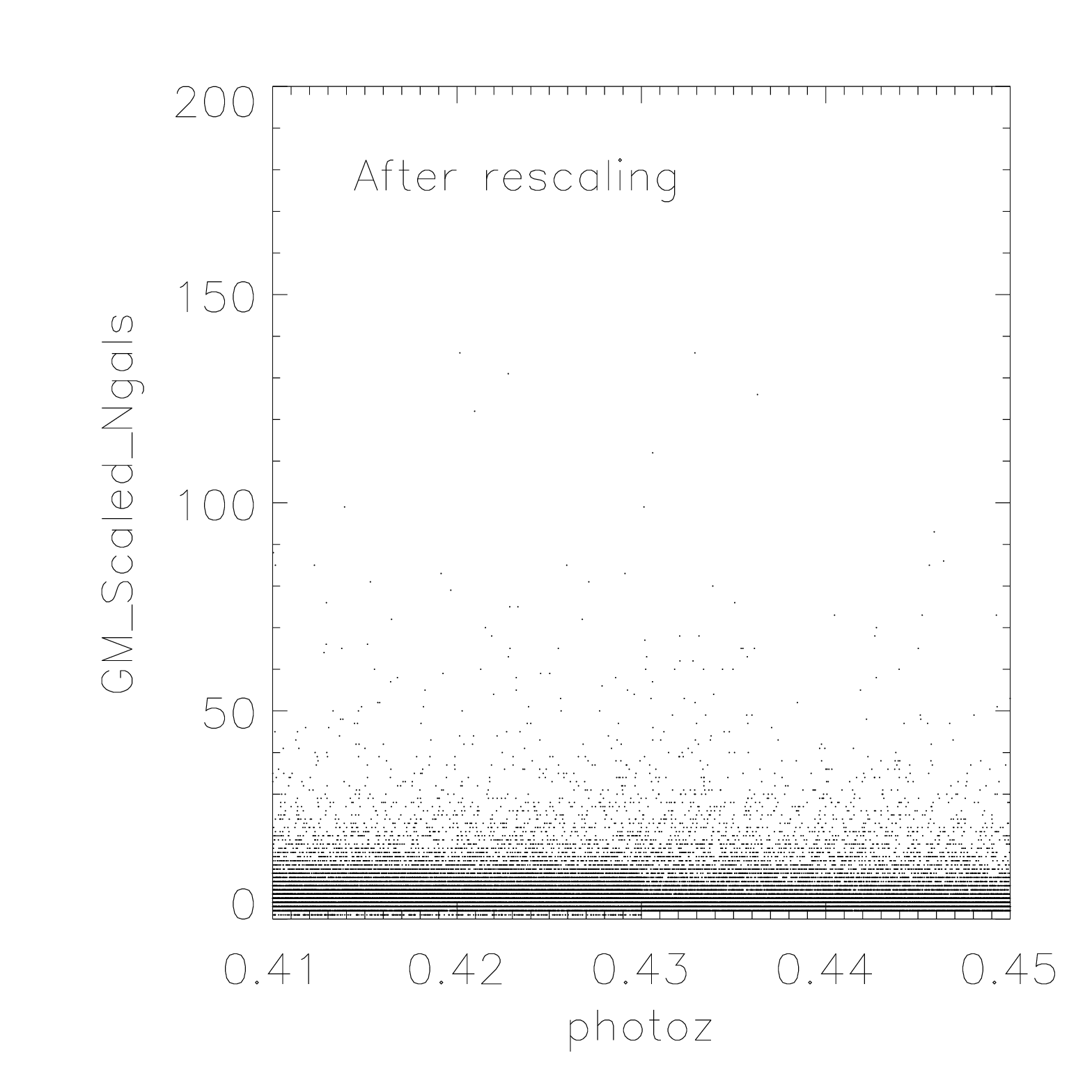}
\end{center}
\caption[Richness re-scaling]{Richness ($N_{gal}^{scaled}$) before and after the re-scaling. This demonstrates that rescaling removes much of the difference in richness measurements between the g-r and r-i bands.}
\label{fig:richness_rescaling}
\end{figure*}

In the redshift range $0.1 \sim 0.55$, only the $g-r$ or $r-i$
ridgeline colors are used, and the switch between them is determined
by the photo-$z$ of the candidate BCG. Since we measure the richness
by counting the number of galaxies falling within $2 \sigma$ of the
ridgeline, the resulting richness from $g-r$ or $r-i$ are not directly
comparable. In part this is due to a changing degree of background
contamination as the ridgeline moves through color space (see
Figure~\ref{fig:bimodal}). Generally, the richness measured from $r-i$
is higher than that measured from $g-r$. To make the richnesses more
consistent across the whole redshift range, we rescale those measured
from $r-i$ color. Clearly, mass is the only true parameter with which
we should relate the two different richness. Therefore, a complete
resolution of this problem requires a carefully mapping of the
mass-richness relation for richness in both redshift ranges.  However,
for the moment, we settle for the simpler first order approach. That
is, we require the statistical distribution of richness measured from
$g-r$ color at redshift range [0.41 - 0.43] and richness measured from
$r-i$ color at redshift range [0.43, 0.45] to be the same since the
true richness of the clusters in these narrow redshift ranges should
vary only mildly. The scaling relation that matches the two
distributions is not necessarily linear. To ensure the distribution to
be the same, we match the richness at different percentile bins of the
two distributions and re-scale them linearly in each bin. Then, we fit
a polynomial to the scaling relation across all the bins to derive a
``continuous'' scaling relation. The richness from the $r-i$ color
will be accordingly re-scaled by this relation.  In
Figure~\ref{fig:richness_rescaling}, we show the richness distribution
before and after the re-scaling. Since the scaling relation is
monotonously increasing, the scaled richness will not alter the
cluster ranking based on the original richness in the $r-i$ region (it
will affect the global ranking for sure). In a similar fashion, we
also re-scale the weighted richness and the clustering strength. In
the following, unless noted otherwise, the richness and clustering
strength all refer to the rescaled values.

\subsubsection{Catalog Cleaning and Masking}
\label{section:cleaning}

\begin{figure}
\begin{center}
\includegraphics[width=4in,height=4in]{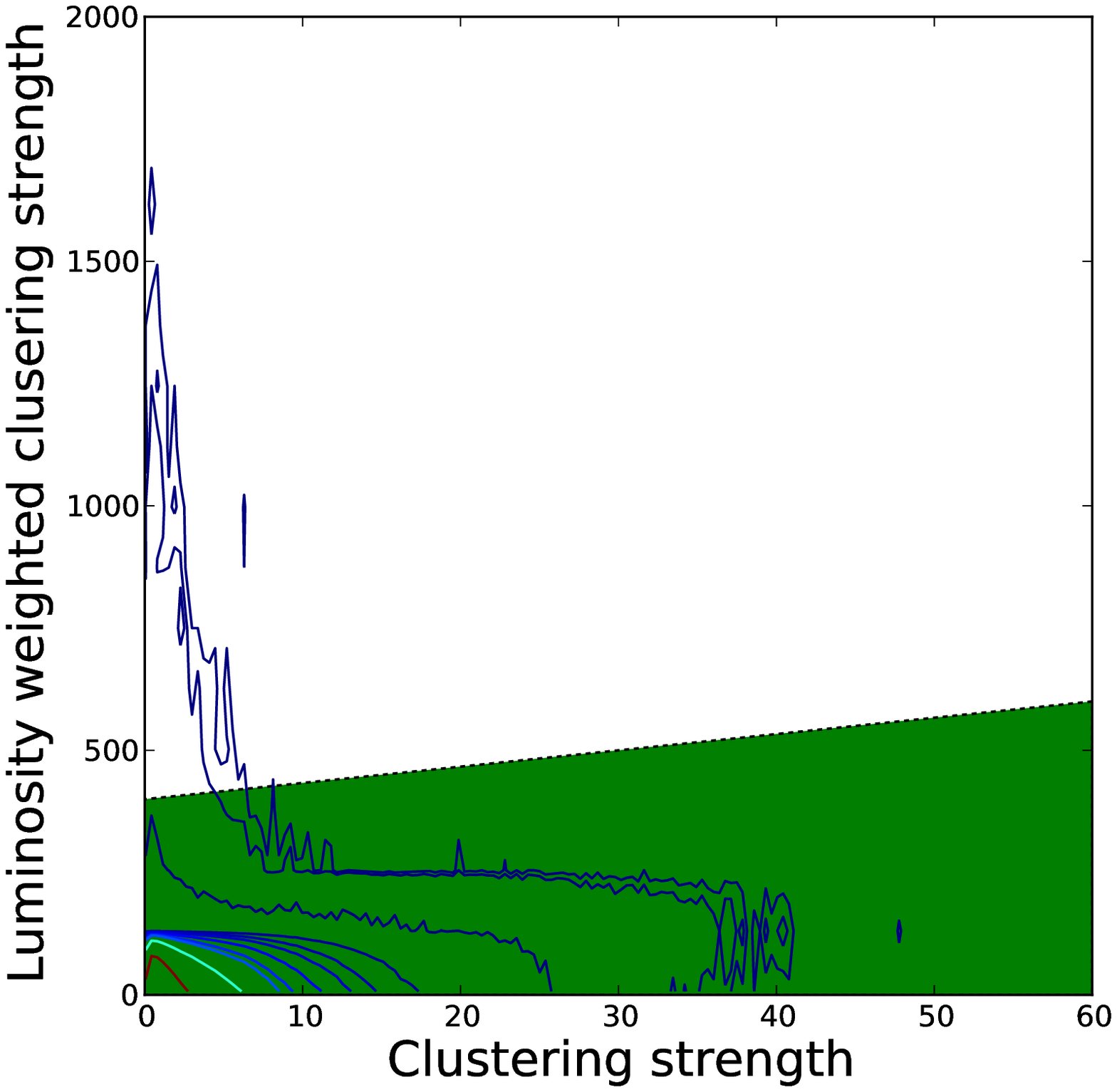}
\caption{Density contour of BCGs in the space of luminosity weighted and non-luminosity weighted clustering strength.  Blue contours show the results for all candidate BCGs. The green region shows cuts applied to candidate BCGs, as described in \S \ref{section:cleaning}, which removes bright stars that pass the star/galaxy separation in the SDSS data processing pipeline.}
\label{fig:weightedlh}
\end{center}
\end{figure}

We apply the GMBCG algorithm to the input catalog and generate a full
catalog of galaxy clusters for the SDSS DR7. We search clusters from
redshift $0.05 < z < 0.60$, but only include in the final catalog the
redshift range $0.1 < z < 0.55$ to reduce redshift range edge
effects. The luminosity weighted and non-luminosity weighted
clustering strength (see above) are employed.  For stars, the
luminosity weighted clustering strength is much greater than its
non-luminosity weighted counterpart. By hand scanning the corresponding
images, we found the cuts as shown in Figure~\ref{fig:weightedlh} are good for removing the contaminated stars.

In addition to the above cuts, we also mask out those clusters that
are close to the brightest stars. We apply the bright star mask from
the NYU VAGC (valued added galaxy catalog) release for SDSS
DR7~\citep{blantonvagc} and mask out all clusters that fall inside the bright star mask polygons.

\subsubsection{Catalog Facts}\label{section:fact}
\begin{figure}
\begin{center}
\includegraphics[width=5in, height=3in]{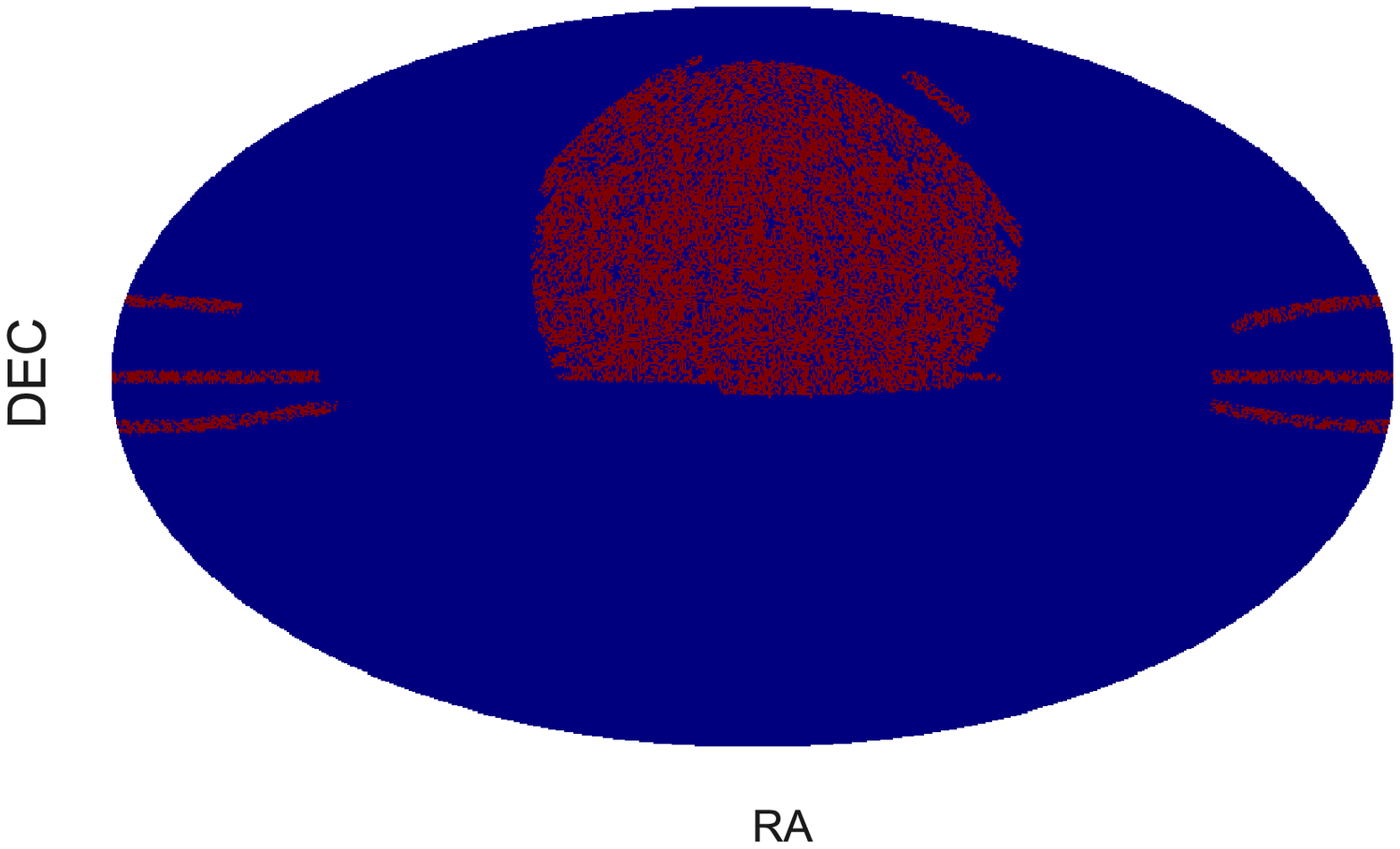}
\caption{Sky coverage in the GMBCG public catalog based on SDSS DR7. Each point shows the position of one cluster on the sky.}
\label{fig:coverage}
\end{center}
\end{figure}

\begin{figure*}
\begin{center}
\includegraphics[width=3in, height=3in]{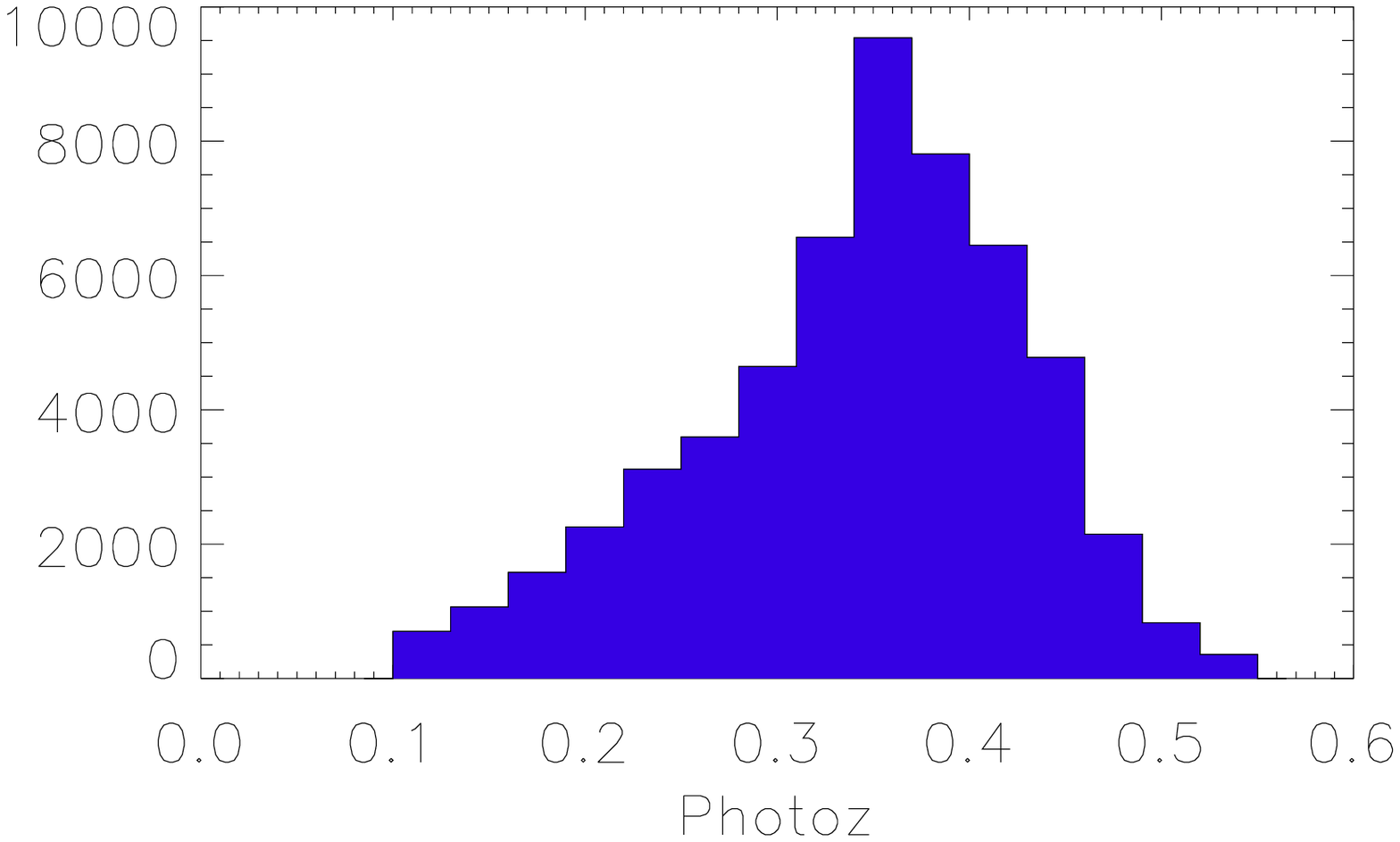}
\includegraphics[width=3in, height=3in]{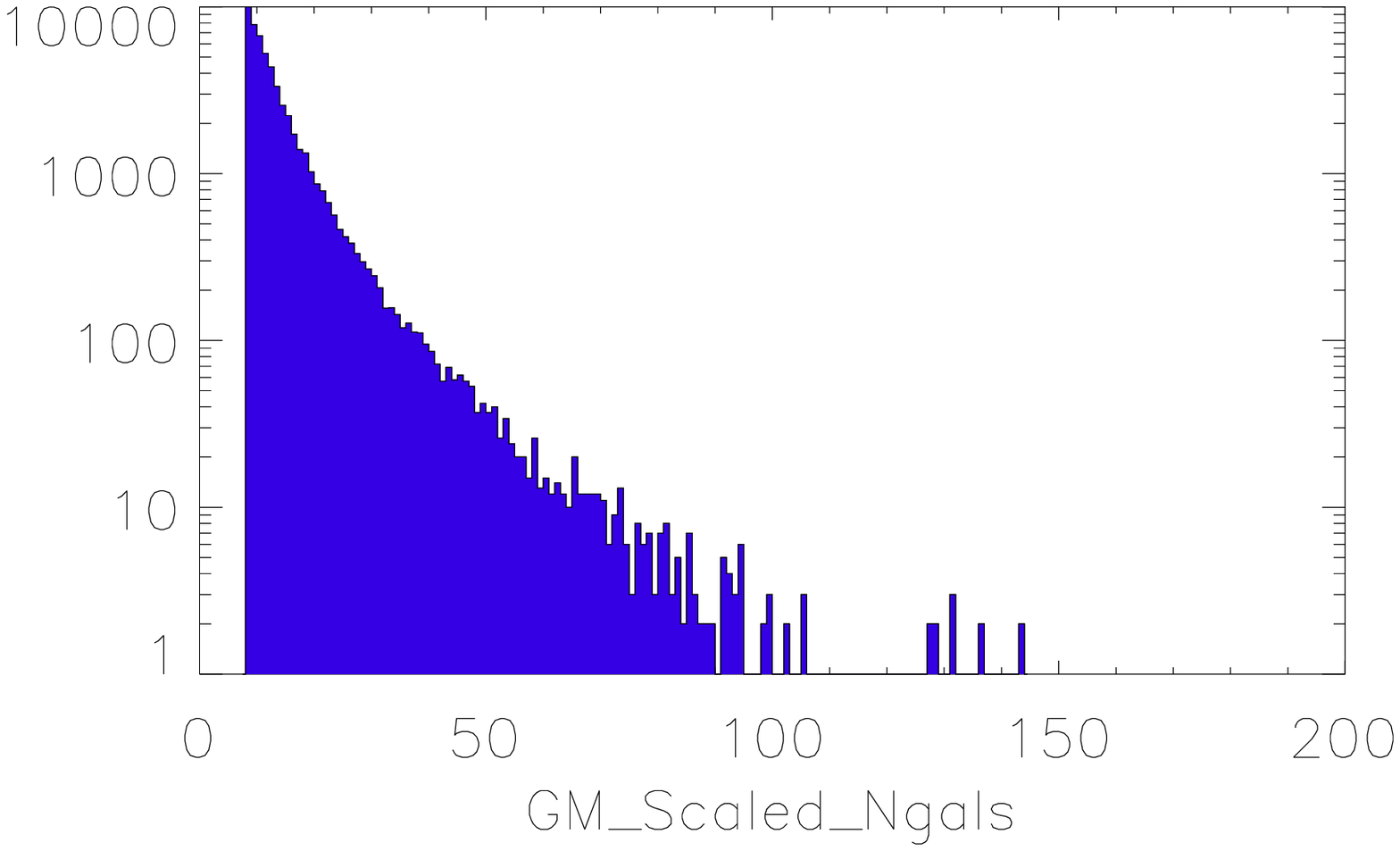}
\caption{Redshift and richness distribution of GMBCG clusters in the
  public catalog. Left panel shows the redshift distribution of
clusters, cut at $0.1 < photo-$z$ < 0.55$.  Right panel shows the
  scaled richness distribution, GM\_scaled\_Ngals, for clusters with GM\_scaled\_Ngals
$> 8$. }
\label{fig:photoz_ngals_distribution}
\end{center}
\end{figure*}

\begin{figure*}
\includegraphics[width=2.in,height=2in]{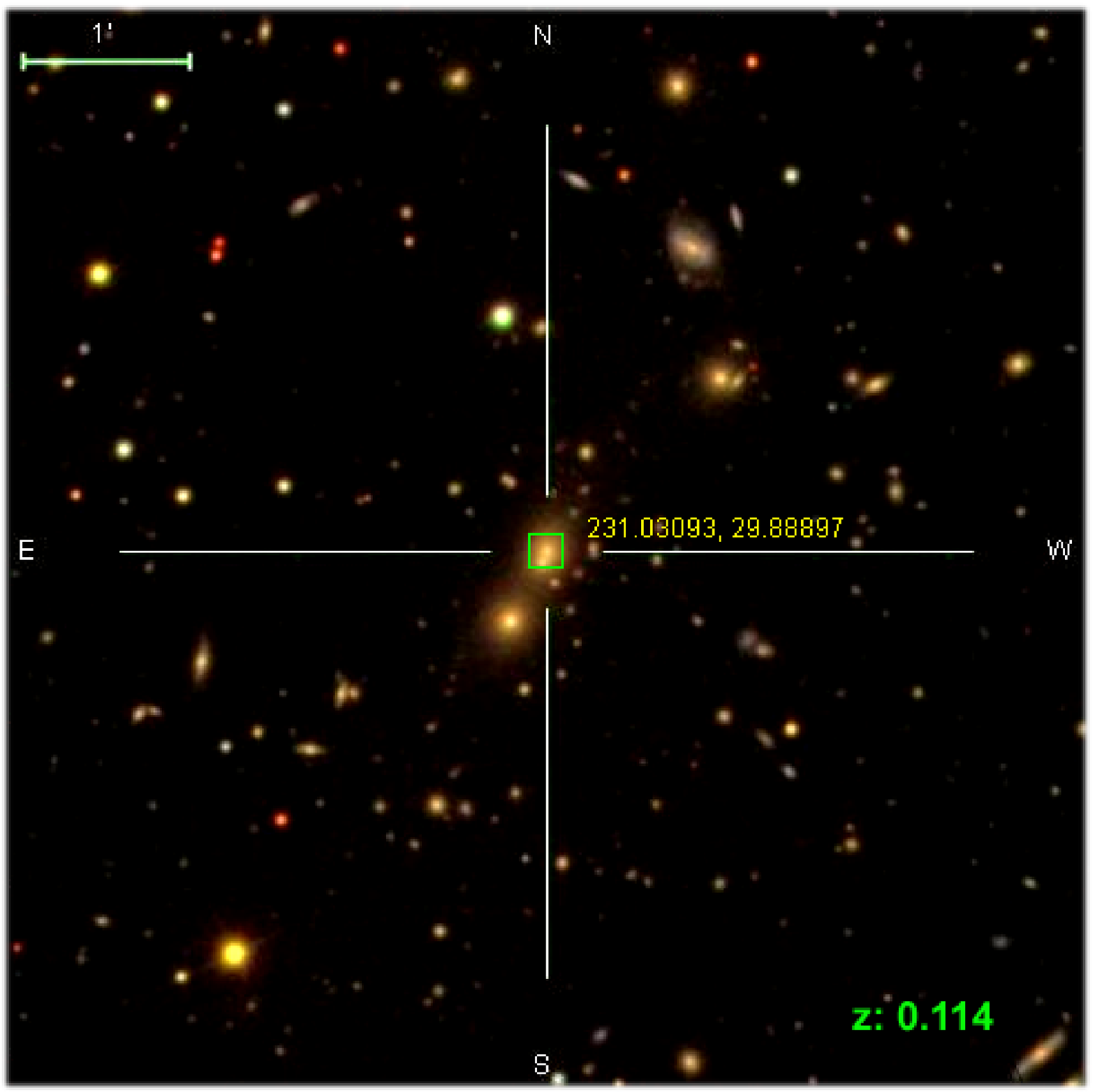}\hfill
\includegraphics[width=2in,height=2in]{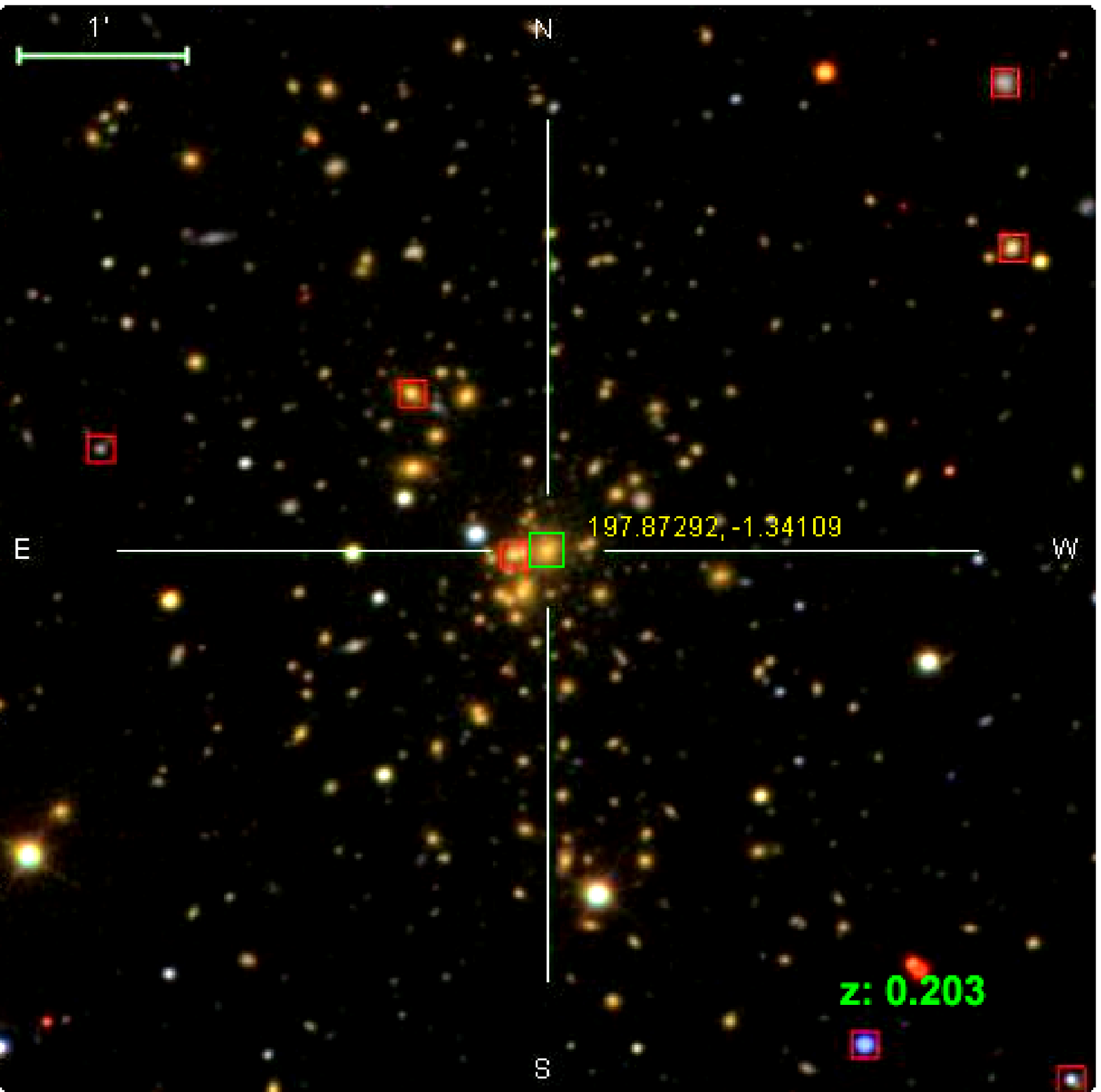}\hfill
\includegraphics[width=2in,height=2in]{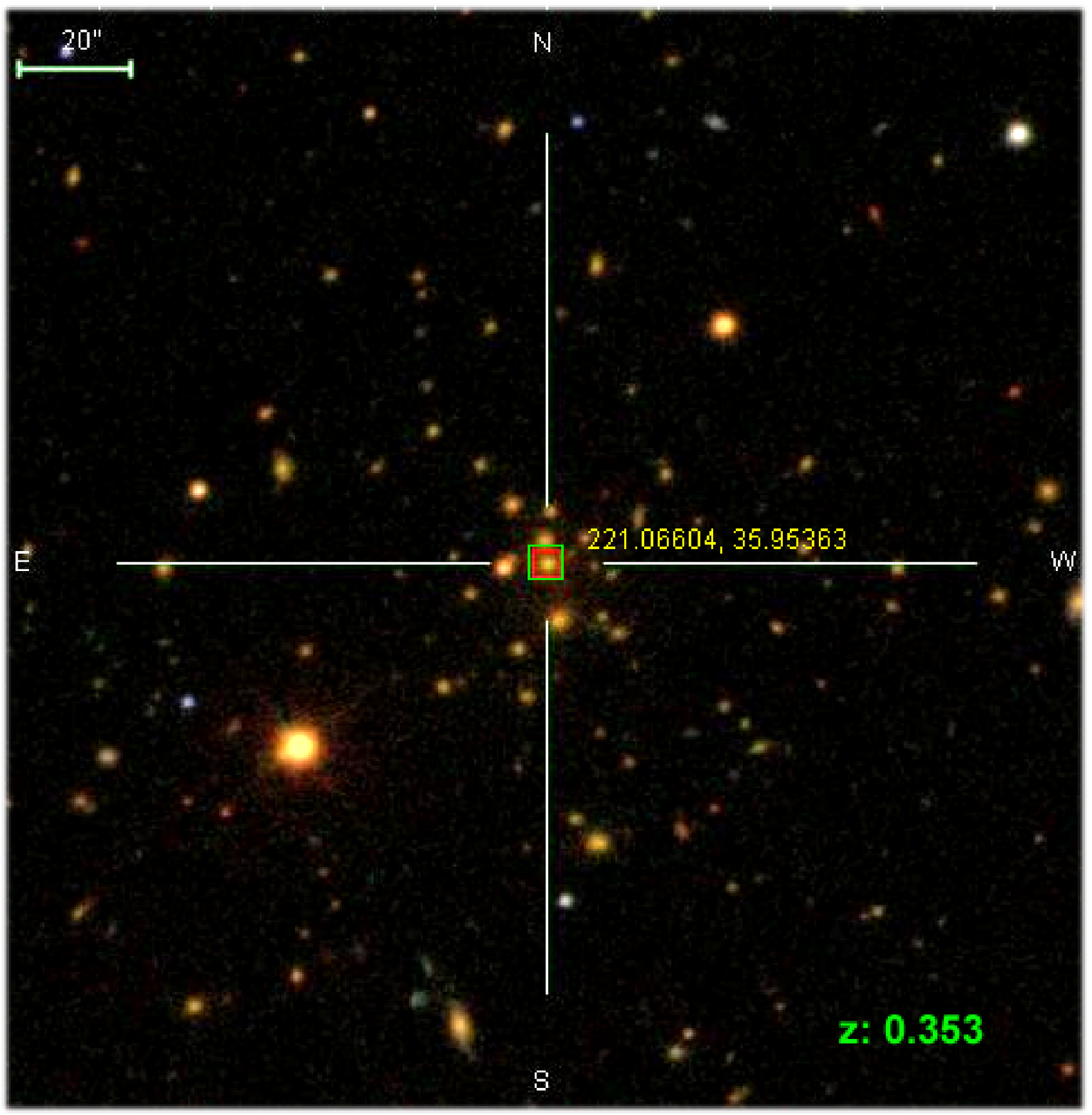}\hfill
\includegraphics[width=2in,height=2in]{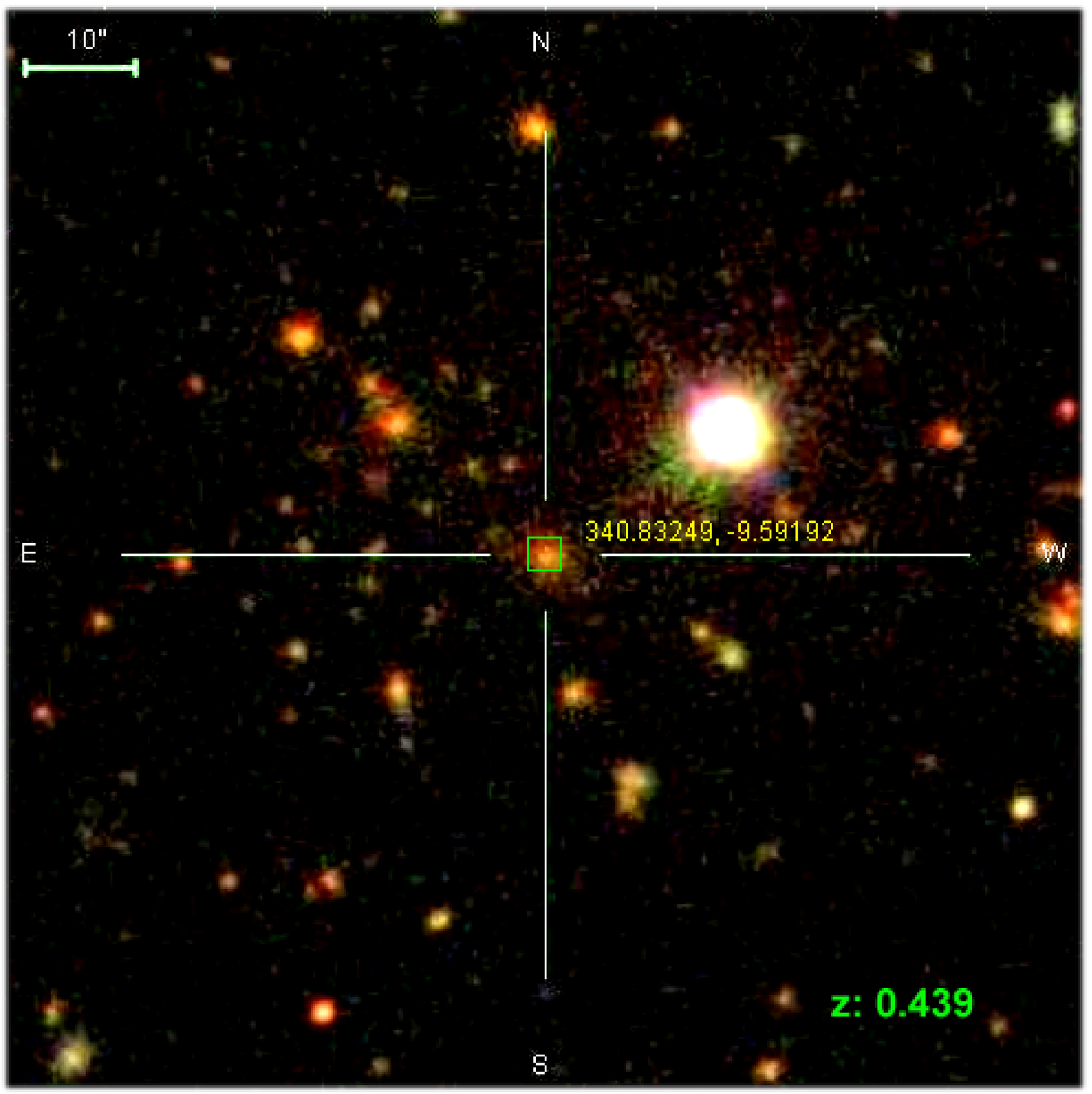}\hfill
\includegraphics[width=2in,height=2in]{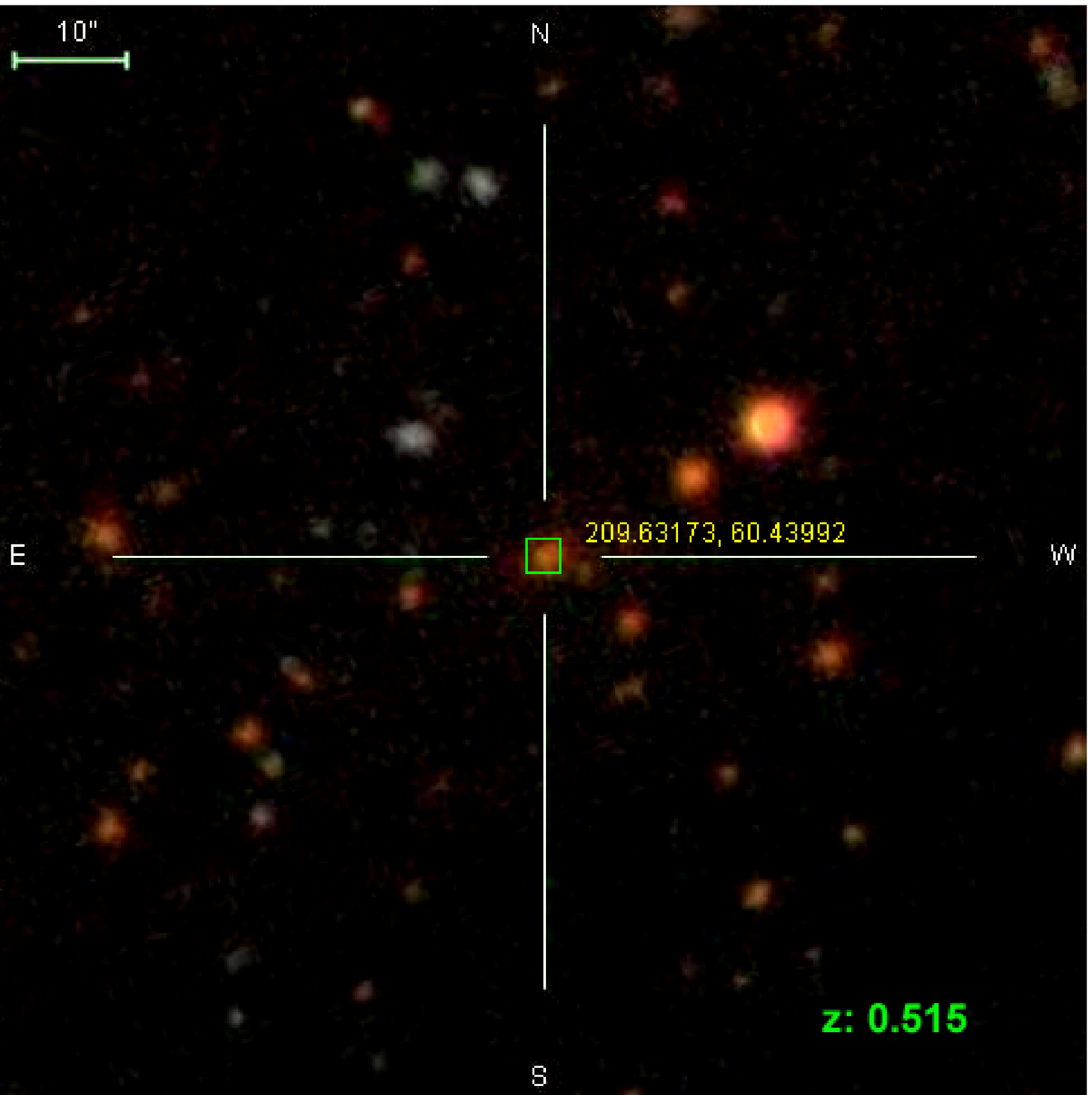}\hfill
\includegraphics[width=2in,height=2in]{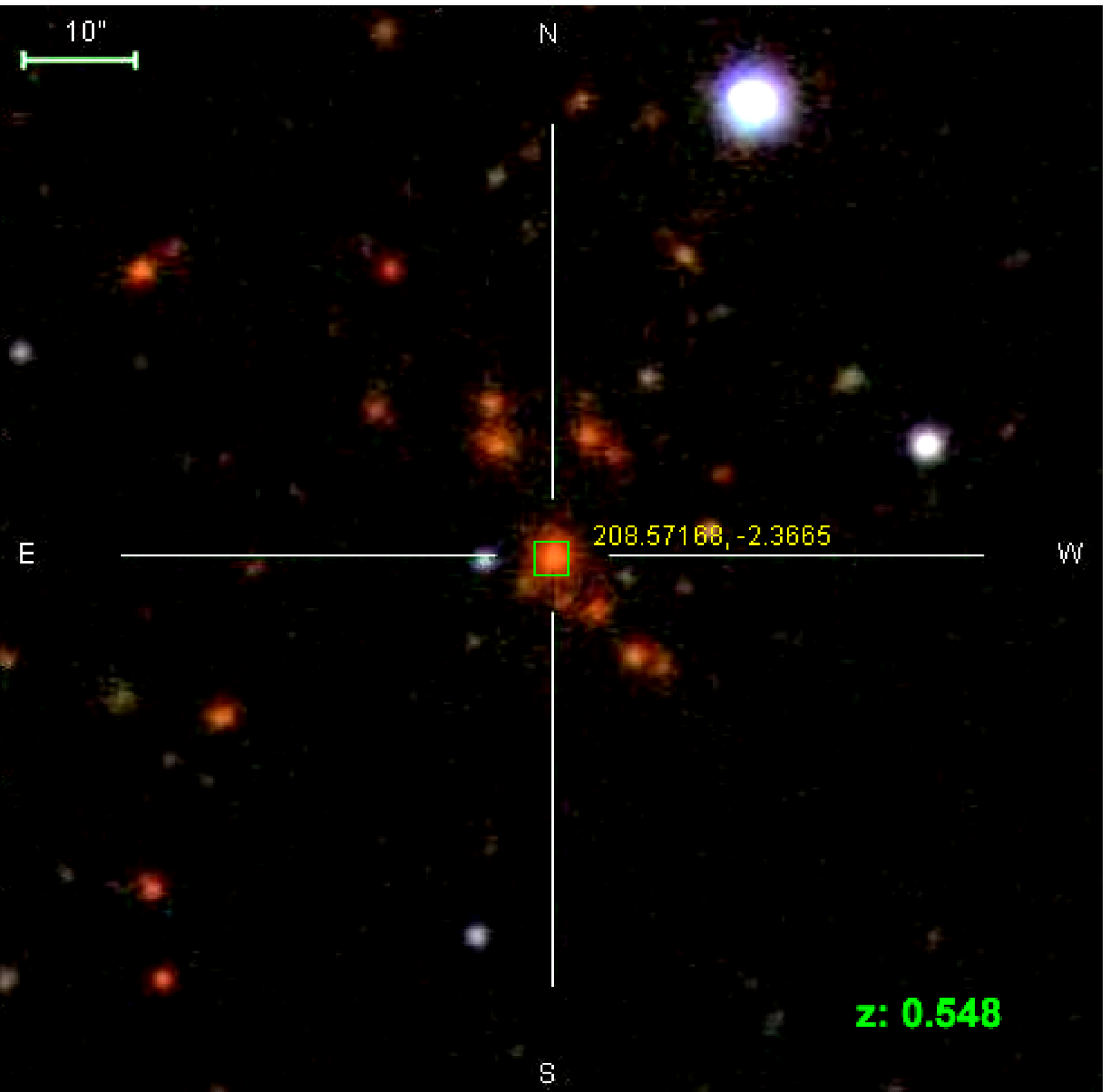}\hfill
\caption{Sample cluster images from SDSS DR7 cluster catalog.  The BCG
  spectroscopic redshift is given in green.}
\label{fig:clusterexample}
\end{figure*}

Cleaning and masking trims the final catalog down to 380,000 clusters,
which we will refer as full catalog. When we apply a richness cut
$N_{gals}^{scaled} \ge 8$, we are left with about 55,000 rich
clusters, which we release with this paper. We refer this as the
``public catalog'' and its sky coverage is shown in
Figure~\ref{fig:coverage}. In Table \ref{table:catalog}, we list the
tags in the public cluster catalog and their corresponding
definitions. The redshift and richness distributions of the clusters
in the public catalog are shown in
Figure~\ref{fig:photoz_ngals_distribution}.  Images of example clusters
at different redshifts are shown in Figure~\ref{fig:clusterexample}.

\begin{table*}
\caption{Tags in the cluster catalog}\label{table:catalog}
\begin{minipage}[b]{1\linewidth}
\small
\begin{tabular}{l l}
\hline\hline Tag Name in catalog & Definition\\
\hline
   OBJID & Unique ID of each galaxy in SDSS DR7 \\
   RA &  Right Ascention\\
   DEC & Declination\\
   PHOTOZ & photo-$z$ from the photo-$z$ table in DR7\\
   PHOTOZ\_ERR & Errors of photo-$z$ \\
   SPZ & Spectroscopic redshift\\
   GMR & $g-r$ color\footnote{All colors are calculated using model magnitude}\\
   GMR\_ERR & Error of $g-r$ color\\
   RMI & $r-i$ color \\
   RMI\_ERR & Error of $r-i$ color\\
   MODEL\_MAG & Dust extinction corrected model magnitude\footnote{For details, see http://www.sdss.org/DR7/algorithms/photometry.html}\\
   MODEL\_MAG\_ERR & Error of model magnitude\\
   S\_CLUSTER & Clustering strength, $S_{cluster}$\\
   GM\_SCALED\_NGALS & Number of member galaxies inside GM\_SCALEDR from BCG\\
   GM\_NGALS\_WEIGHTED & Weighted richness.\\
   WEIGHTOK & If it is set to 1, we recommend the use of weighted richness for this cluster\\
\hline
\end{tabular}
 \vspace{-7pt}\renewcommand{\footnoterule}{}
 \end{minipage}
\end{table*}

\begin{figure*}
\begin{center}
\includegraphics[width=2.5in, height=2.5in]{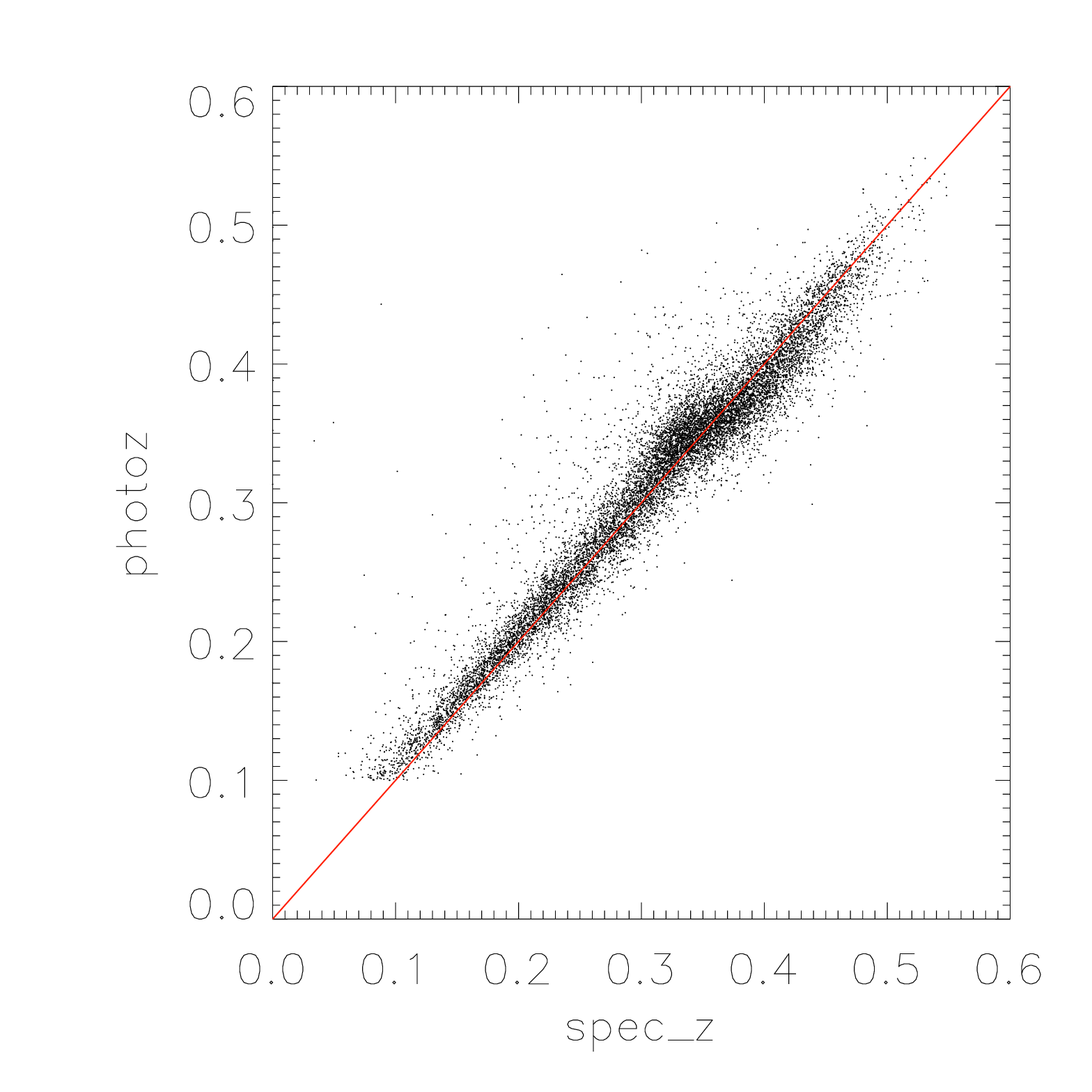}
\includegraphics[width=2.5in, height=2.5in]{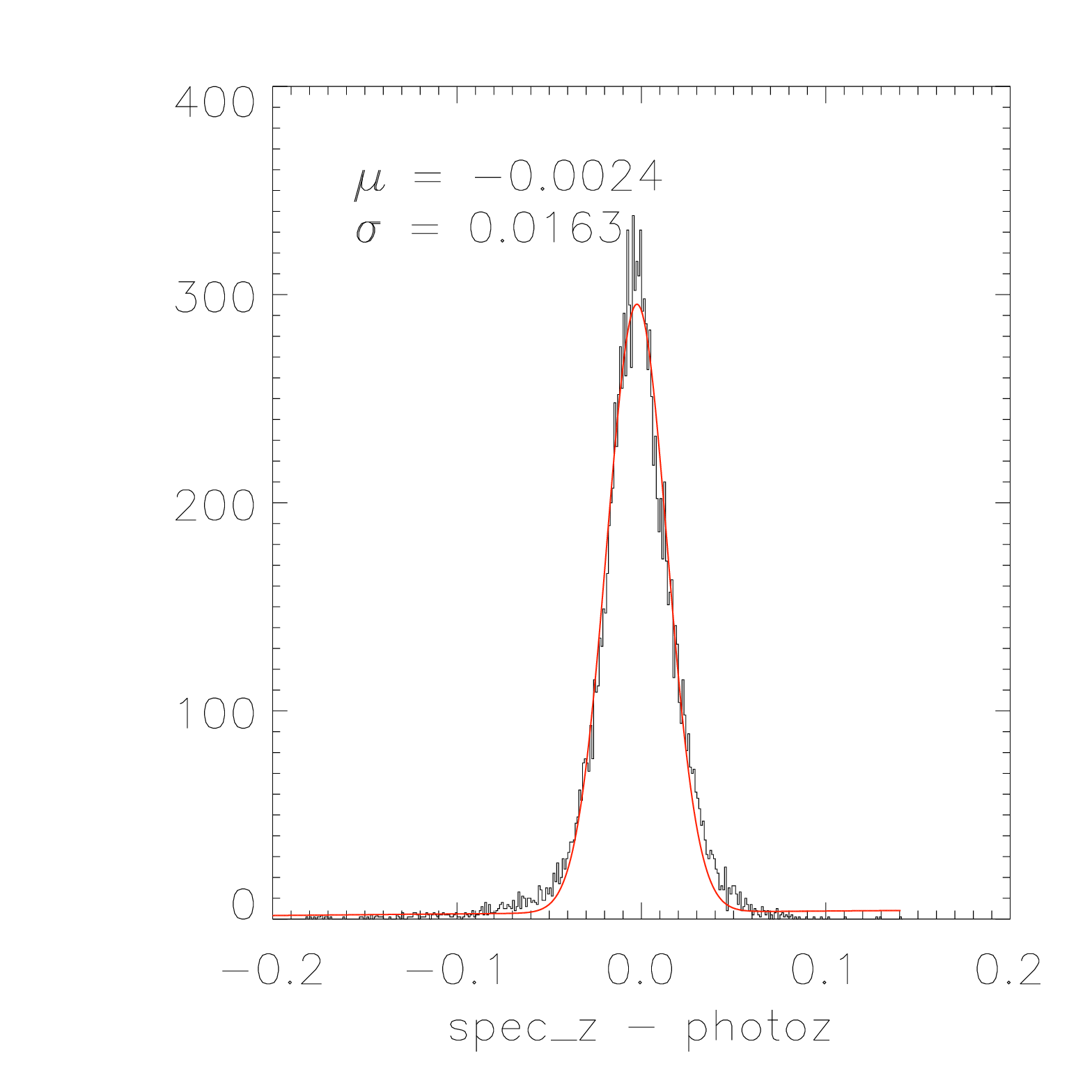}
\caption{The difference between photo-$z$ and spec-z for the BCGs in the public catalog.} \label{fig:photoz_specz}
\end{center}
\end{figure*}

An inherent assumption in GMBCG is that the BCG's photo-$z$ should be
determined much better than the rest of galaxies. We now test that
assumption. In the public catalog, about 20,000 BCGs have
spectroscopic redshift. In Figure~\ref{fig:photoz_specz}, we show the
performance of photo-$z$ for BCGs. The rms of the difference between
BCG and photo-$z$ is $\sim 0.015$, which is almost the same as the
photo-$z$s from maxBCG clusters~\citep{koester07cat}, an indication
that the assumption is secure.

\subsection{Bimodality in color Space}
As we have shown in previous sections, the apparent color distribution
around a cluster generally shows bi-modality. However, there are
situations where the cluster is so big that its members completely
dominate the field within the aperture we impose; in this case, the
color distribution may be uni-modal. In our implementation of the GMBCG
algorithm, we also consider this situation as a potential cluster as
long as the width of the dominant uni-modal distribution is narrow
enough (width $<0.16$).

\begin{figure*}
\begin{center}
\includegraphics[width=6in,height=7in]{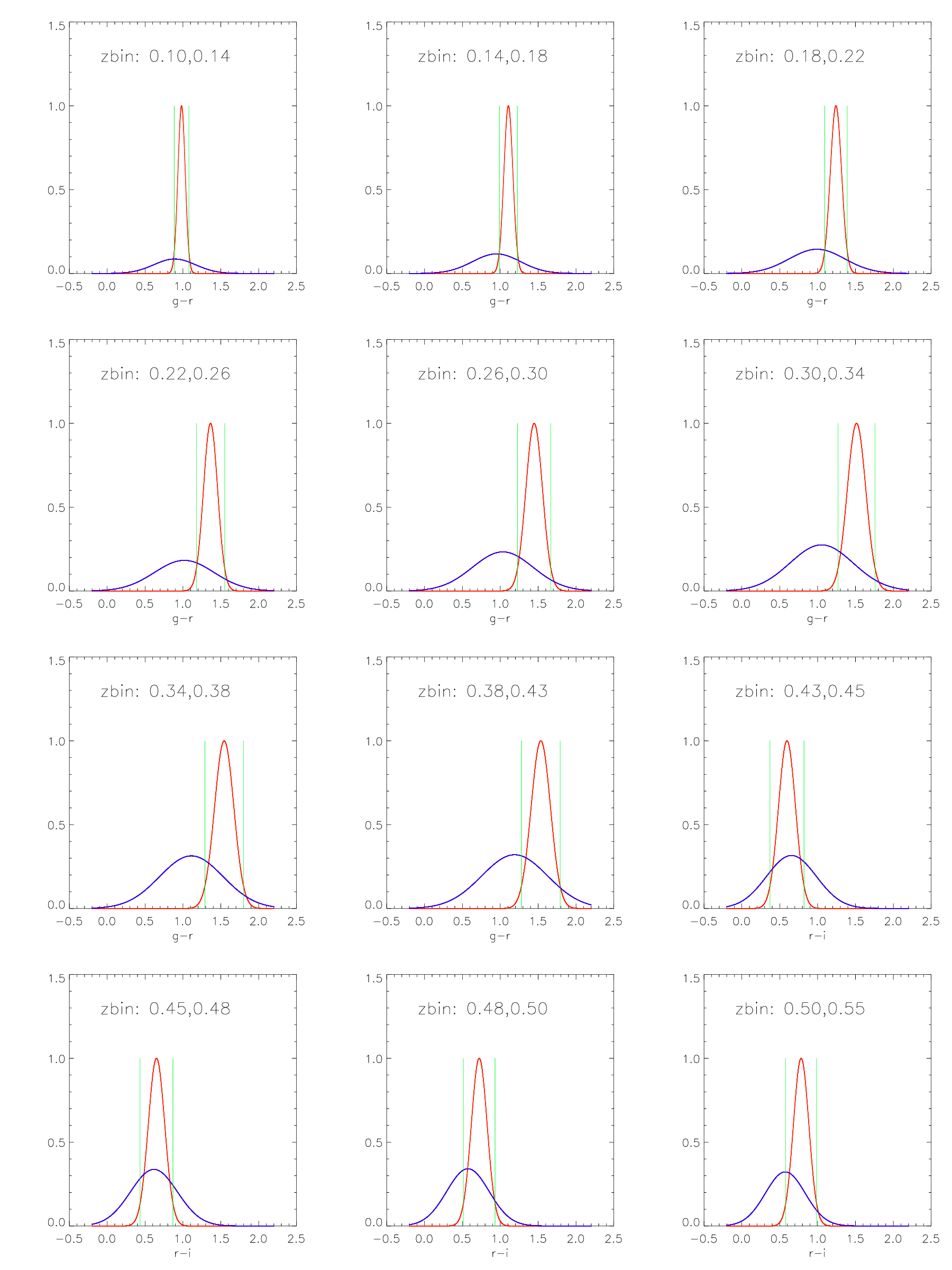}
\caption[Bimodality of color space]{The bimodal distribution of red
sequence galaxy colors and background/blue galaxies. The results are
based on the average results of clusters falling in each redshift bin as indicated in
the plots. The green vertical lines are the $2\sigma$ clip of the
red sequence peak. } \label{fig:bimodal}
\end{center}
\end{figure*}

In the case of a bimodal color distribution, the separation between
the two Gaussian components will vary as redshift changes, leading to
different degrees of overlap. This overlap of the two Gaussian
components measures the fraction of projected galaxies when we impose
the color cuts on the red sequence galaxies.  Therefore, the richness
for the clusters should be appropriately weighted to account for the
projection. In Figure~\ref{fig:bimodal}, we show the color
distribution of clusters at different redshifts.  From the plot, the
$2\sigma$ ($\sigma$ is the standard deviation of the Gaussian
component corresponding to red sequence) cut we imposed for selecting
red sequence members coincides with point at which the likelihood of
red sequence galaxy becomes equal to that of background/blue galaxies.

This information is important for getting consistent richness
estimates across the redshift range. The 2 $\sigma$ clip we use to
select member galaxies will lead to different levels of background
galaxy contamination at different redshifts. The weighted richness
introduced in \S~\ref{wrichness} takes this overlap into account
automatically and thus is a better richness estimator than the direct
cluster member counts based on the top-hat $2\sigma$ color
cuts. However, the weighted richness is not always better than the
direct number counts. There are two cases that demand caution when the
weighted richness is used. In first case there is only one Gaussian
component, which does not permit a weighted richness. The second case
is that there are situations where the relative weight estimates from
the ECGMM is not reliable, e.g. very small, leading to a very small
weighted richness. In this case, we recommend the direct richness
counts, i.e. $N_{gal}^{scale}$. To make this more clear in the public catalog, 
we have a tag ``WEIGHTOK''. The weighted richness is recommended if ``WEIGHTOK'' is set to one.

\section{Evaluating the catalog}\label{cattest}

Any cluster finding algorithm can be evaluated by two simple criteria:
completeness and purity. Completeness quantifies whether the cluster
finder can find all true clusters, while purity quantifies whether the
clusters found by the cluster finder are real clusters. However,
calculating the completeness and purity requires that we know in
advance what is a true cluster. Ideally, the true cluster here should
correspond to a dark matter halo. This issue can only be completely
resolved when we have a high resolution simulation that can properly
reflect the galaxies' colors as well as their interaction with dark
matter halos. However, creating a realistic galaxy catalog from the
N-body simulation has proven to be very challenging, complicated by
various factors such as unknown physics processes, limited resolution of
simulation, unknown behaviour of galaxies at high redshift, and other
complications that affect the evolution of galaxy colors and
distribution.   Therefore, in
practice, we need to slightly change the definition of true cluster to
certain model clusters we defined in terms of observational features.

In this section, we introduce a simple but realistic mock catalog to
test our cluster finder. The result can tell us the purity and
completeness of our cluster catalog with respect to the model clusters
we put in. In addition, as a check of completeness of the cluster
catalog, we also cross match our clusters to X-ray clusters and
clusters from maxBCG catalog. Considering uncertainties in cluster
richness measurement, we will use the full catalog in this section to
accommodate the richness variances.

\subsection{Mock catalog}

Inserting model clusters into a realistic background is a widely used
method to create mock catalogs for evaluating cluster finding
algorithms~\citep{diaferio99,adami00,postman02,kim02,goto02,
  koester07cat}. In practice, there are different schemes to make the
mock catalog as realistic as possible. In this paper, we develop a
Monte Carlo scheme that is similar to those used in~\citep{goto02,
  koester07cat}, but with additional features. We construct mock
catalogs in four steps:
\begin{enumerate}
\item \emph{The background galaxy distribution}: To make a realistic
  background, we consider 25 stripes from our input galaxy catalogs
  from SDSS DR7.  We remove the rich clusters (richness greater than
  20 in our cluster catalog, about 4\% of the total galaxy in the input catalog) and shuffle the remaining galaxies'
  positions (ra/dec), while keeping their colors and other properties
  unchanged, creating a 'base' catalog.
\item \emph{Model cluster selection}: We select 49 rich clusters whose
  redshift ranges from 0.1 to 0.55 from our cluster catalog. About
  60\% of these clusters have a match with known x-ray clusters (see
  \S \ref{rosatmatch}) and all of them have been visually checked
  to be very rich. Each cluster has a BCG and about 30-100 member
  galaxies brighter than $0.4L^*$.
\item \emph{Model mock clusters re-sampling}: Pick a BCG randomly from
  the 49 model clusters and then select a fixed number of member
  galaxies from the corresponding model cluster's members. The fixed
  number is randomly chosen from [10, 15, 20, 25, 30, 35, 40, 45,
  50]. The relative positions, colors and luminosities of these
  galaxies all remain unchanged with respect to BCG. In this way, we
  can generate a re-sampled model cluster of a given richness.
\item \emph{Putting re-sampled model clusters into base catalog:} For
  every stripe of the base catalog, we select 500 re-sampled model
  clusters (roughly the number of clusters removed in step 1) and put
  them into the background galaxy catalog so that their corresponding
  BCGs replace 500 randomly chosen galaxies in the base catalog. Then,
  we will have a Monte Carlo catalog that are based on the real
  photometry of the SDSS DR7 data.
\end{enumerate}
By construction, the Monte Carlo catalog is based on actual SDSS
photometry, and produces a mock catalog with reasonably realistic
background galaxies.

\subsection{Completeness and Purity}

\begin{figure*}
\begin{center}
\includegraphics[width=5.5in, height=3in]{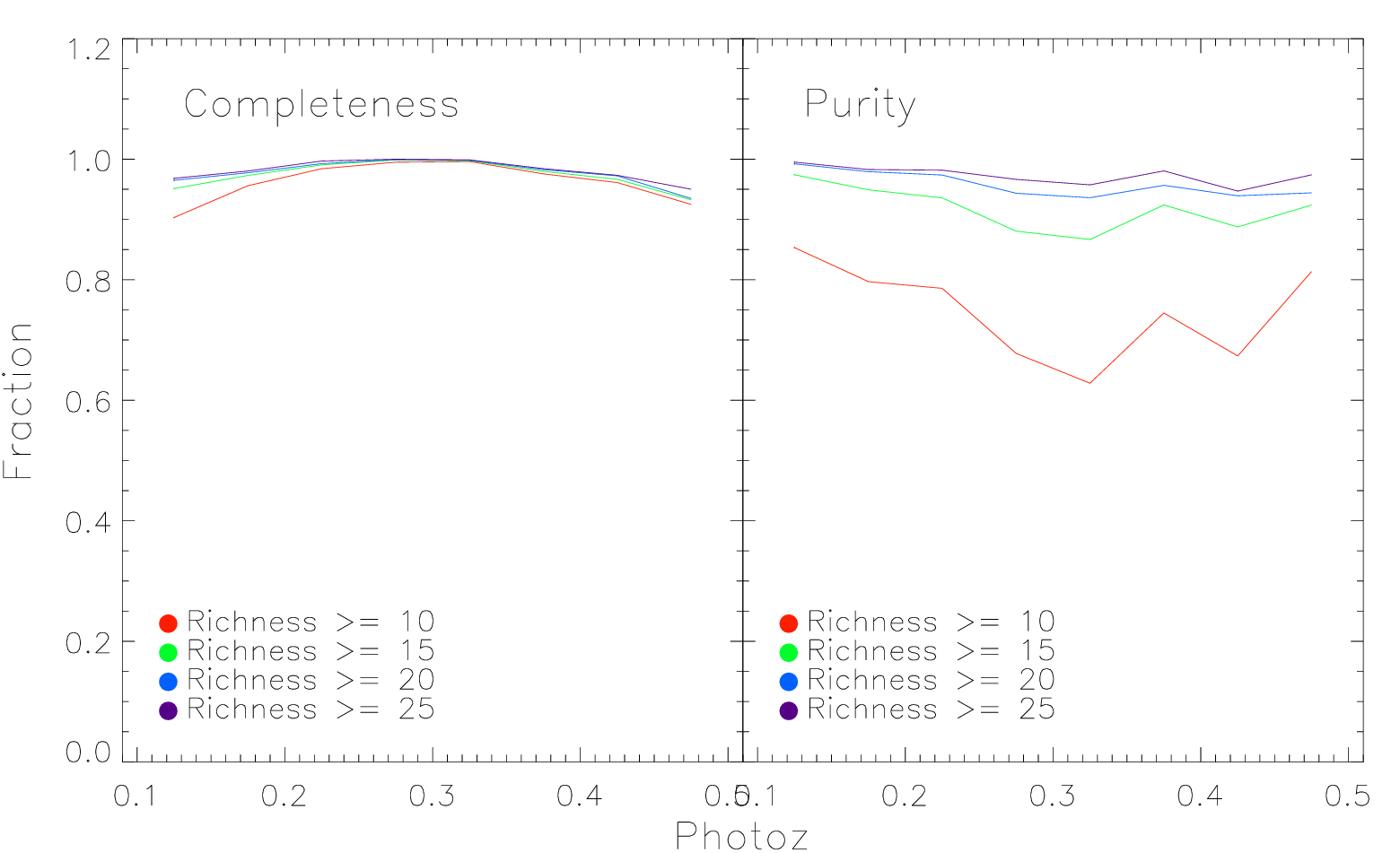}
\caption[Completeness and purity]{The completeness and purity of the
GMBCG catalog based on the Monte Carlo catalog. In the completeness
plot, ``Richness'' is the number of member galaxies of our input model
clusters. In the purity plot, ``Richness'' is the number of
member galaxies measured by the cluster finder. } \label{fig:comp_purity}
\end{center}
\end{figure*}

To test the completeness and purity of our cluster finder, we run it
on the mock catalog created above. Then, we cross match the detected
clusters and the model clusters using a simple cylinder matching,
i.e. searching in a cylinder of $R_{scale}$ in radius and $\pm 0.05$
in redshift. When we test the completeness, we firstly sort the model
cluster list by the cluster richness and then match the detected
clusters to them through the cylinder match. While we test the purity,
we sort the detected cluster list by their richness and then match the
model clusters to them via the cylinder match.  In both cases, we will
consider only those unique and exclusive matches, meaning that a model
cluster will not be used any more once it is matched to a detected
cluster for purity test and a detected cluster will not be used any
more once it is matched to a model cluster for the completeness
test. If more than one cluster falls in the cylinder, we choose the
richest one. After doing the matching, at a given redshift bin and
above a given $N_{gal}$, the completeness and purity can then be
defined as
\begin{equation}
{\rm completeness} = \frac{N_{model}^{match}(z, N_{gal})}{N_{model}(z,
N_{gal})}
\end{equation}
\begin{equation}
{\rm purity} = \frac{N_{found}^{match}(z, N_{gal}^{scaled})}{N_{found}(z,
N_{gal}^{scaled})}
\end{equation}
\noindent where $N_{model}^{match}(z, N_{gal})$ denote the number of
model clusters that are matched to the found clusters by ,
$N_{model}(z, N_{gal})$ is the total number of model clusters ,
$N_{found}^{match}(z, N_{gal}^{scaled})$ is the number of found
clusters that are matched to model clusters and $N_{found}(z,
N_{gal}^{scaled})$ is the total number of found clusters. The results
of the completeness and purity are plotted in
Figure~\ref{fig:comp_purity}. The plot show that the GMBCG algorithm
can yield a highly complete and pure cluster catalog.

\subsection{Richness Recovery}

\begin{figure*}
\begin{center}
\includegraphics[width=3in, height=3in]{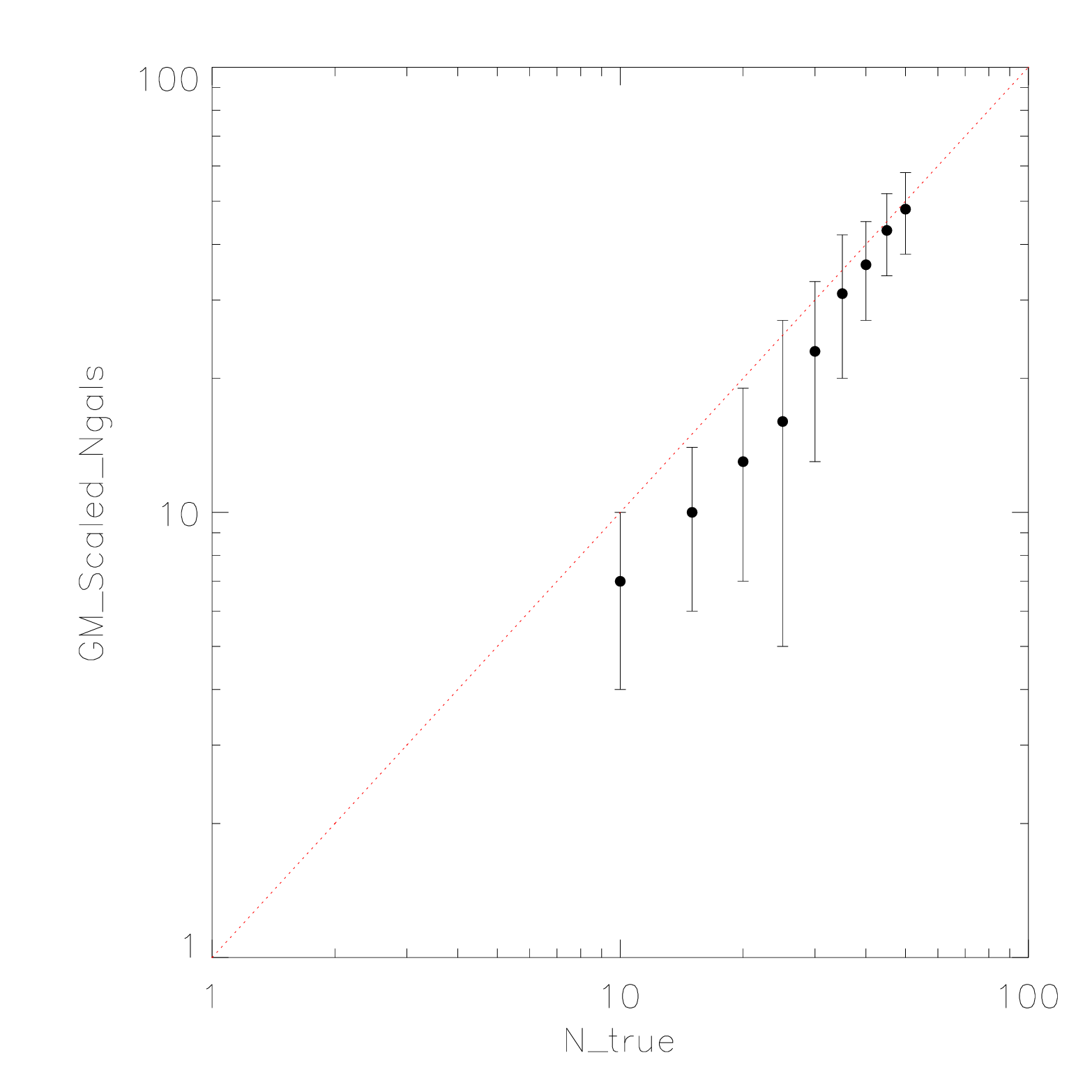}
\caption[Richness recovery]{The recovered richness (GM\_Scaled\_Ngals) from GMBCG vs. the input richness (N\_true). The error bars are the scatters of the recovered richness for the given input richnesses (N\_true). The red dotted line is the 45 degree line.} \label{fig:rich_mc}
\end{center}
\end{figure*}

In addition the the completeness and purity, it is also important to compare the richness estimated from GMBCG and the input richness of the mock clusters. Note that when we create the mock clusters, the mock cluster richness is randomly sampled from [10, 15, 20, 25, 30, 35, 40, 45, 50]. That is, the possible input richness of the mock clusters are only the above numbers. Therefore, for a given input richness, we look at the distribution of the recovered richness from GMBCG. The results are shown in Figure~\ref{fig:rich_mc}. From the plot, GM\_Scaled\_Ngals well recover the input N\_true, though systematically underestimate the true richness for low richness clusters. The reason for this underestimation is primarily due to an artifact of our mock cluster creation. Our low richness mock clusters are essentially re-sampled from real rich clusters but their relative positions are retained. Therefore, some low richness cluster may have members locate outside of the $R_{scaled}$ based on the low richness, leading to an underestimated richness from the cluster finder.

\subsection{Cross-Matching of GMBCG clusters to MaxBCG Clusters}

As a further test of the completeness of the GMBCG catalog, we make
a comparison to the maxBCG catalog~\citep{koester07cat}. The maxBCG
catalog consists of 13,823 clusters in the redshift range
$0.1<z<0.3$ with a threshold on richness set at $N_{200}=10$.
It is derived from DR5 of the Sloan Digital Sky Survey and covers a
slightly smaller area than the new GMBCG catalog.

Several complications arise in the process of performing
cluster-to-cluster matches between catalogs, namely redshift
uncertainties, centring differences between the two algorithms, and
scatter in the richness measurements. Although many similarities
exist between the maxBCG and GMBCG algorithms, it is not always the
case that they choose the same central galaxy for a given cluster.
When matching clusters, a careful cut must be made in the
two-dimensional physical separation in order to allow for this
centering ambiguity, while at the same time minimizing 
matches due to random projection. Uncertainty in the photometric
redshifts can yield a similar problem along the line of sight;
a cut in $\Delta z=|z_{maxBCG}-z_{GMBCG}|$ must be made to
accommodate these errors.  Finally, the richness measurements
themselves have large scatter, i.e. clusters that appear in one
catalog may have richness values below the richness threshold 
of the counterpart catalog,
rendering them unavailable to match. These problems 
ultimately will determine a reasonable matching scheme to
that can be used to quantify the agreement between the GMBCG
catalog and the maxBCG catalog. We now consider these effects.

The uncertainty in redshift estimates for maxBCG clusters is
$\sigma_{z} \sim 0.015$~\citep{koester07cat}. In the GMBCG catalog,
the uncertainty of the photo-$z$s at redshift below 0.3 is $\sim
0.016$ (Figure\ref{fig:photoz_specz}). Therefore, a redshift difference of $\sim
0.05$ between the two catalogs is an appropriate selection window
for matching. As for the radial separation, given the fact that
the maxBCG clusters are percolated within a separation of $R200 \sim
1.0 - 2.0$ Mpc~\citep{koester07alg}, a radial separation of $\sim
2.0$ Mpc is appropriate for our matching search. Generally speaking, the smaller the matching separation, the higher
the probability of real matches. Also, the lower the richness of the maxBCG
cluster, the less likely they are true clusters. Therefore, we will represent our matching with respect to both the separation and the richness of maxBCG
clusters.

\begin{figure*}
\begin{center}
\includegraphics[width=2.5in, height=2.5in]{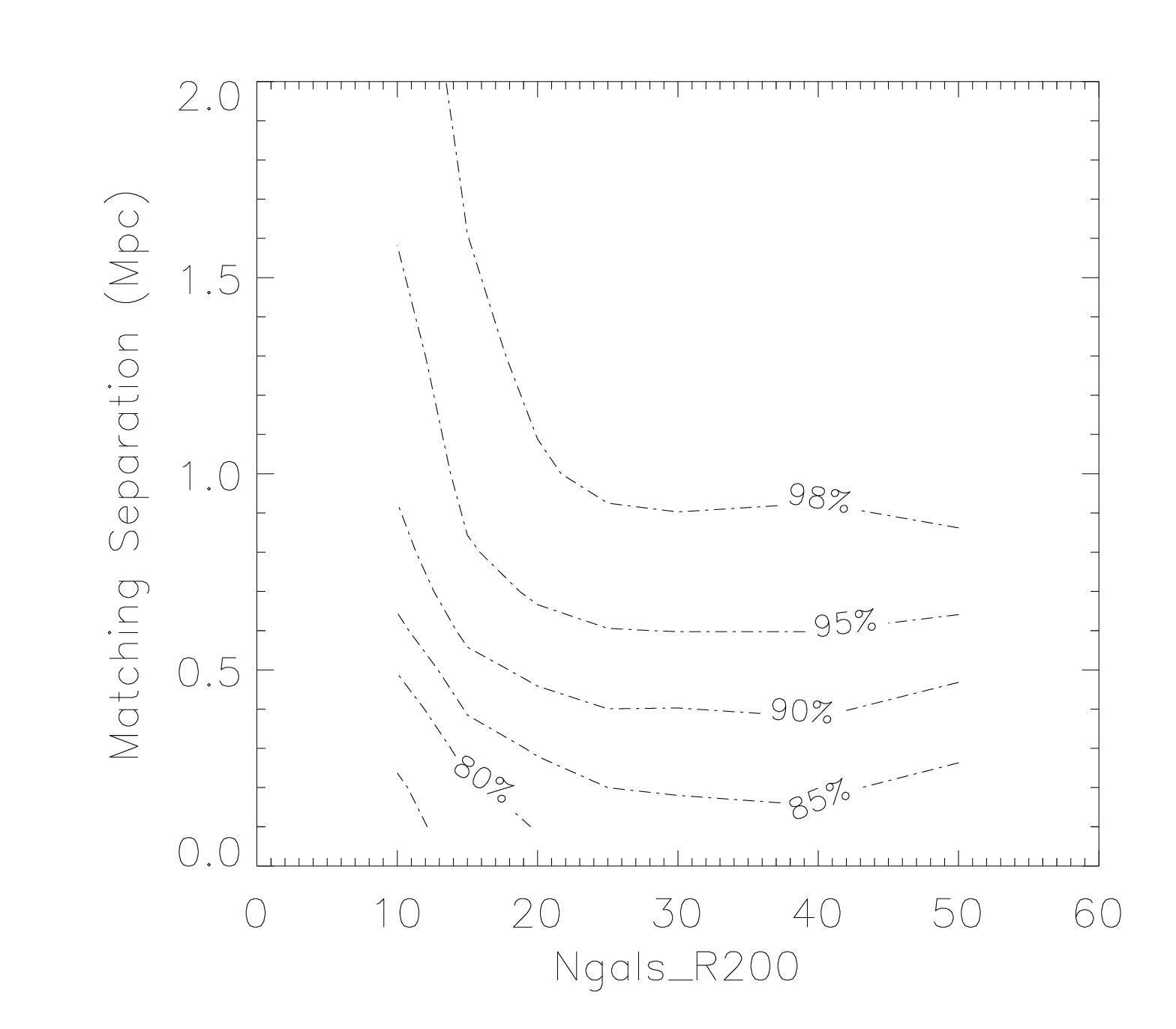}
\includegraphics[width=2.5in, height=2.5in]{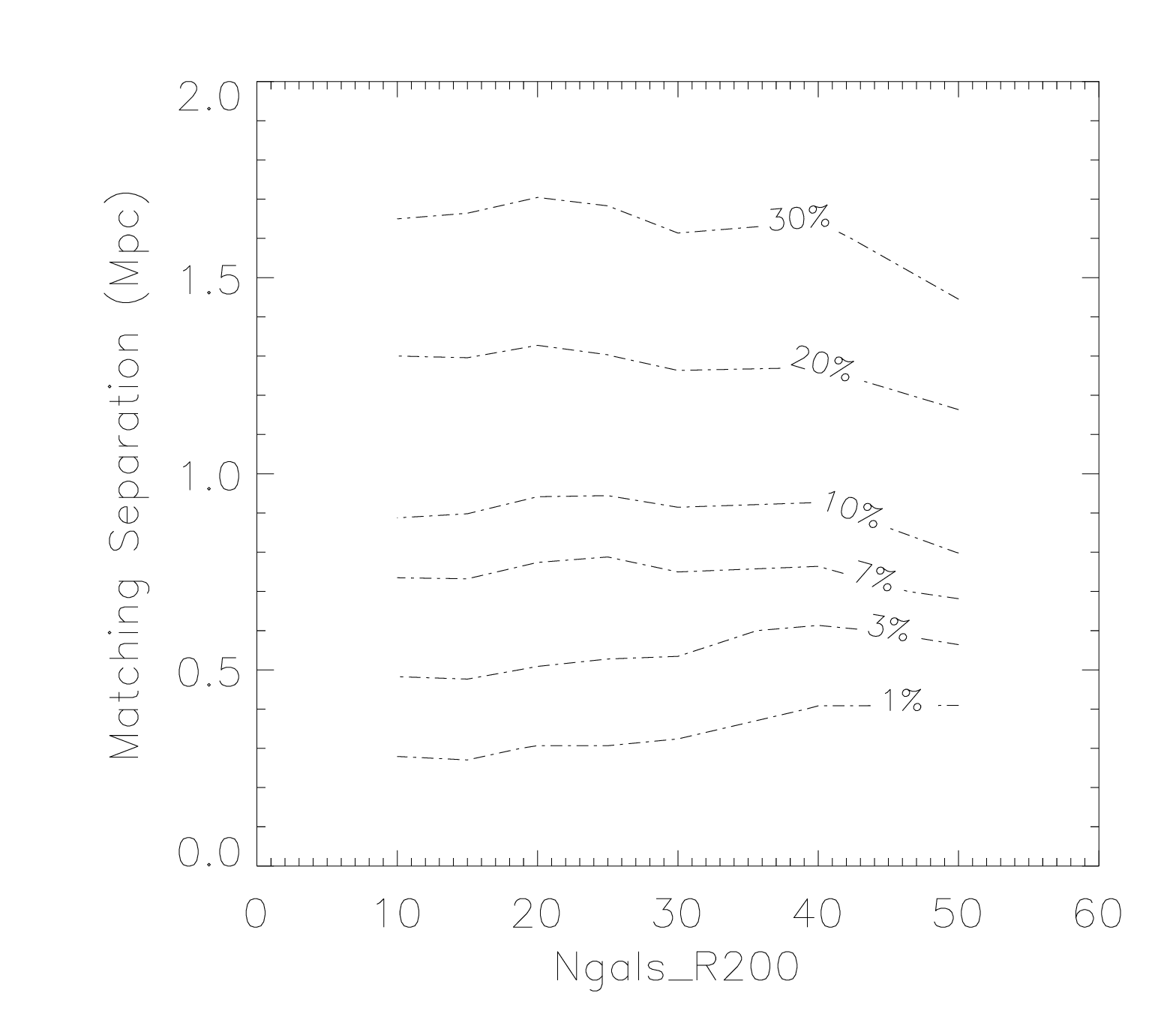}
\caption[Matching to maxBCG clusters]{Left panel is the contour of
matching fraction of the maxBCG clusters to the GMBCG clusters as a function of richness (Ngals\_R200) and separations. From the plot, we can read that for clusters with richness above 20 in maxBCG catalog, 90\% of them can be matched to GMBCG clusters with separations less than 0.5 Mpc. In the right panel, we show the matching results from the control catalog of random positions.}
\label{fig:maxbcg_gmbcg_contour}
\end{center}
\end{figure*}

We hold the maxBCG clusters as target and match the clusters from our
full GMBCG catalogs to them. In other words, it is essentially a
completeness test of the GMBCG catalog. We then execute the
cylindrical matching algorithm described above. The matching yields
that 13,374 out of 13,823 ($\sim 96.8\%$) clusters in maxBCG catalog
have a match in the GMBCG catalog. Those non-matched clusters are
mostly at low richness end, which is mainly due to the low end cuts
placed on the catalog. There are also 8,818 of the 13,374 matched
clusters ($\sim 65.9\%$) that have identical BCGs in both catalogs. In
the left panel of Figure~\ref{fig:maxbcg_gmbcg_contour}, we show the
matching fraction of the GMBCG clusters to maxBCG clusters at
different maxBCG richness and separation. As a comparison, we create a
control catalog of the same size as the GMBCG catalog, but with the ra
and dec randomized. The matching results of this control catalog to
maxBCG clusters are shown in the right panel of Figure~\ref{fig:maxbcg_gmbcg_contour}.

\begin{figure*}
\begin{center}
\includegraphics[width=2.5in, height=2.5in]{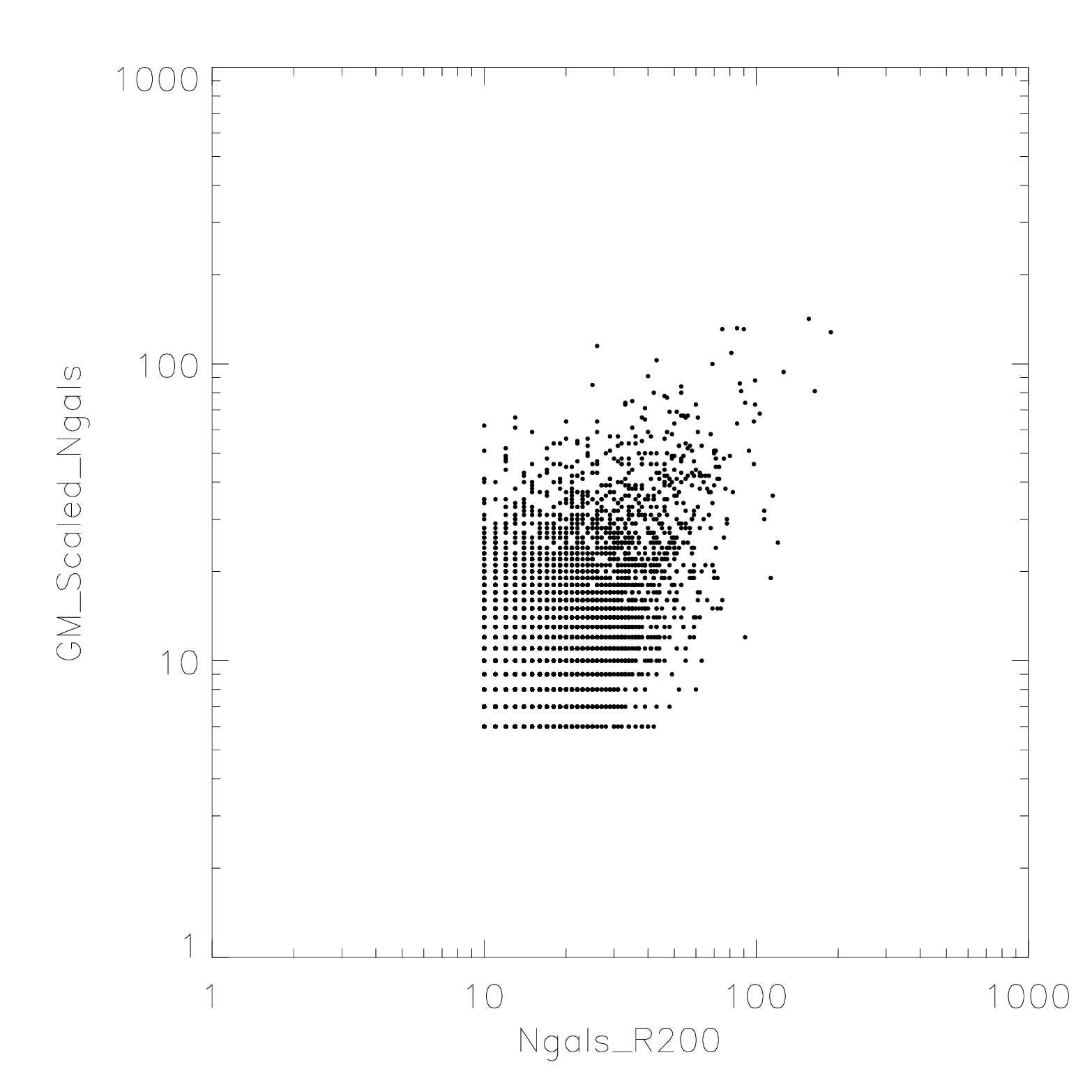}
\includegraphics[width=2.5in, height=2.5in]{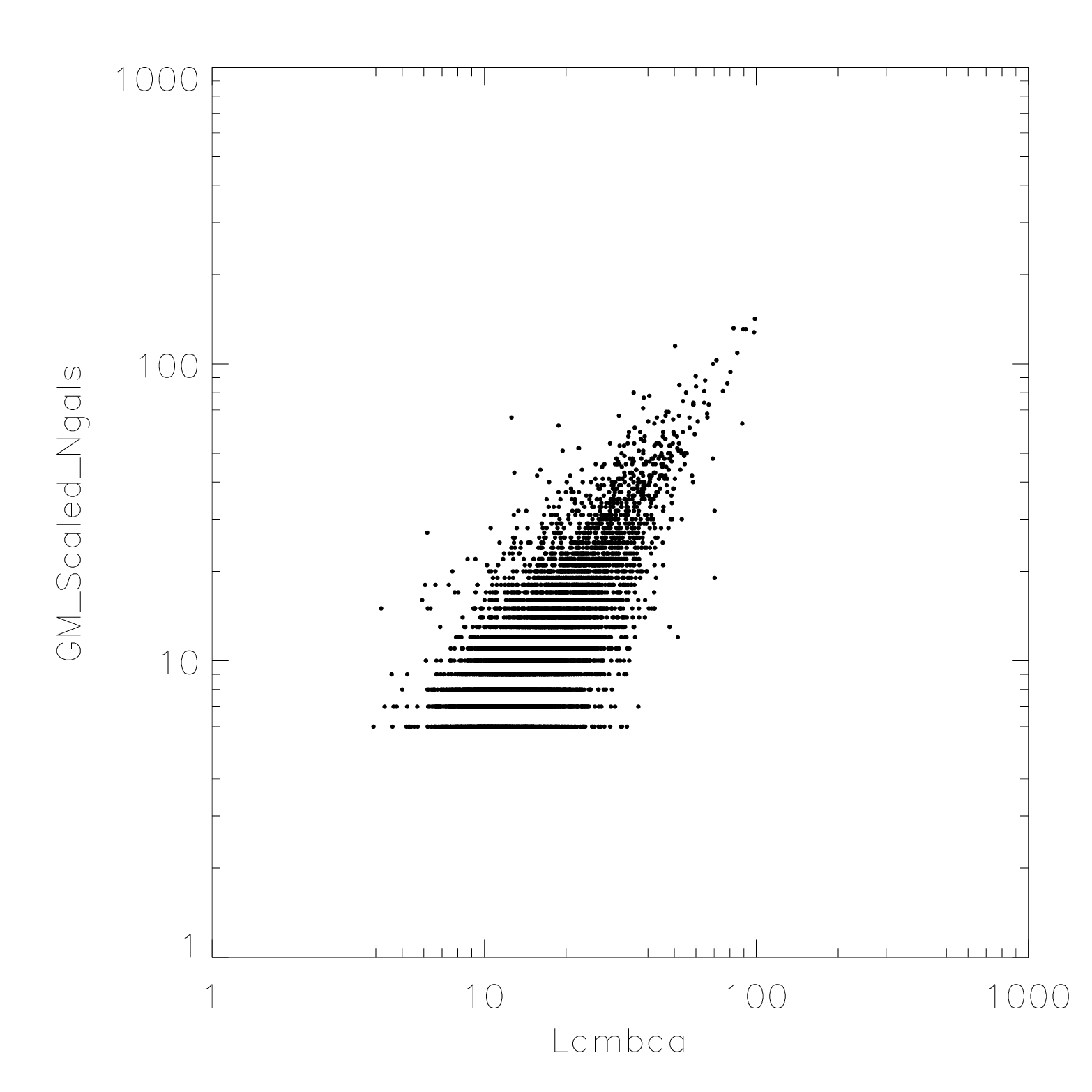}
\caption{Left panel is the comparison of richness measured in maxBCG (Ngals\_R200) and in GMBCG (GM\_Scaled\_Ngals) for those matched clusters. In the
right panel, we show the comparison of GM\_Scaled\_Ngals and Lambda for the same sample of matched clusters.}
\label{fig:richness_max_gm}
\end{center}
\end{figure*}

On the other hand, it is interesting to compare the richness estimate for the cross matched clusters. Note that in maxBCG, the cluster members are counted from the BCG color while in GMBCG, members are selected from the ridgeline. This will lead to large scatters among the two richness estimates. However, in \citet{rozo0809}, a new richness estimator, Lambda, was proposed and it out-performs the original richness estimator (Ngals\_R200) in \citet{koester07cat}. So, the richness estimate from the GMBCG catalog should have tighter correlation with the Lambda than with the Ngals\_R200. In Figure~\ref{fig:richness_max_gm}, we plot the richness comparisons.

\subsection{Matching with WHL clusters}

In addition to maxBCG clusters, we also match the GMBCG clusters with clusters from the catalog by \citet{wen09} (WHL catalog hereafter), which is built based on SDSS DR6 and ranges from 0.05 to 0.6 in redshift. We apply the same matching codes we used for maxBCG catalog to the WHL catalog. There are about 22,000 clusters in WHL catalog can find matched clusters in GMBCG catalog by cylindrical matching (2 Mpc, 0.05 redshift uncertainty). Among these matches, 13,531 of them have identical BCGs detected in both catalogs. 

\begin{figure*}
\begin{center}
\includegraphics[width=6in, height=2in]{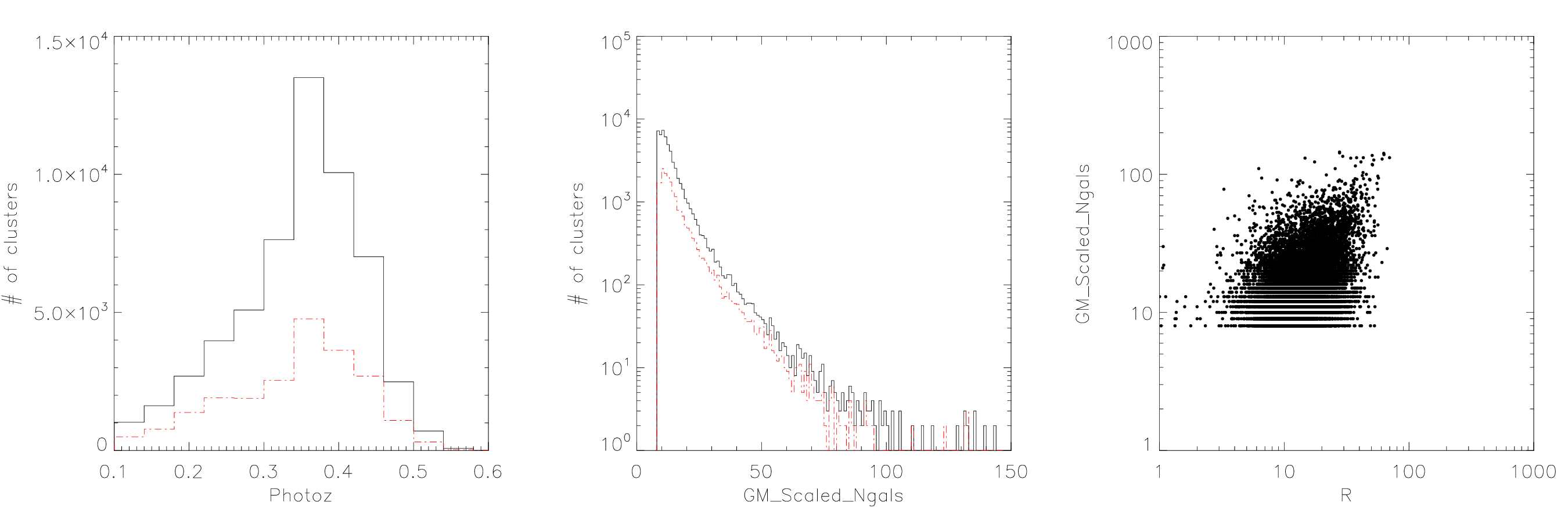}
\caption{Left panel is the distribution of GMBCG clusters and the matched clusters with WHL catalog. Middle panel is the same for cluster richness distribution. The red dashed lines in both panel denotes the distribution of those cross matched clusters. Right panel is the comparison of richness measured in WHL catalog (R) and in GMBCG catalog (GM\_Scaled\_Ngals) for those matched clusters. }
\label{fig:wen_gm_rich}
\end{center}
\end{figure*}

In Figure~\ref{fig:wen_gm_rich}, we show the matching results. The richness scales reasonably in the two catalogs. 

\subsection{Cross-Matching of GMBCG to ROSAT X-ray Clusters}\label{rosatmatch}

Optical identification of peaks in the galaxy distribution represents
only one of many methods used to find clusters. Other observables
employed in cluster detection include thermal emission of x-rays from
the hot intra-cluster medium, weak-lensing distortion of background
sources, and the Sunyaev-Zeldovich effect of hot gas on the cosmic
microwave background. Each method has certain advantages and
disadvantages. Each also provides a distinct proxy for the mass of a
cluster, which can be used to probe cosmological constraints. It is
important that our cluster finding algorithm be able to detect those
clusters found by alternative means.  X-ray cluster catalogs are the
most appealing candidate for exploring this question.  Numerous x-ray
catalogs exist with large sky coverage overlapping the DR7 survey
area. Follow up optical examination is frequently performed on these
catalogs to confirm their identity as clusters and is required to
obtain redshifts.

Matching complications likewise arise when comparing to X-ray catalogs. 
It is not always the case that the BCG lies
exactly on the X-ray peak. There also exists significant scatter in the
x-ray luminosity-richness relation~\citep{rykoff08}. Furthermore,
the DR7 catalog contains clusters down to a richness threshold much
lower then current x-ray catalogs can detect. The main goal of this
subsection is to test the extent to which our algorithm is able to
identify the most luminous x-ray clusters.

We compare the DR7 catalog to three x-ray identified cluster catalogs:
NORAS ~\citep{bohringer00}, REFLEX ~\citep{bohringer04} and 400
deg$^{2}$ ~\citep{Burenin07}. NORAS and REFLEX consist of clusters
identified from extended sources on the ROSAT all-sky survey x-ray
maps. Together they cover the northern and southern galactic caps and
are flux limited at $3 \times 10^{-12}$ ergs s$^{-1}$cm$^{-2}$ in the
0.1 - 2.4 KeV energy band. The 400 deg$^{2}$ catalog is composed of
serendipitous clusters found in the high galactic latitude ROSAT
pointings. It is flux limited at 1.4 s$^{-1}$cm$^{-2}$ in the 0.5 -
2.0 KeV energy band. Sources from all three catalogs have been
confirmed as clusters through follow up optical
identification. Combining these catalogs yields 229 unique clusters in
the survey area spanned by DR7.

\begin{figure*}
\begin{center}
\includegraphics[width=2.5in, height=2.5in]{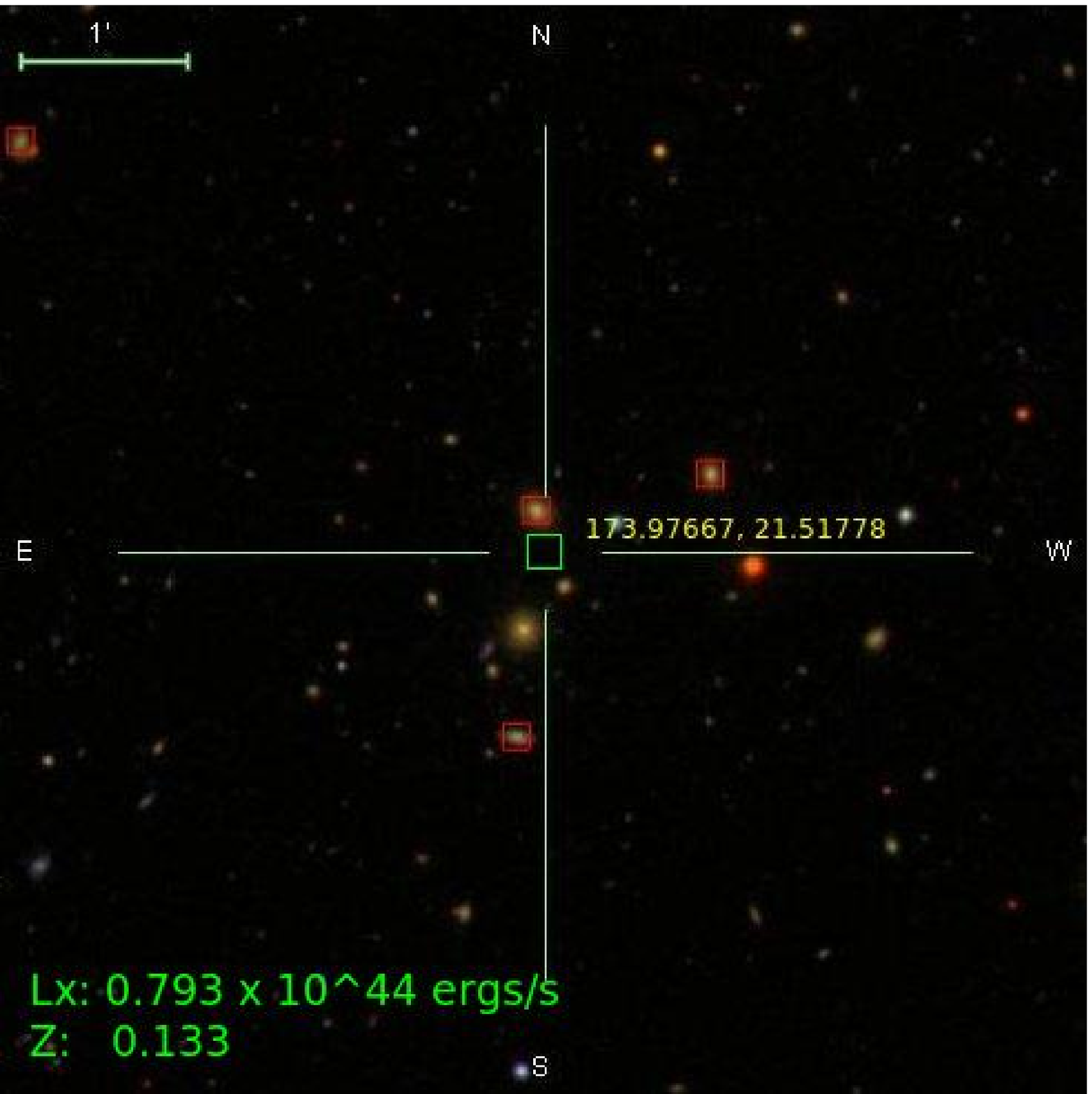}
\includegraphics[width=2.5in, height=2.5in]{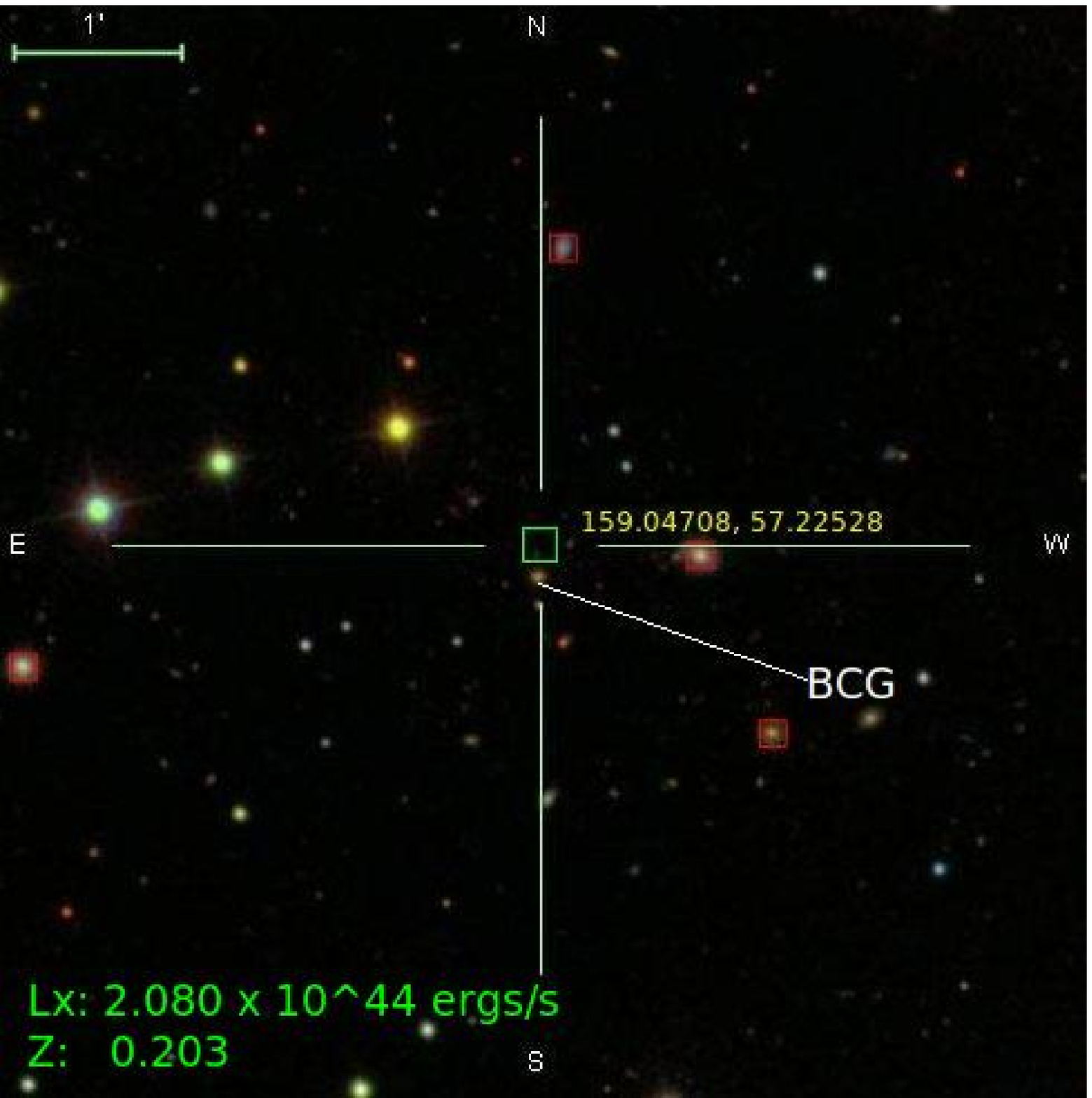}
\caption{Two non-matched X-Ray clusters. The cluster on right panel actually
has a BCG identified in the GMBCG catalog, but it is not recorded as a match
because of the photo-$z$ of the BCG is assigned as 0.549, falling outside of
our redshift matching envelope.} \label{fig:rosat_match_nomatch}
\end{center}
\end{figure*}

\begin{figure*}
\begin{center}
\includegraphics[width=5in, height=3in]{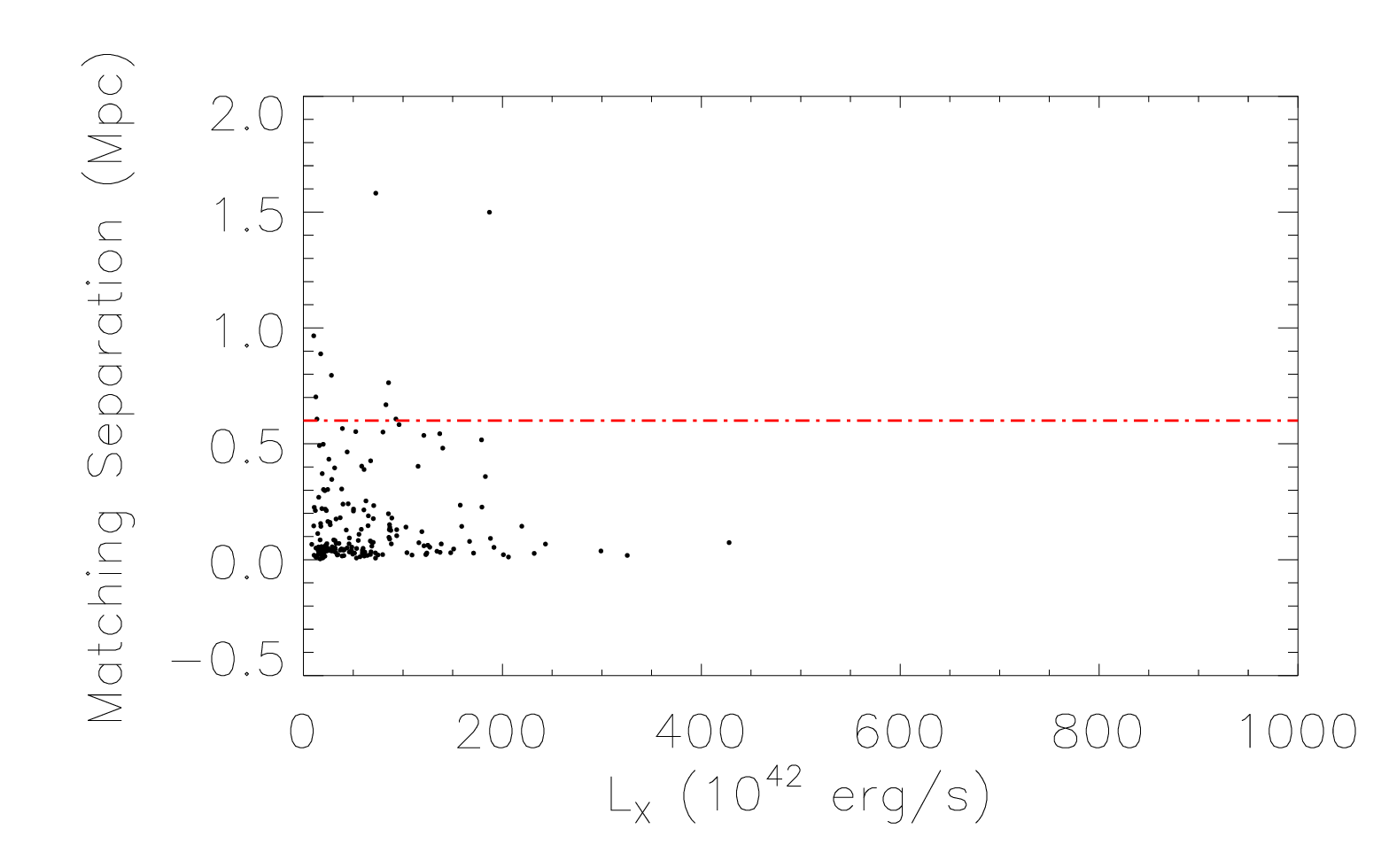}
\includegraphics[width=5in, height=3in]{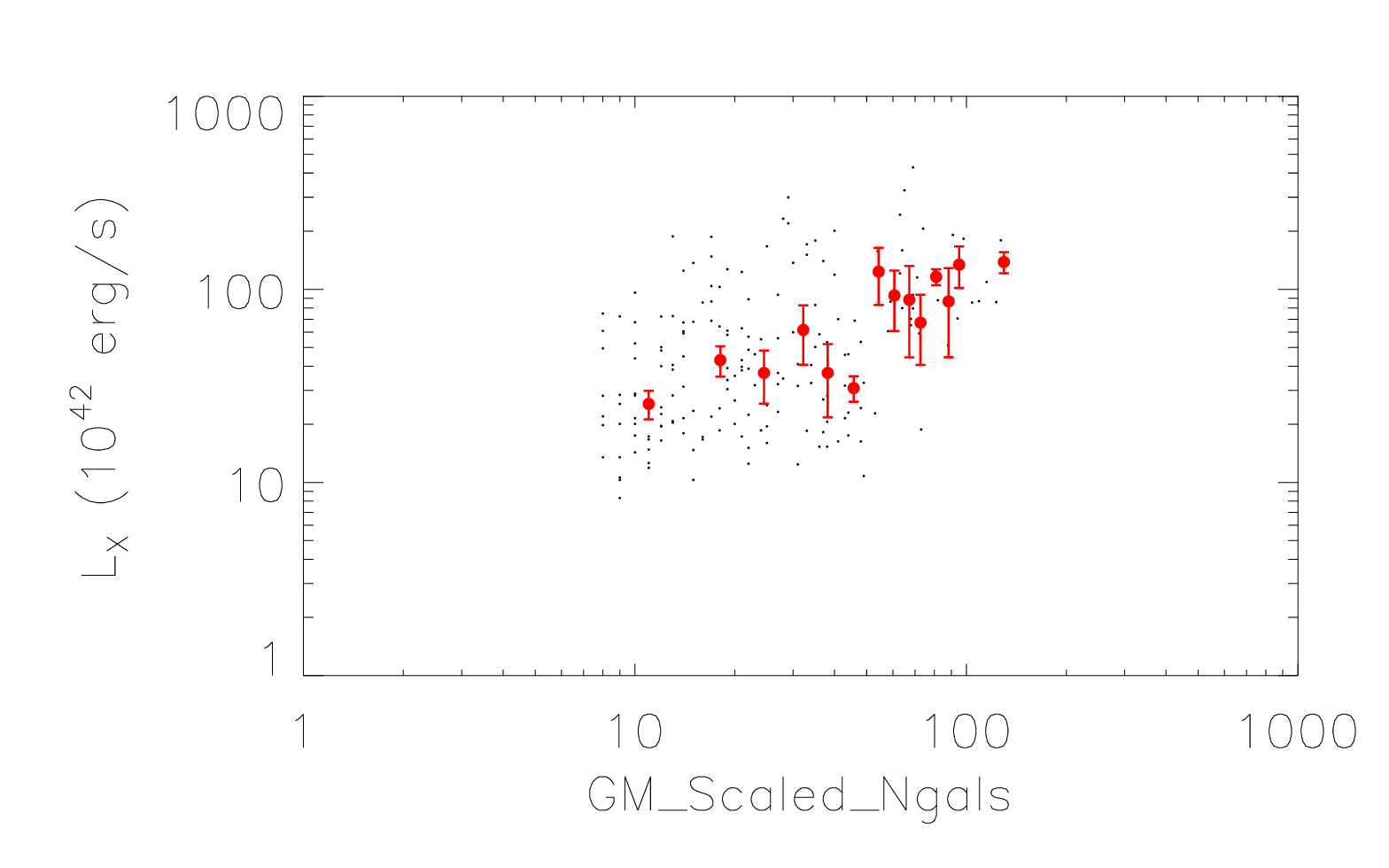}
\caption{Matched ROSAT clusters in GMBCG catalog. Top panel shows the location of matched clusters in the phase plane of X-Ray luminosity and matching separation. 90\% of the matched X-Ray clusters are within a matching separation less than 0.6 Mpc. Bottom panel shows the scatter plot of cluster richness vs. X-Ray luminosity for those matched clusters. The over-plotted red dots and error bars are the median relation and scatter in each richness bin of size 10.}
\label{fig:rosat_gmbcg}
\end{center}
\end{figure*}

A cylindrical search is performed on the combined x-ray catalogs in
order to determine if these clusters were found by the GMBCG
algorithm, effectively a completeness test of GMBCG. We consider two
clusters a match if they have a physical separation in the projected
plane $sep < 2.0$ Mpc and a redshift difference
$|z_{xray}-z_{photo}|<0.05$. By this criteria, 227 out of 229 X-ray
clusters are matched with at least one GMBCG cluster.  In
Figure~\ref{fig:rosat_match_nomatch}, we show the images of the two
non-matched X-Ray clusters. In Figure~\ref{fig:rosat_gmbcg}, we show the scatter plot of the richness and X-Ray luminosity as well as the matching separation vs. X-Ray luminosity for those matched clusters. The results show that we can reliably recover about 90\% of the X-Ray clusters with separation less than 0.6 Mpc.

\section{Discussion}

In this paper, we present a new cluster finding algorithm, GMBCG, and
publish the largest ever optical cluster catalog, with more than
55,000 rich clusters. Compared to the public maxBCG cluster catalog that goes from redshift 0.1 to 0.3, the current GMBCG catalog covers a wider redshift range from 0.1 to 0.55. GMBCG identifies galaxy clusters by detecting
the BCG plus red sequence feature that exists only in galaxy clusters
and is not possessed by field galaxies. This feature provides a
powerful means for detection of galaxy clusters with minimal
line-of-sight projection contamination. The effectiveness of this
algorithm is based on the assumption that a BCG plus red sequence
feature is ``universal" among galaxy clusters. Though this feature is
preserved in almost all clusters known to us, we cannot exclude the
possibility that there are some clusters that do not have this
feature. In particular, this is more likely at very high redshift
where clusters are forming. However, even if such ``blue" clusters
exist, it will be very challenging to detect them using photometric
data in optical bands in a consistent way across a range of richness unless they also exhibit a tight blue-sequence. But it is not very likely for the blue galaxy to be tightly clustered in color since their spectra are not as regular as red galaxy and the effect of 4000~\AA break in their color normally shows large scatters.

The GMBCG algorithm uses the BCG's photo-$z$ to determine the metric
aperture size, and uses the red sequence color to select member
galaxies. It separates the process of getting photo-$z$s and detecting
clusters. This differentiates it from matched filter algorithms
(including maxBCG). For the SDSS data the precision of the photo-$z$s for the BCGs
from the machine learning algorithms are within a factor of 2 of the photo-$z$'s from
maxBCG, which means there is not a serious disadvantage in this choice. Using existing photo-$z$s significantly boosts the computation
efficiency. GMBCG can produce a cluster catalog for the full SDSS DR7
within 23 hours on a DELL computer with a single quad core CPU and 8G
RAM.  As long as the photo-$z$ is not catastrophically bad, GMBCG can
detect the BCG plus red sequence feature of clusters; though richness
measurements may be affected by imprecise redshift estimates.

It is worth noting that for cosmological application, we generally
want to know the best mass proxy. Recent work has shown that weighted
richnesses are among the best optical mass proxies, rather than the
direct counts of member galaxies~\citep{rozolambda}. However, this
does not mean that we should abandon the direct member galaxy count
and identification. On the contrary, it will be very interesting to
have the member galaxies explicitly determined for cluster science,
i.e., the formation and evolution of clusters.

Though GMBCG works very well for the current SDSS DR7 data, there is
still room for improvement, especially for deeper data. For example,
GMBCG does not work well for very low richness clusters, say richness
less than 4 for SDSS DR7 data. This is mainly because GMM/ECGMM will
not reliably detect the red sequence at such low richness. GMBCG
relies on the good photo-$z$s for BCGs, which may be risky at very
high redshift where photo-$z$ precision is not guaranteed. The current
GMBCG implementation relies on the photo-$z$ to decide the color to
search for the red sequence.  This is not a serious issue for the
current data set, but will be preferable to perform a more
comprehensive analysis on color space spanned by all colors. These are
beyond the scope of this paper and additional improvements are left to
future work on deeper data, such as SDSS co-added data and the
incoming Dark Energy Survey data~\citep{des}.

\bibliography{gmbcg}

\section*{Acknowledgments}
JH and TM gratefully acknowledge support from NSF grant AST 0807304
and DoE Grant DE-FG02-95ER40899. JH thanks Brian Nord, Jeffery Kubo, Marcelle Soares-Santos and Heinz Andernach for helpful conversation. AEE acknowledges support from NSF AST-0708150 and NASA NNX10AF61G. This work was supported in part by a Department of Energy contract DE-AC02-76SF00515. This project was made possible by
workshops support from the Michigan Center for Theoretical Physics.

Funding for the SDSS and SDSS-II has been provided by the Alfred P.
Sloan Foundation, the Participating Institutions, the National
Science Foundation, the U.S. Department of Energy, the National
Aeronautics and Space Administration, the Japanese Monbukagakusho,
the Max Planck Society, and the Higher Education Funding Council for
England. The SDSS Web Site is http://www.sdss.org/.

The SDSS is managed by the Astrophysical Research Consortium for the
Participating Institutions. The Participating Institutions are the
American Museum of Natural History, Astrophysical Institute Potsdam,
University of Basel, University of Cambridge, Case Western Reserve
University, University of Chicago, Drexel University, Fermilab, the
Institute for Advanced Study, the Japan Participation Group, Johns
Hopkins University, the Joint Institute for Nuclear Astrophysics,
the Kavli Institute for Particle Astrophysics and Cosmology, the
Korean Scientist Group, the Chinese Academy of Sciences (LAMOST),
Los Alamos National Laboratory, the Max-Planck-Institute for
Astronomy (MPIA), the Max-Planck-Institute for Astrophysics (MPA),
New Mexico State University, Ohio State University, University of
Pittsburgh, University of Portsmouth, Princeton University, the
United States Naval Observatory, and the University of Washington.

\label{lastpage}

\end{document}